\documentclass{cmslatex}
\usepackage{latexsym, amssymb, enumerate, amsmath}
\usepackage{graphicx}
    \graphicspath{{./pix/}}
\usepackage{subfigure}

\renewcommand{\theequation}{\arabic{section}.\arabic{equation}}

\begin{document}

\title{The adaptive patched cubature filter and its implementation}

\author{
Wonjung Lee
\thanks {
Stochastic Analysis Group and 
Oxford Centre for Collaborative Applied Mathematics (OCCAM),
Mathematical Institute
and
Oxford-Man Institute of Quantitative Finance,
University of Oxford, 
U.K.
(leew@maths.ox.ac.uk).
}
\and Terry Lyons
\thanks {
Stochastic Analysis Group,
Mathematical Institute 
and
Oxford-Man Institute of Quantitative Finance,
University of Oxford,
U.K. 
(tlyons@maths.ox.ac.uk).
}
}

\pagestyle{myheadings} 
\markboth{The adaptive patched cubature filter and its implementation}{Wonjung Lee and Terry Lyons}
\maketitle

\begin{abstract}
There are numerous contexts where one wishes to describe the state of a randomly evolving system. Effective solutions combine models that quantify the underlying uncertainty with available observational data to form scientifically reasonable estimates for the uncertainty in the system state. Stochastic differential equations are often used to mathematically model the underlying system. 

The Kusuoka-Lyons-Victoir (KLV) approach is a higher order particle method for approximating the weak solution of a stochastic differential equation that uses a weighted set of scenarios to approximate the evolving probability distribution to a high order of accuracy. The algorithm can be performed by integrating along a number of carefully selected bounded variation paths. The iterated application of the KLV method has a tendency for the number of particles to increase. This can be addressed and, together with local dynamic recombination, which simplifies the support of discrete measure without harming the accuracy of the approximation, the KLV method becomes eligible to solve the filtering problem in contexts where one desires to maintain an accurate description of the ever-evolving conditioned measure. 

In addition to the alternate application of the KLV method and recombination, we make use of the smooth nature of the likelihood function and high order accuracy of the approximations to lead some of the particles immediately to the next observation time and to build into the algorithm a form of automatic high order adaptive importance sampling. 

We perform numerical simulations to evaluate the efficiency and accuracy of the proposed approaches in the
example of the linear stochastic differential equation driven by three dimensional Brownian motion. Our
numerical simulations show that, even when the sequential Monte-Carlo methods poorly perform, the KLV method and recombination can together be used to approximate higher order moments of the filtering solution in a moderate dimension with high accuracy and efficiency.
\end{abstract}


\section{Introduction}
Filtering
is an approach for
calculating the probability distribution of an evolving system 
in the presence of
noisy observations. 
The problem
has many significant and practical applications in science and engineering,
for example
navigational and guidance systems, radar tracking, sonar ranging, 
satellite and airplane orbit determination,
the spread of hazardous plumes or pollutants,
prediction of weather and climate in atmosphere-ocean dynamics
\cite{Kalman60, 
kalman1961new,
  kushner1967approximations, 
  jazwinski1970stochastic, 
  gelb1974applied,
anderson1979optimal, 
gordon1993novel,
evensen2009data}.
If both the underlying system 
and the observation process satisfy linear equations,
the solution of the filtering problem 
can be obtained from the Kalman filter
\cite{Kalman60,
kalman1961new}.
For nonlinear filtering problems in finite dimension,
there occasionally exist analytic solutions
but the results are too narrow in applicability
\cite{benevs1981exact}.
As a result,
a number of 
numerical schemes 
have been developed
toward an aim
to accurately describe
the fundamental object of interest in filtering, i.e.,
the conditioned measure,
in terms of
collection of weighted Dirac masses
\cite{gordon1993novel,
evensen2009data,
doucet2001sequential}.

When the underlying dynamics is a continuous process and the available observations
are intermittent in time, 
the standard approach of filtering is 
to perform a forward uncertainty quantification
and then to 
incorporate data into
this 
predicted measure
using Bayes' rule
in a sequential fashion.
The 
former
prediction step
corresponds
to solving the Kolmogorov forward equation
when the system is driven by Brownian motions.
For the numerical integration of a stochastic differential equation,
the sequential Monte-Carlo method
uses sampling from random variables
whose distribution agrees with 
the law of the
truncated
strong Taylor expansion of 
the solution of an Ito diffusion.
The algorithm usually gives lower order strong convergence
of the probability distribution
\cite{kloeden2011numerical}.

Instead of randomly simulating Wiener measure as in the sequential Monte-Carlo method,
the KLV method
at the path level
replaces Brownian motion
by a weighted combination of bounded variation paths
while making sure that expectations of the 
iterated integrals 
with respect to
these two measures on Wiener space agree up to a certain degree. 
Then the particles are 
pushed 
forward 
along the 
deterministically chosen
paths
to yield a weighted discrete measure.
The KLV method 
is
of higher order with effective and transparent error bounds
obtained from
the Stratonovich-Taylor expansion of the solution of a stochastic differential equation
\cite{lyons2004cubature}.

It is intrinsic to the KLV method that the number of particles increases 
when the algorithm is iterated.
Therefore
its successive application
without an efficient suppression of the growth of the number of particles
cannot be used to filter the ever-evolving dynamics.
Given a family of test functions, one can replace the original discrete measure 
by a simpler measure with smaller support 
whose integrations against these test functions agree with those against the original measure.
Recombination achieves the reduction of particles 
in this way
using the polynomials as test functions
\cite{litterer2012high}.
One advantage of recombination is its local applicability in space.
Therefore one can divide the set of particles into 
a number of disjoint subsets and recombine 
each 
subset of
discrete measure separately,
a process which we call the patched recombination.
The dynamic property of patched recombination,
if an efficient classification method is provided,
leads to
a competitive high order reduction algorithm
whose error bound
can be obtained from the
Taylor expansion of the test function.

One can use 
the alternate application of
the KLV method
and patched recombination 
as an algorithm for 
the prediction step
in filtering.
However the cost of this non-adaptive method
would become extremely high particularly in high dimension.
Therefore
we 
further
modify 
the algorithm so that it can
significantly reduce the computational efforts.
More precisely,
we exploit the internal smoothness of the likelihood 
to allow some particles to immediately leap to the next observation time
provided 
certain conditions are fulfilled.
The bootstrap reweighting is subsequently applied 
to obtain our non Monte-Carlo particle approximation of the optimal filter.

The paper is organised as follows.
Section~\ref{sec:BF}
introduces 
the filtering problem and 
the Bayesian filter as its formal solution.
In section~\ref{sec:PF},
a prototypical sequential Monte-Carlo filtering algorithm and 
one of its 
clever variants that adapts importance sampling
are 
described.
The rest of the paper is devoted to
develop two non Monte-Carlo particle filtering algorithms
that retain the strengths and mitigate the weaknesses of 
these 
existing
Monte-Carlo methods.
In order to do that,
two essential
building blocks,
cubature measure on 
(infinite dimensional)
Wiener space
and
cubature measure on 
a finite dimensional space,
are introduced
in sections~\ref{sec:PCF}
and 
\ref{sec:simplification},
respectively.
In section~\ref{sec:PCFAPCF}
we define 
the main algorthms
and
in section~\ref{sec:numerical}
we perform
numerical simulations
to validate
the algorithms.
Concluding discussions are in section~\ref{sec:discussion}.

\section{Bayesian filter}
\label{sec:BF}
Suppose that
the 
$N$-dimensional
underlying Markov process 
$X(t),$ $t \in \mathbb{R}^+ \cup \{0\}$,
and
the
$N'$-dimensional
observation process
$Y_n$, $n\in \mathbb{N}$,
associated with 
$X_n = X(n T)$
are given,
for some inter-observation time $T>0$.
Let $Y_{1:n'} \equiv \lbrace Y_1,\cdots,Y_{n'} \rbrace$ 
be the path of the observation process
and
$y_{1:n'} \equiv \lbrace y_1,\cdots,y_{n'} \rbrace$ 
be a generic point in the space of paths.
We define
the measure of the conditioned variable
$X_n | Y_{1:n'}$ 
by
$\pi_{n|n'}(dx_n) = \mathbb{P}(X_n\in dx_n \vert Y_{1:n'}=y_{1:n'})$.
Assuming
the law of $X(0)$ is given,
filtering is to find $\pi_{n|n}$ for all $n \geq 1$.

This intermittent data assimilation problem can in principle be solved by
the alternate application of the prediction,
to obtain the prior measure ${\pi}_{n|n-1}$ from
${\pi}_{n-1|n-1}$,
and 
the updating,
to obtain
the posterior measure ${\pi}_{n|n}$ from ${\pi}_{n|n-1}$.
If the transition kernel $K(dx_n |x_{n-1})$ 
and the likelihood function $g(y_n| x_n)$,
satisfying
\begin{equation*}
  \begin{split}
	\mathbb{P}(X_n \in A\vert X_{n-1}=x_{n-1}) & =\int_A K(dx_n\vert x_{n-1}), \\
	\mathbb{P}(Y_n \in B\vert X_{n}=x_{n}) &=\int_B g(y_n\vert x_{n})\,dy_n,
  \end{split}
\end{equation*}
for all
$A\in \mathcal{B}(\mathbb{R}^N)$, the Borel $\sigma$-algebra,
and
$B\in \mathcal{B}(\mathbb{R}^{N'})$,
are given,
the prediction and the updating are achieved by
\begin{align}
\pi_{n|n-1}(dx_n) 
& =\int K(dx_{n}|x_{n-1}) \pi_{n-1|n-1}(dx_{n-1}),
	\label{eq:pred}
\\
\pi_{n|n}(dx_n)
& =\frac{g\left(y_n|x_n\right) \pi_{n|n-1}(dx_n)}{\int g\left(y_n|x_n\right) \pi_{n|n-1}(dx_n)},
	\label{eq:bayes}
\end{align}
respectively.
Eq.~(\ref{eq:bayes}) is Bayes' rule and the recursive scheme 
(\ref{eq:pred}),
(\ref{eq:bayes})
is called a Bayesian filter.

\section{Particle filtering}
\label{sec:PF}
\subsection{Weak approximation}
The closed form of
$\pi_{n|n'}$ 
is in general not available.
In many cases the essential properties of a probability measure we are interested can
accurately be described 
by the expectation of 
test functions. 
If the class of test functions is specified, we can replace the
original measure with a simpler measure that integrates the test functions
correctly and hence still 
keeps the right properties of the original measure.
Therefore
efforts have been devoted to weakly
approximating $\pi_{n|n'}$
by finding an efficient way to compute
$\mathbb{E}(f(X_n) | Y_{1:n'}) = \int f(x_n) \pi_{n|n'}(dx_n)$
accurately for a sufficiently large class of 
$f:\mathbb{R}^N \to \mathbb{R}$.
We mention that the class of test functions is not given in the filtering problem.
However their choice is quite critical as it affects the notion of an optimal algorithm 
and controls the detailed description of the conditioned measure.

One of the methodologies
for the weak approximation
is to employ
\emph{particles}
whose locations and weights
characterise the approximation of the conditioned measure.
More precisely, a particle filter is a recursive algorithm that produces
\begin{equation}
  \label{eq:weighteddiscretemeasure}
  {\pi}^{\text{PF}}_{n|n'} = \sum_{i=1}^{M_{n|n'}} \lambda^i_{n|n'} \delta_{x^i_{n|n'}}
\end{equation}
approximating ${\pi}_{n|n'}$,
where $\delta_x$ denotes the Dirac mass centred at $x$.
One approximates
$(\pi_{n|n'},f)$
by
$({\pi}^{\text{PF}}_{n|n'},f) = \sum_{i=1}^{M_{n|n'}} \lambda^i_{n|n'} f(x^i_{n|n'})$
where
the notation $(\pi,f)=\int f(x) \pi(dx)$ is used.

\subsection{Sequential Monte-Carlo methods}
Particle approximation is widely used in Monte-Carlo frameworks.
We here introduce 
two representative algorithms, 
the 
sampling importance resampling (SIR) suggested in
\cite{gordon1993novel}
and the sequential importance sampling and resampling (SISR) algorithm
\cite{liu1998sequential, pitt1999filtering, doucet2000sequential}.
The number of particles does not have to be equal in each step,
but it is here fixed by $M_{n|n'} = M$ for simplicity
\cite{del2013mean}.

\subsubsection{Sampling importance resampling (SIR)}
The prediction step is achieved by using
$(\pi_{n|n-1},f)=(\pi_{n-1|n-1},Kf)$
from Eq.~(\ref{eq:pred}).
Given the empirical measure
${\pi}_{n-1|n-1}^{\text{SIR}}=\frac{1}{M} \sum_{i=1}^M \delta_{x^i_{n-1|n-1}}$
approximating $\pi_{n-1|n-1}$,
one performs
independent and identically distributed (i.i.d.)
sampling
$\bar{x}^{i}_{n|n-1}$ drawn from $K(dx_n| x^{i}_{n-1|n-1})$.
Then
${\pi}_{n|n-1}^{\text{SIR}} = \frac{1}{M} \sum_{i=1}^M \delta_{\bar{x}^{i}_{n|n-1}}$
is an empirical measure with respect to
$\pi_{n|n-1}$.

For the updating step,
Eq.~(\ref{eq:bayes}) implies
$(\pi_{n|n},f)=(\pi_{n|n-1},fg^{y_n})/(\pi_{n|n-1},g^{y_n})$
where the notation 
$g^{y_n}(\cdot) \equiv g(y_n| \cdot)$ is used.
Define the bootstrap reweighting operator
\begin{equation}
	\label{eq:bootstrapreweighting}
\text{REW}\left( \sum_{i=1}^{n} \kappa_i \delta_{x^i},g^{y_n}\right)
\equiv
\frac{\sum_{i=1}^{n} \kappa_i g^{y_n}({x}^i)\delta_{{x}^i}}{\sum_{i=1}^{n} \kappa_i g^{y_n}({x}^i)}
\end{equation}
then $\bar{\pi}^{\text{SIR}}_{n|n}= \text{REW}\left( \pi^{\text{SIR}}_{n|n-1},g^{y_n}\right)$
is an approximation of
$\pi_{n|n}$.

In order to prevent degeneracy in the weights,
caused by a successive application of Eq.~(\ref{eq:bootstrapreweighting}),
one approximates the weighted discrete measure
$\bar{\pi}^{\text{SIR}}_{n|n}$
by an equally weighted discrete measure
\cite{doucet2001sequential}.
Random resampling performs
$M$ independent samples $\{ x^i_{n|n} \}_{i=1}^M$ from $\bar{\pi}^{\text{SIR}}_{n|n}$.
This process might introduce
a large Monte-Carlo variation 
and work has been done to reduce the variance
\cite{carpenter1999improved,
crisan2002minimal}.
The resulting one
${\pi}_{n|n}^{\text{SIR}} = \frac{1}{M} \sum_{i=1}^M \delta_{{x}^{i}_{n|n}}$
is an empirical measure with respect to
$\pi_{n|n}$.

The SIR algorithm can be displayed by
\begin{equation}
  \label{eq:SIR}
{\pi}_{n-1|n-1}^{\text{SIR}} 
\mapsto 
{\pi}_{n|n-1}^{\text{SIR}} 
\Rightarrow
\bar{\pi}_{n|n}^{\text{SIR}} 
\to
{\pi}_{n|n}^{\text{SIR}} 
\end{equation}
where the notation $\mapsto$ is used for moving particles forward in time,
$\Rightarrow$ for reweighting and
$\to$ for random resampling.
The algorithm is very intuitive and straightforward to implement.
Further, it produces an approximation that converges toward to the truth posterior measure
as the number of particles increases
\cite{crisan2002survey}.
However, SIR might be inaccurate when 
${\pi}_{n|n-1}^{\text{SIR}}$ is far from $\pi_{n|n}$ in the sense that
bootstrap reweighting generates importance weights distributed with a high variance.
The following 
algorithm modifies SIR to 
get over this degeneracy problem to some extent.

\subsubsection{Sequential importance sampling and resampling (SISR)}
Given the unweighted measure
${\pi}_{n-1|n-1}^{\text{SISR}}=\frac{1}{M} \sum_{i=1}^M \delta_{x^i_{n-1|n-1}}$
that approximates $\pi_{n-1|n-1}$,
one performs
i.i.d. 
sampling
$\widetilde{x}^{i}_{n|n-1} \sim \widetilde{K}(dx_n| x^{i}_{n-1|n-1},y_n)$
instead of
$\bar{x}^{i}_{n|n-1} \sim K(dx_n| x^{i}_{n-1|n-1})$.
Here the new transition kernel $\widetilde{K}$ 
can depend on the instance $y_n$
and should be chosen in a way that
the distribution of 
${\pi}_{n|n-1}^{\text{SISR}}=\frac{1}{M} \sum_{i=1}^M \delta_{\widetilde{x}^i_{n|n-1}}$
is closer to $\pi_{n|n}$ than
${\pi}_{n|n-1}^{\text{SIR}}$
in the above-mentioned sense
\cite{doucet2000sequential}.

Note that
${\pi}_{n|n-1}^{\text{SISR}}$ is not distributed according to
$\pi_{n|n-1}$.
To account for the effect of this discrepancy,
the expression
\begin{equation}
  \begin{split}
& \mathbb{P}(X_{n-1} \in dx_{n-1},X_n \in dx_{n}|Y_{1:n} = y_{1:n}) \\
& \quad = \frac{ w(x_{n-1},x_n,y_n) \widetilde{K}(dx_n|x_{n-1},y_n)\pi_{n-1|n-1}(dx_{n-1}) }
  { \int w(x_{n-1},x_n,y_n) \widetilde{K}(dx_n|x_{n-1},y_n)\pi_{n-1|n-1}(dx_{n-1}) }
  \end{split}
	\label{eq:marginal}
\end{equation}
where
\begin{equation*}
  w(x_{n-1},x_n,y_n) \propto 
  \frac{g (y_n|x_n) K(dx_n|x_{n-1})}{\widetilde{K}(dx_n|x_{n-1},y_n) }
\end{equation*}
is used.
Replacing $\widetilde{K}(dx_n|x_{n-1},y_n)\pi_{n-1|n-1}(dx_{n-1})$ in 
Eq.~(\ref{eq:marginal}) by its empirical approximation and
integrating over $x_{n-1}$,
one obtains
$\widetilde{\pi}_{n|n}^{\text{SISR}} = \sum_{i=1}^M w^i \delta_{ \widetilde{x}^i_{n|n-1} }$ 
where $w^i \propto  w(x^i_{n-1|n-1},\widetilde{x}^i_{n|n-1},y_n)$.
A random resampling from
$\widetilde{\pi}_{n|n}^{\text{SISR}}$
yields 
the empirical measure 
with respect to $\pi_{n|n}$, denoted by
${\pi}_{n|n}^{\text{SISR}}$.

If $\widetilde{K}(dx_n|x_{n-1},y_n)$ and $w(x_{n-1},x_n,y_n)$ have better theoretical properties
than ${K}(dx_n|x_{n-1})$ and $g(y_n|x_n)$ such as 
better mixing properties of 
$\widetilde{K}(dx_n|x_{n-1},y_n)$ 
or flatter likelihood,
then the algorithm can produce a better approximation.
Because one needs to integrate an evolution equation of a Markov process with transition kernel 
$\widetilde{K}$ in any practical implementation, 
designing efficient particle filtering methods is equivalent to 
building an appropriate dynamic model 
that has good theoretical properties while keeping the same filtering distributions. 
The SISR algorithm
\begin{equation}
  \label{eq:SIS}
{\pi}_{n-1|n-1}^{\text{SISR}} 
\mapsto 
{\pi}_{n|n-1}^{\text{SISR}} 
\Rightarrow
\widetilde{\pi}_{n|n}^{\text{SISR}} 
\to
{\pi}_{n|n}^{\text{SISR}} 
\end{equation}
might use fewer particles 
than SIR
to achieve 
a similar accuracy
\cite{van2010nonlinear}.
One can find a considerable study illustrating the difference in performance of 
SISR
using different proposal distributions 
in \cite{del2004genealogical}.

\section{Kusuoka-Lyons-Victoir (KLV) method}
\label{sec:PCF}
Suppose that
a random vector $X(t) \in \mathbb{R}^N$ 
evolves according to a Stratonovich 
stochastic differential equation (SDE)
\begin{equation}
  \label{eq:dynamics}
  dX(t)  = V_0(X(t))\,dt +\sum_{i=1}^d V_i(X(t)) \circ dW_i(t)
\end{equation}
where $\{ V_i 
\in C_b^\infty( \mathbb{R}^N,\mathbb{R}^{N})
\}_{i=0}^d$ 
is a family of smooth vector fields 
from $\mathbb{R}^N$
to $\mathbb{R}^N$
with bounded derivatives of all orders,
and $W =(W_1,\cdots, W_d)$ denote a set of Brownian motions, independent of one another.
The KLV method enables to deterministically approximate the law of $X(T)$
in terms of discrete measure.

\subsection{Cubature on Wiener space on path level}
\label{sec:cubonws}
Let us use the notations
$W_0(t) = t$,
$\omega_{T,0}(t) = t$
and 
$I = (i_1,\cdots,i_l) \in \{0,\cdots,d\}^l$.
Consider 
the iterated integral with respect to 
$W =(W_1,\cdots, W_d)$,
\begin{equation*}
\mathcal{J}^I_{0,T} (\circ W) 
\equiv \int_{0<t_1<\cdots<t_l<T} \circ\, dW_{i_1}(t_{1}) \cdots \circ dW_{i_l}(t_{l}),
\end{equation*}
and 
the iterated integral with respect to
a continuous path of bounded variation
$\omega_T=(\omega_{T,1},\cdots,\omega_{T,d}): [0,T] \to \mathbb{R}^d$,
\begin{equation*}
\mathcal{J}^I_{0,T} (\omega_{T}) 
 \equiv \int_{0<t_1<\cdots<t_l<T} d\omega_{T,i_1}(t_{1}) \cdots d\omega_{T,i_l}(t_{l}).
\end{equation*}
Recall that Wiener space
$C^0_{0}\left( [0,T], \mathbb{R}^{d}\right)$ is
the set of continuous functions starting at zero.
We define
a discrete measure 
$\mathbb{Q}^{m}_T = \sum_{j=1}^{n_m} \lambda_j \delta_{\omega^{j}_{T}}$
supported on continuous paths of bounded variation 
to be a cubature on Wiener space on path level of degree $m$
with respect to the Wiener measure
$\mathbb{P}$
provided the equation
\begin{align}
  \label{eq:momentmatching}
  \mathbb{E}_\mathbb{P} \left( \mathcal{J}^I_{0,T}(\circ W)  \right)
& = \mathbb{E}_{\mathbb{Q}^{m}_T} \left( \mathcal{J}^I_{0,T}(\circ W)  \right) \nonumber \\
& = \sum_{j=1}^{n_m} \lambda_j \mathcal{J}^I_{0,T}(\omega^{j}_{T})
\end{align}
holds for all 
$I$
satisfying
$||I|| \equiv l+\text{card}\{j,i_j=0\} \leq m$.
Note
$\mathbb{Q}^{m}_T$ is obtained from $\mathbb{Q}^{m}_1$ via a suitable rescaling
and
the existence of
$\mathbb{Q}^{m}_{1}$
with $n_m \leq \text{card} \{ I: \|I \| \leq m \}$
is proved in
\cite{lyons2004cubature}.

The cubature measure on Wiener space 
can be used 
to approximate 
$\mathbb{E}_\mathbb{P}(f(X^x_{T}))$
for the random process $X^x_{t}$ in $N$ dimension 
satisfying
\begin{equation}
  \label{eq:dynamic}
dX^x_{t} = V_0(X^x_{t})\,dt + \sum_{i=1}^d V_i(X^x_{t}) \circ dW_i(t)
\end{equation}
and $X^x_{0}=x$.
The expectation of $f(X^x_T)$ against Wiener measure
can be viewed as an integral with respect to infinite dimensional Wiener space.

Let $t \mapsto X_t^{x,\omega^{j}_{\Delta}}$ for $t \in [0,\Delta]$ 
be the deterministic process satisfying
\begin{equation}
  \label{eq:nonauto}
dX_{t}^{x,\omega^{j}_\Delta} = \sum_{i=0}^d V_i(X_{t}^{x,\omega^{j}_\Delta}) \,d\omega^{j}_{\Delta,i}(t)
\end{equation}
and $X_{0}^{x,\omega^{j}_{\Delta}}=x$.
The ordinary differential equations (ODEs) of Eq.~(\ref{eq:nonauto})
are obtained from replacing the Brownian motions
$W$ 
in Eq.~(\ref{eq:dynamic})
by the bounded variation path $\omega^{j}_{\Delta}$.
The measure
$\sum_{j=1}^{n_m} \lambda_j \delta_{X_{T}^{x,\omega^{j}_{T}}}$
on $\mathbb{R}^N$ is called
the cubature approximation 
of the law of $X^x_T$ at the path level.

An error estimate for 
the weak approximation of this particle method
can be derived from
the Stratonovich-Taylor expansion
of a smooth function $f$,
\begin{equation}
  \label{eq:staylor}
   f(X_T^x) =
   \sum_{||I||\leq m} V_If(x)
	\mathcal{J}^I_{0,T}(\circ W)
   +R_m(x,T,f)
\end{equation}
where the remainder $R_m(x,T,f)$ satisfies
\begin{equation}
  \label{eq:remainder}
   \sup_{x\in \mathbb{R}^N} 
    \sqrt{\mathbb{E}_{\mathbb{P}}(R_m(x,T,f)^2)} 
 \leq C \sum_{i=m+1}^{m+2} T^{i/2} \sup_{\parallel I \parallel= i}  \parallel V_If \parallel_\infty 
\end{equation}
for a constant $C$ depending on $d$ and $m$
\cite{kloeden2011numerical}.
Here
the vector field $V_i=(V_{i,1},\cdots,V_{i,N})$
is used as the differential operator
$V_i \equiv \sum_{j=1}^N V_{i,j} \partial x_j$
and 
$V_I$ denotes $V_{i_1} \cdots V_{i_k}$.

The process $R_m(x,T,f)$ further satisfies
\begin{equation}
  \label{eq:cubremainder}
   \sup_{x\in \mathbb{R}^N} 
    {\mathbb{E}_{\mathbb{Q}^m_T}( \lvert R_m(x,T,f) \rvert )} 
 \leq C \sum_{i=m+1}^{m+2} T^{i/2} \sup_{\parallel I \parallel= i}  \parallel V_If \parallel_\infty 
\end{equation}
for a constant $C$ depending on $d$, $m$ and $\mathbb{Q}^m_1$
\cite{lyons2004cubature}.
Let the operators 
$P_T$ and $Q^m_T$
be defined 
by
$P_{T}f(x)\equiv  \mathbb{E}_\mathbb{P}(f(X^x_{T}))$
and 
$Q^{m}_{T}f(x) \equiv \mathbb{E}_{\mathbb{Q}^{m}_T}(f(X^x_{T}))$.
Then the error bound of the cubature approximation at the path level
is given by
\begin{align}
	\label{eq:errorbound}
  \sup_{x \in \mathbb{R}^N} 
  \left\lvert 
\mathbb{E}_\mathbb{P}(f(X^x_{T}))
-\sum_{j=1}^{n_m} \lambda_j f(X_{T}^{x,\omega^{j}_{T}}) 
  \right\rvert \nonumber 
& = 
\parallel  (P_T-Q^{m}_T)f \parallel_\infty \\
& \leq C \sum_{i=m+1}^{m+2} T^{{i}/{2}}  \sup_{\parallel I \parallel= i} \parallel V_I f \parallel_\infty
\end{align}
for smooth $f$,
from
Eq.~(\ref{eq:momentmatching})
and
Eqs.~(\ref{eq:staylor}),
(\ref{eq:remainder}),
(\ref{eq:cubremainder}).

The algorithm
was developed by Lyons, Victoir
\cite{lyons2004cubature},
following the work of Kusuoka
\cite{kusuoka2001approximation,
kusuoka2004approximation},
so it is referred to as
the KLV method.
Eq.~(\ref{eq:errorbound}) leads to define 
\begin{equation}
  \label{eq:klvop}
\text{KLV}^{(m)}
  \left(\sum_{i=1}^n \kappa_i \delta_{x^i}, \Delta \right)
  \equiv \sum_{i=1}^n \sum_{j=1}^{n_m} \kappa_i \lambda_j \delta_{X_{\Delta}^{x^i, \omega^{j}_{\Delta}}}
\end{equation}
that may be interpreted as a Markov operator acting on discrete measure on $\mathbb{R}^N$.

In the following, assume $T\in (0,1)$ for simplicity.
One may take a higher degree $m$ 
to achieve a given degree of accuracy
in Eq.~(\ref{eq:errorbound}).
An alternative method to improve the accuracy of the particle approximation 
is a successive application of the KLV operator.
Let 
$\mathcal{D} = \{ 0=t_0 < t_1 < \cdots < t_k = T \}$
be a partition of $[0,T]$ with $s_j=t_j-t_{j-1}$.
Instead of
$Q^m_{T}f(x) = ({\text{KLV}^{(m)}\left(\delta_x,T \right)},f)$,
the value of $P_Tf(x) = P_{s_1}P_{s_2} \cdots P_{s_k}f(x)$
can accurately be approximated 
by a multiple step algorithm
$Q^m_{s_1}Q^m_{s_2} \cdots Q^m_{s_k}f(x)$.

Given a discrete measure ${\mu^0}$,
we define a sequence of discrete measure by
\begin{equation}
  \begin{split}
	\label{eq:sequentialklv}
\Phi_{\mathcal{D}}^{m,0}({\mu^0}) &={\mu^0},  \\
\Phi_{\mathcal{D}}^{m,j}({\mu^0})
&=\text{KLV}^{(m)}(\Phi_{\mathcal{D}}^{m,j-1}({\mu^0}),s_j) \quad 1\leq j\leq k
  \end{split}
\end{equation}
that can be viewed as Markov chain.
The inequality
\begin{align}
  \label{eq:inequality1}
	 \left\lvert P_Tf(x) -(\Phi^{m,k}_{\mathcal{D}}(\delta_x),f) \right\rvert  \nonumber 
	& = \left\lvert \sum_{j=1}^k \left(\Phi^{m,j-1}_{\mathcal{D}}(\delta_x),P_{T-{t_{j-1}}}f \right)- 
	\left(\Phi^{m,j}_{\mathcal{D}}(\delta_x),P_{T-{t_{j}}}f \right) \right\rvert \nonumber \\
	& = \left\lvert \sum_{j=1}^k \left(\Phi^{m,j-1}_{\mathcal{D}}(\delta_x),
	(P_{s_j}-Q^{m}_{s_j}) P_{T-{t_{j}}}f \right) \right\rvert \nonumber \\
	& \leq \sum_{j=1}^k \parallel (P_{s_j}-Q^{m}_{s_j})P_{T-t_j}f \parallel_\infty
\end{align}
obtained from the Markovian property of the KLV operator
shows that
the total error of the repeated 
KLV application
is bounded above by the sum of 
the errors over the subintervals in the partition.
Applying Eq.~(\ref{eq:errorbound})
to estimate the upper bound of
Eq.~(\ref{eq:inequality1}),
we need $P_{T-t_j}f$ to be smooth.
When $f$ is smooth,
this is true and 
the error bound
\begin{equation*}
\sup_{x \in \mathbb{R}^N} 
\left\lvert 
P_Tf(x)
-({\Phi^{m,k}_{\mathcal{D}}(\delta_x)},f) \right\rvert 
\leq C \sum_{i=m+1}^{m+2}\sum_{j=1}^k s_j^{{i}/{2}} 
 \sup_{\parallel I \parallel = i}  \parallel V_IP_{T-t_j}f \parallel_\infty 
\end{equation*}
is obtained from
Eqs.~(\ref{eq:errorbound}), (\ref{eq:inequality1}).

The case of Lipschitz continuous $f$ is of particular interest
because 
$P_tf$ is indeed smooth 
in the direction of $\{V_i\}_{i=0}^d$
with additional conditions for these vector fields
\cite{crisan2006convergence}.
In this case,
the regularity estimate
\begin{equation}
  \label{eq:ugf}
\parallel V_I P_tf \parallel_\infty 
\leq \frac{C}{t^{(\parallel I\parallel-1)/2}}
\parallel \nabla f \parallel_\infty
\end{equation}
holds
for all $t\in (0,1]$,
where $C$ is a constant independent of 
$f$
\cite{kusuoka1987applications, kusuoka2003malliavin}.
Combining Eqs.~(\ref{eq:errorbound}), (\ref{eq:inequality1})
and Eq.~(\ref{eq:ugf}),
we obtain an error estimate for the KLV method in terms of the gradient of 
the Lipschitz continuous $f$,
\begin{align}
	  \label{eq:totalerror}
 \sup_{x \in \mathbb{R}^N} \left\lvert 
P_Tf(x)
-({\Phi^{m,k}_{\mathcal{D}}(\delta_x)},f) \right\rvert 
\leq C
\parallel \nabla f \parallel_\infty 
\left(
s_k^{{1}/{2}}+\sum_{i=m+1}^{m+2}\sum_{j=1}^{k-1} 
\frac{s_j^{{i}/{2}}}{(T-{t_j})^{{(i-1)}/{2}}} 
\right)
\end{align}
where $C$ is a constant independent of $k$.
Here the final term in the upper bound of
Eq.~(\ref{eq:inequality1})
is estimated by
$\parallel (P_{s_k}-Q^m_{s_k})f \parallel_\infty 
\leq \parallel P_{s_k}f-f \parallel_\infty 
+ \parallel f-Q^m_{s_k}f \parallel_\infty 
\leq Cs_k^{1/2}\parallel \nabla f \parallel_\infty$
using
the boundedness of 
$\{V_i\}_{i=0}^d$.

Let $\mathcal{D}(\gamma) = \{ t_j\}_{j=0}^k$ be
the Kusuoka partition
\cite{kusuoka2001approximation}
given by
\begin{equation}
  \label{eq:Kpartition}
 	t_j= T\left( 1-\left( 1-\frac{j}{k}\right)^\gamma\right)
\end{equation}
then the error estimate
\begin{equation}
  \label{eq:KLVconvergece}
 \sup_{x \in \mathbb{R}^N} 
 \left\lvert 
 P_Tf(x)
-({\Phi^{m,k}_{\mathcal{D}(\gamma)}(\delta_x)},f) \right\rvert  
 \leq 
C \parallel \nabla f \parallel_\infty 
T^{1/2}
k^{-(m-1)/2}
\end{equation}
is satisfied
for a Lipschitz continuous $f$
when $\gamma>m-1$.

Eq.~(\ref{eq:KLVconvergece}) is
obtained from substituting 
the non-equidistant time discretisation
$\mathcal{D}(\gamma)$
into
Eq.~(\ref{eq:totalerror}).
Using this particular choice of partition ensures that the bound of the KLV error is of high order
in the number of iterations $k$.

Before concluding this subsection,
we here mention that
$u(x,t) \equiv \mathbb{E}_\mathbb{P}(f(X^x_{T-t}))$
satisfies the partial differential equation (PDE)
\begin{equation}
  \begin{split}
  \label{eq:parabolicpde}
  \frac{\partial}{\partial t}u(x,t) & = - \left(V_0 + \frac{1}{2}\sum_{i=1}^d V_i^2 \right)u(x,t), \\
  u(x,T) & =f(x).
  \end{split}
\end{equation}
where $\{V_i\}_{i=0}^d$ are used as differential operators
\cite{watanabe1981stochastic}.
Therefore $P_Tf(x)$, the heat kernel applied to $f$, is equal 
to the solution $u(x,0)$ of
Eq.~(\ref{eq:parabolicpde}).
Due to this inherent relationship between parabolic PDEs and SDEs,
one can apply any well-known algorithm for the solution of 
Eq.~(\ref{eq:parabolicpde})
to the prediction step of the filtering problem
where the underlying system is given by
Eq.~(\ref{eq:dynamics}).
However
it is very important to understand the critical difference between these two problems.
One needs to weakly approximate 
the law of $X(T)$,
when $X(0)$ is given by $\delta_x$,
that accurately integrate the test function $f$
for the PDE problem
while the filtering problem requires one to approximate the conditioned measure of
$X_{n}|Y_{1:n}$
for all $n \geq 1$,
in which the test function 
is not at all specified.

\subsection{Cubature on Wiener space on flow level}
\label{sec:flows}
We here 
study the construction of cubature formula
$\mathbb{Q}^{m}_T$.
Meanwhile
cubature on Wiener space on flow level
is defined
in terms of Lie polynomial
and 
used to develop an approximation 
based on the autonomous ODEs at flow level.

Let $\{e_i\}_{i=0}^d$
be the standard basis of $\mathbb{R} \oplus \mathbb{R}^d$.
Let $\mathcal{T}$ denote the associative and non-commutative tensor algebra 
of polynomials
generated by $\{e_i\}_{i=0}^d$.
The exponential and logarithm on $\mathcal{T}$ are defined by
\begin{equation}
  \begin{split}
	\label{eq:explog}
	\text{exp}(a) & \equiv \sum_{i = 0}^\infty \frac{{a}^{\otimes i}}{i !}, \\
  \text{log}(a) & \equiv \log(a_0)+\sum_{i = 1}^\infty
  \frac{(-1)^{i-1}}{i} \left(\frac{a}{a_0} -1 \right)^{\otimes i},
  \end{split}
\end{equation}
where $a=\sum_{I} a_I e_I$
and
$ e_I = e_{i_1}\otimes \cdots \otimes e_{i_l}$
for a multi-index
$I = (i_1,\cdots,i_l) \in \{0,\cdots,d\}^l$.
Here $\otimes$ denotes the tensor product.
Let the operators
$\text{exp}^{(m)}(\cdot)$ and $\text{log}^{(m)}(\cdot)$
be defined
by the truncation
of Eq.~(\ref{eq:explog})
leaving the case $\parallel I \parallel \leq m$.

The signature of a continuous path of bounded variation
$\omega_T:[0,T]\to \mathbb{R}^d$ by
\begin{equation*}
  \begin{split}
\mathcal{S}_{0,T}(\omega_T)
& \equiv \sum_{l=0}^\infty \int_{0<t_1<\cdots <t_l<T} d\omega_{T}(t_1)\otimes\cdots\otimes
d\omega_{T}(t_l)\\
& =  \sum_{I}   
\mathcal{J}^I_{0,T} (\omega_{T})\,e_I  
  \end{split}
\end{equation*}
and similarly the signature of a Brownian motion $W$ by
\begin{equation*}
\mathcal{S}_{0,T}(\circ W) \equiv \sum_{I }   \mathcal{J}^I_{0,T} (\circ W) \,e_I.
\end{equation*}
In view of Eq.~(\ref{eq:momentmatching}),
the definition of cubature on Wiener space of degree $m$ can be rephrased by
\begin{equation}
\label{eq:signaturemomentmatching}
\mathbb{E}_{\mathbb{P}}\left( \mathcal{S}^{(m)}_{0,T}(\circ W)\right)
=
\mathbb{E}_{\mathbb{Q}_T^m}\left( \mathcal{S}^{(m)}_{0,T}(\circ W)\right)
\end{equation}
where
$\mathcal{S}^{(m)}_{0,T}(\cdot)$ 
is the truncation of 
$\mathcal{S}_{0,T}(\cdot)$ 
to the case $\parallel I \parallel \leq m$.

Define $\mathcal{L}$ to be the space of Lie polynomials, i.e.,
linear combinations of finite sequences of Lie brackets
$[e_i,e_j]=e_i\otimes e_j - e_j \otimes e_i$.
Because
Chen's theorem ensures
that the logarithm of signature is a Lie series
\cite{reutenauer2003free},
its truncation
\begin{equation}
  \label{eq:liepoly}
\mathcal{L}_T^{j} \equiv \text{log}^{(m)}( \mathcal{S}_{0,T}(\omega^{j}_{T}) )
\end{equation}
is 
a Lie polynomial and
an element of $\mathcal{L}$.
Then the measure 
$\widetilde{\mathbb{Q}}^{m}_{T} = \sum_{j=1}^{n_m} \lambda_j \delta_{\mathcal{L}^{j}_T}$
supported on Lie polynomials
satisfies
\begin{align}
\label{eq:liepolysignaturemomentmatching}
	  \mathbb{E}_{\mathbb{P}}\left( S^{(m)}_{0,T}(\circ W)\right)
	  &= \mathbb{E}_{\widetilde{\mathbb{Q}}^m_T}\left( \text{exp}^{(m)}(\mathcal{L})  \right)
	  \nonumber \\
	  &= \sum_{j=1}^{n_m} \lambda_j \text{exp}^{(m)}(\mathcal{L}_T^{j}).
\end{align}
Conversely, for any Lie polynomials $\mathcal{L}_T^j$, 
there exists continuous bounded variation paths $\omega_T^j$
whose truncated logarithmic signature is $\mathcal{L}_T^j$.
Moreover if 
$\widetilde{\mathbb{Q}}^{m}_{T}$ satisfies
Eq.~(\ref{eq:liepolysignaturemomentmatching}),
then
${\mathbb{Q}}^{m}_{T}$ satisfies
Eq.~(\ref{eq:signaturemomentmatching}).
Therefore 
Eq.~(\ref{eq:signaturemomentmatching})
and
Eq.~(\ref{eq:liepolysignaturemomentmatching})
are equivalent.
The discrete measure
$\widetilde{\mathbb{Q}}^{m}_{T}$ is defined as 
cubature on Wiener space on flow level.

The expectation of the truncated Brownian signature is
\begin{equation}
  \label{eq:expsig}
\mathbb{E}_{\mathbb{P}}\left( \mathcal{S}^{(m)}_{0,1}(\circ W) \right) = 
\text{exp}^{(m)}\left( e_0 + 
\frac{1}{2}
\sum_{i=1}^d e_i\otimes e_i \right)
\end{equation}
which is proved
in \cite{lyons2004cubature}.
It is immediate from Eq.~(\ref{eq:expsig}) that
cubature formulae on Wiener space for $m=2n-1$ and $m=2 n$ are equal to one another.
A formula
$\{\lambda_j,\mathcal{L}^j_1\}_{j=1}^{n_m}$
satisfying Eq.~(\ref{eq:liepolysignaturemomentmatching})
is found when $m=3$ and $m=5$ for any $d$
\cite{lyons2004cubature}.
In some cases of $m \geq 7$,
cubature formula of Lie polynomial is available when $d=1, 2$
(See \cite{gyurko2011efficient}).

From 
this $\widetilde{\mathbb{Q}}^{m}_{1}$
and 
Eq.~(\ref{eq:liepoly}),
one can construct 
${\mathbb{Q}}^{m}_{1}$
(See 
\cite{lyons2004cubature,
gyurko2011efficient}).
It follows from the scaling property of the Brownian motion that
$\omega^{j}_{T,0}(t) = \omega^{j}_{1,0}(t)$ and
$\omega^{j}_{T,i}(t) = \sqrt{T}\omega^{j}_{1,i}(t/T)$ for $1 \leq i \leq d$.
The paths define
a cubature formula $\widetilde{\mathbb{Q}}^{m}_{T}$.
Using
$\mathcal{J}^I_{0,T} (\circ W) \triangleq T^{\parallel I\parallel/2}\mathcal{J}^I_{0,1} (\circ W)$
and Eq.~(\ref{eq:liepoly}),
the scaling of the Lie polynomial 
is 
$\mathcal{L}^{j}_{T} = \langle T, \mathcal{L}^{j}_{1} \rangle$
where
$\left\langle t, \sum_{I} a_I e_I \right\rangle \equiv \sum_{I} t^{\parallel I \parallel/2}a_I e_I$.
The Lie polynomials define
a cubature formula $\mathbb{Q}^m_T$.

We next study 
the approximation
based on the flows of autonomous ODEs.
It is in fact
corresponds to a version of Kusuoka's algorithm
\cite{kusuoka2001approximation}.
Let $\Gamma$ denote the algebra homomorphism generated by $\Gamma(e_i)= V_i$ for
$i=0,\cdots,d$. 
For a vector field 
$V \in C_b^\infty( \mathbb{R}^N,\mathbb{R}^{N})$,
we define 
the flow $\text{Exp}\left(tV\right)(x) \equiv \xi_t^x$
to be 
the solution of the 
ODE
$  {d\xi_t^x} = V(\xi_t^x)\,dt$
with $\xi_0^x=x$.
By interchanging the algebra homomorphism $\Gamma$ with the exponentiation (so far taken in the
tensor algebra) we arrive at an approximation operator in which the exponentiation is understood as
taking the flow of autonomous ODEs.
More precisely,
one has
\begin{equation*}
  \begin{split}
\mathbb{E}_{\mathbb{P}}\left( \Gamma \left(S^{(m)}_{0,T}(\circ W) \right)\right) f(x)
& = \sum_{j=1}^{n_m} \lambda_j \Gamma \left(\exp^{(m)}(\mathcal{L}_T^{j}) \right) f(x) \\
& \simeq \sum_{j=1}^{n_m} \lambda_j f \left( \text{Exp}\left( \Gamma (\mathcal{L}_T^{j})\right)(x) \right)
  \end{split}
\end{equation*}
using Eq.~(\ref{eq:liepolysignaturemomentmatching}).
The error introduced while interchanging $\text{exp}$ and $\Gamma$ operators
turns out to be of the similar order with the error 
in the cubature approximation of the path level
as shown below.

Formally the cubature approximation at the flow level is defined as follows.
Let $t \mapsto X_t^{x,\mathcal{L}^{j}_{\Delta}}$ for $t \in [0,1]$ 
be the deterministic process satisfying
\begin{equation}
  \label{eq:auto}
dX_{t}^{x,\mathcal{L}^{j}_\Delta} = \Gamma(\mathcal{L}^{j}_\Delta)(X_{t}^{x,\mathcal{L}^{j}_\Delta}) \,dt
\end{equation}
and 
$X_0^{x,\mathcal{L}^{j}_{\Delta}}=x$.
Define the operator 
\begin{equation}
  \label{eq:KLVflow}
\widetilde{\text{KLV}}^{(m)}
\left(\sum_{i=1}^n \kappa_i \delta_{x^i}, \Delta \right)
\equiv \sum_{i=1}^n \sum_{j=1}^{n_m} \kappa_i \lambda_j \delta_{X_{1}^{x^i, \mathcal{L}^{j}_{\Delta}}}
\end{equation}
and a sequence of discrete measure
\begin{equation*}
  \begin{split}
\widetilde{\Phi}_{\mathcal{D}}^{m,0}({\mu^0}) &={\mu^0},  \\
\widetilde{\Phi}_{\mathcal{D}}^{m,j}({\mu^0})
&=\widetilde{\text{KLV}}^{(m)}(\widetilde{\Phi}_{\mathcal{D}}^{m,j-1}({\mu^0}),s_j) 
  \end{split}
\end{equation*}
for $1 \leq j \leq k$.

Let
$\widetilde{Q}^m_{T}f(x) \equiv (\widetilde{\text{KLV}}^{(m)}(\delta_x,T),f)$
be a flow level cubature approximation,
then
the Taylor expansions of 
Eq.~(\ref{eq:nonauto})
and
Eq.~(\ref{eq:auto})
lead to
\begin{equation}
  \label{eq:flowlevelapp}
\parallel (Q^{m}_{T}-\widetilde{Q}^{m}_{T})f \parallel_\infty
\leq C \sum_{m+1 \leq \parallel I \parallel \leq 2m} T^{\parallel I \parallel/{2}} 
\parallel V_I f \parallel_\infty
\end{equation}
for a smooth $f$, where $C$ is a constant depending on $m$, $d$, 
$\mathbb{Q}^m_1$
and
$\widetilde{\mathbb{Q}}^m_1$
\cite{kusuoka2001approximation}.

The error estimate
\begin{equation}
  \label{eq:KLVconvergeceflow}
 \sup_{x \in \mathbb{R}^N} 
 \left\lvert 
 P_Tf(x)
-({\widetilde{\Phi}^{m,k}_{\mathcal{D}(\gamma)}(\delta_x)},f) \right\rvert 
\leq
C \parallel \nabla f \parallel_\infty 
T^{1/2}
k^{-(m-1)/2}
\end{equation}
is satisfied
for a Lipschitz continuous $f$
when $\gamma>m-1$.

Eq.~(\ref{eq:KLVconvergeceflow}) 
is obtained using Eq.~(\ref{eq:flowlevelapp})
and
demonstrates that
for a suitable partition
the bounds for the approximation at flow and path level have the same rate of convergence
in view of Eq.~(\ref{eq:KLVconvergece}).

\section{Simplification of particle approximation}
\label{sec:simplification}
A successive application of the KLV operator gives rise to geometric growth of the number of
particles in view of 
Eqs.~(\ref{eq:klvop}) and (\ref{eq:KLVflow}).
Except some cases of PDE problems
in which the KLV method can produce an accurate approximation
with small number of iterations,
this geometric growth of particle number
prohibits an application of the KLV method
particularly to the filtering problem where
to maintain an accurate description of the ever-evolving measure
with reasonable computational cost is 
the key requirement.
It is therefore necessary to add a simplification 
algorithm
between 
two consecutive
iterations,
which suppresses the growth of
the number of particles
in the KLV framework.
Though it is possible to achieve
the simplification
through 
one of several Monte-Carlo methods,
we here 
make use of 
cubature measure on a finite dimensional space
to efficiently reduce the support of discrete measure.
This will let the entire algorithm
consistently
step outside of the Monte-Carlo paradigm.
Furthermore, its proper applications
never harm the accuracy of the KLV approximation
as we shall see.

\subsection{Cubature on a finite dimensional space}
\label{sec:cubonRN}
Let $\nu$ be a (possibly unnormalised) positive measure on $\mathbb{R}^N$.
A discrete measure
$\widehat{\nu}^{(r)}=\sum_{j=1}^{n_r} w_j \delta_{y^j}$
is called a cubature 
(quadrature when $N=1$)
of degree $r$ 
with respect to $\nu$
provided
$\text{supp}(\widehat{\nu}^{(r)})\subseteq \text{supp}(\nu)$
and
$ (\nu,q)$ equals $(\widehat{\nu}^{(r)},q) = \sum_{j=1}^{n_r} w_j q(y^j)$
for all 
polynomials
$q$
whose total degree is less than or equal to $r$.
It is proved that 
a cubature $\widehat{\nu}^{(r)}$ with respect to an arbitrary positive measure $\nu$
satisfying 
$n_r \leq \binom{N+r}{r}$
exists
\cite{putinar1997note}.
As a result,
one can adopt
a cubature measure on $\mathbb{R}^N$ with respect to the original measure
as
the reduced measure.

Importantly,
an error bound of 
$(\nu,F)-(\widehat{\nu}^{(r)},F) \equiv (\nu-\widehat{\nu}^{(r)},F)$
for a smooth function $F:\mathbb{R}^N \to \mathbb{R}$
can be obtained from 
the Taylor expansion.
The value of 
$F$ 
at $x=(x_1,\cdots,x_N)$ is written as
\begin{equation}
  \label{eq:taylorex}
F(x) = \sum_{|\alpha| \leq r} \frac{D^\alpha F(x_0)}{\alpha!}(x-x_0)^{ \alpha}+R^r(x,x_0,F)
\end{equation}
where 
$\alpha\equiv (\alpha_1,\cdots,\alpha_N)$,
$|\alpha| \equiv \alpha_1+\cdots+\alpha_N$,
$\alpha! \equiv \alpha_1!\cdots\alpha_N!$,
$D^\alpha \equiv \partial x_1^{\alpha_1} \cdots \partial x_N^{\alpha_N}$,
$x^{ \alpha} \equiv x_1^{\alpha_1}\cdots x_N^{\alpha_N}$
and
\begin{equation}
  \label{eq:rem}
  R^r(x,x_0,F)=\sum_{|\alpha|=r+1} \frac{D^\alpha F(x^*)}{\alpha!}(x-x_0)^\alpha
\end{equation}
for some $x^* \in \mathbb{R}^N$.
If the support of $\nu$ is in a closed ball of centre $x_0$ and radius $u$,
denoted by $B(x_0,u)$,
then we have
\begin{align}
  \label{eq:taylorremainder}
\lvert(\nu-\widehat{\nu}^{(r)},F) \rvert
= \lvert (\nu-\widehat{\nu}^{(r)},R^r) \rvert  
& \leq 2 (\nu,1)\parallel R^r\parallel_{L_\infty({B(x_0,u)})} \nonumber \\
& \leq \frac{ Cu^{r+1}}{(r+1)!} \sup_{|\alpha|=r+1} {\parallel D^{\alpha}F
\parallel_{L_\infty({B(x_0,u)})}}.
\end{align}
Eq.~(\ref{eq:taylorremainder}) reveals that
cubature on a finite dimensional space 
is an approach for the numerical integration of functions on finite dimensional space
with a clear error bound.

\subsection{Local dynamic recombination}
\label{sec:recomb}
Instead of using a cubature of higher degree 
to reduce the entire family of particles all at once,
we 
improve the performance
by dividing a given discrete measure into locally supported unnormalised 
positive measures
and replacing each separated measure by the cubature of lower degree
\cite{litterer2012high}.
This so-called local dynamic recombination can be a competitive algorithm 
because each reduction can be performed in a parallel manner to save computational time
and the error bound from the Taylor approximation remains of higher order. 

Let $U=(U_i)_{i=1}^R$ be a collection of balls of radius $u$
that covers the support of discrete measure $\mu$ on $\mathbb{R}^N$,
then one can find unnormalised measures $(\mu_i)_{i=1}^R$ such that $\mu=\bigsqcup_{i=1}^R \mu_i$
($\mu_i$ and $\mu_j$ have disjoint support for $i\neq j$)
and $\text{supp}(\mu_i)\subseteq U_i \cap \text{supp}(\mu)$.
In this case, 
we define the patched recombination operator by
\begin{equation}
  \label{eq:recomb}
\text{REC}^{(u,r)}
\left( \mu \right)
\equiv
\bigsqcup_{i=1}^R 
\widehat{\mu}^{(r)}_i
\end{equation}
where
$\widehat{\mu}^{(r)}_i$ denotes a cubature of degree $r$ with respect to $\mu_i$.

Given a discrete measure ${\mu^0}$,
we define a sequence of discrete measure by
\begin{equation}
  \begin{split}
  \label{eq:sequenceofdiscretemeasure}
\Phi^{m,0}_{\mathcal{D},(u,r)}({\mu^0}) & ={\mu^0}, \\
\widehat{\Phi}_{\mathcal{D},(u,r)}^{m,j-1}({\mu^0}) & = 
\text{REC}^{(u_{j-1},r_{j-1})}\left( {\Phi}_{\mathcal{D},(u,r)}^{m,j-1}({\mu^0}) \right), \\
\Phi_{\mathcal{D},(u,r)}^{m,j}({\mu^0}) &= \text{KLV}^{(m)}
\left(\widehat{\Phi}_{\mathcal{D},(u,r)}^{m,j-1}({\mu^0}),s_j \right),
  \end{split}
\end{equation}
for $1 \leq  j \leq k$.
An application of Eq.~(\ref{eq:sequenceofdiscretemeasure})
with initial condition $\delta_x$
yields a weak approximation for the law of $X^x_T$.
One obtains the estimate
\begin{align}
	  \label{eq:errorineq}
\left\lvert P_Tf(x) -(\Phi^{m,k}_{\mathcal{D},(u,r)}(\delta_x),f) \right\rvert  
& = \Bigg\lvert \sum_{j=1}^k
\left(\widehat{\Phi}^{m,j-1}_{\mathcal{D},(u,r)}(\delta_x),P_{T-{t_{j-1}}}f\right)- 
\left({\Phi}^{m,j}_{\mathcal{D},(u,r)}(\delta_x),P_{T-{t_{j}}}f \right) \nonumber \\
& \quad\,\, + \left({\Phi}^{m,j-1}_{\mathcal{D},(u,r)}(\delta_x),P_{T-{t_{j-1}}}f \right)- 
\left(\widehat{\Phi}^{m,j-1}_{\mathcal{D},(u,r)}(\delta_x),P_{T-{t_{j-1}}}f \right) \Bigg\rvert
\nonumber \\
& = \Bigg\lvert \sum_{j=1}^k
\left(\widehat{\Phi}^{m,j-1}_{\mathcal{D},(u,r)}(\delta_x),
(P_{s_j}-Q^m_{s_j}) P_{T-{t_{j}}}f \right)  \nonumber\\
& \quad\,\, + \left({\Phi}_{\mathcal{D},(u,r)}^{m,j-1}(\delta_x)
-\widehat{\Phi}_{\mathcal{D},(u,r)}^{m,j-1}(\delta_x),P_{T-t_{j-1}}f \right)
  \Bigg\rvert \nonumber\\
& \leq \sum_{j=1}^k \parallel (P_{s_j}-Q^m_{s_j})P_{T-t_j}f \parallel_\infty\nonumber\\
 & \quad \,\, +\sum_{j=0}^{k-1} 
 \left\lvert  
\left({\Phi}_{\mathcal{D},(u,r)}^{m,j}(\delta_x)
-\widehat{\Phi}_{\mathcal{D},(u,r)}^{m,j}(\delta_x),P_{T-t_{j}}f\right)
 \right\rvert 
\end{align}
where the first sum of the upper bound is
due to the KLV approximation.
The second sum is the error 
caused
by 
the recombination.

Suppose that $f$ is Lipschitz continuous.
The smoothness of $P_tf$ leads to
\begin{align}
	\label{eq:errorbound3}
 \sup_{x \in \mathbb{R}^N} 
 \left\lvert 
\left({\Phi^{m,j}_{\mathcal{D},(u,r)}(\delta_x)}
-\widehat{\Phi}^{m,j}_{\mathcal{D},(u,r)}(\delta_x),P_{T-t_j}f \right)
\right\rvert  
\leq C u_j^{r_j+1} \sup_{|\alpha|=r_j+1} \parallel D^{\alpha}P_{T-t_j}f \parallel_\infty
\end{align}
for $0 \leq j \leq k-1$,
where
Eq.~(\ref{eq:taylorremainder}) 
and the triangle inequality are used.
Like the case of Eq.~(\ref{eq:ugf}),
a suitable condition on
$\{V_i\}_{i=0}^d$
ensures
there exists a positive integer $p\in \mathbb{N}$ 
such that
\begin{equation}
  \label{eq:uh}
  \sup_{|\alpha|=r+1}
  \parallel
  D^\alpha P_tf
  \parallel_\infty
  \leq C t^{-rp/2}
  \parallel
  \nabla f
  \parallel_\infty
\end{equation}
for all $t\in (0,1]$.
When Eqs.~(\ref{eq:ugf}) and (\ref{eq:uh})
are satisfied
\cite{litterer2012high, kusuoka2003malliavin, crisan2006convergence},
one obtains
\begin{align}
	\label{eq:errorbound4}
& \sup_{x \in \mathbb{R}^N} 
 \left\lvert 
P_Tf(x)
-\left( {\Phi^{m,k}_{\mathcal{D},(u,r)}(\delta_x)},f \right) \right\rvert \nonumber \\
& 
\leq 
\Bigg( C_1
\Bigg(
s_k^{{1}/{2}}+\sum_{i=m+1}^{m+2}\sum_{j=1}^{k-1} 
\frac{s_j^{{i}/{2}}}{(T-{t_j})^{{(i-1)}/{2}}} 
\Bigg)  
+ C_2
\sum_{j=1}^{k-1} 
\frac{u_j^{r_j+1}}{(T-t_j)^{{r_jp}/{2}}}  \Bigg)
\parallel \nabla f\parallel_\infty 
\end{align}
from Eqs.~(\ref{eq:totalerror}), (\ref{eq:errorbound3}).
Here $C_1$ and $C_2$ are constants.

The recombination error can be controlled by
the radius of the ball $u_j$
and the cubature on $\mathbb{R}^N$ degree $r_j$.
By choosing an appropriate pair $(u_j,r_j)$, 
one can make the order of the recombination error bound
not dominant over the order of the error bound in the KLV method.
For example,
in the case of
$(u_j,r_j)=(s_j^{p/2-a},\lceil m/p \rceil)$
where $a = (p-1)/(2( \lceil m/p \rceil+1 ))$
($\lceil x \rceil$ denotes the smallest integer greater than or equal to $x$)
or
$(u_j,r_j)=( ( s_j^{m+1}/ (T-t_j)^{m-rp} )^{1/2(r+1)}, m)$,
the error estimate
\begin{equation}
\label{eq:klvconv}
 \sup_{x \in \mathbb{R}^N} 
\left\lvert  
P_Tf(x)
-\left( {\Phi^{m,k}_{\mathcal{D}(\gamma),(u,r)}(\delta_x)},f \right) \right\rvert 
\leq C \parallel \nabla f \parallel_\infty T^{1/2} k^{-(m-1)/2}  
\end{equation}
is satisfied
for a Lipschitz continuous $f$
when $\gamma > m-1$.
Eq.~(\ref{eq:klvconv}) is obtained from substituting 
the partition defined in Eq.~(\ref{eq:Kpartition})
into Eq.~(\ref{eq:errorbound4})
and 
shows the same convergence rate with the ones without recombination,
Eqs.~(\ref{eq:KLVconvergece}) and
(\ref{eq:KLVconvergeceflow}).

\section{Patched cubature filter and Adaptive patched cubature filter}
\label{sec:PCFAPCF}
Recall 
$X(t) \in \mathbb{R}^N$ 
is governed by
\begin{equation}
  \label{eq:dynamics2}
  dX(t)  = V_0(X(t))\,dt +\sum_{i=1}^d V_i(X(t)) \circ dW_i(t).
\end{equation}
Let
the noisy observations $Y_n$ associated with $X_n = X(n T)$ 
satisfy
\begin{equation}
\label{eq:observation}
Y_n  = \varphi(X_n) + \eta_n, \quad \eta_n \sim \mathcal{N}(0,R_n)  
\end{equation}
where
$\varphi \in C_b^\infty( \mathbb{R}^N,\mathbb{R}^{N'})$
and realisations of the noise $\eta_n$ are 
i.i.d. 
random vectors
in $\mathbb{R}^{N'}$.

For a deterministic particle approximation of the optimal filtering
solution of
Eqs.~(\ref{eq:dynamics2}) and (\ref{eq:observation}),
we 
employ 
the KLV method and recombination
to define 
the patched cubature filter (PCF)
in subsection~\ref{sec:algorithm}
and
and the adaptive patched cubature filter (APCF)
in subsection~\ref{sec:aPCF}.
We address
several issues encountered during their practical implementations
in subsection~\ref{sec:implementation}.

\subsection{Patched cubature filter (PCF)}
\label{sec:algorithm}
Let $\pi_{n|n'}$ be the law of the conditioned variable $X_n|Y_{1:n'}$
and
$ {\pi}^{\text{PCF}}_{0|0}$ be a discrete measure 
approximation of
the law of $X(0)$.
We define the 
\emph{patched cubature filter (PCF) at the path level}
by the recursive algorithm
\begin{equation}
  \begin{split}
	\label{eq:patchedparticlefitler}
{\pi}^{\text{PCF}}_{n|n-1} & =
\Phi^{m,k}_{\mathcal{D},(u,r)}({\pi}^{\text{PCF}}_{n-1|n-1}), \\
{\pi}^{\text{PCF}}_{n|n} & = \text{REW}\left( {\pi}^{\text{PCF}}_{n|n-1}, g^{y_n}
\right),
  \end{split}
\end{equation}
for $n \geq 1$.
The algorithm 
can be stated as the following.

\begin{enumerate}
  \item One breaks the measure into patches and performs individual recombination for each one.
  \item One moves given discrete measure forward in time using the KLV method.
  \item One performs data assimilation via bootstrap reweighting at 
	every inter-observation time 
	which might differ from the time step 
	for the numerical integration.
  \item One again applies the patched recombination.
\end{enumerate}

Using ${\pi}^{\text{PCF}}_{n-1|n-1}$ in place of $\delta_x$ in 
Eq.~(\ref{eq:errorineq}),
an error bound of the prior approximation
of the PCF
is given by
\begin{align}
	  \label{eq:priorerrorineq}
\left\lvert (\pi_{n|n-1} - {\pi}^{\text{PCF}}_{n|n-1},f ) \right\rvert  
& \leq \left\lvert  (\pi_{n-1|n-1},P_Tf )-
({\pi}^{\text{PCF}}_{n-1|n-1},P_Tf )\right\rvert \nonumber\\
& \quad + \left\lvert({\pi}^{\text{PCF}}_{n-1|n-1},P_Tf)
-(\Phi^{m,k}_{\mathcal{D},(u,r)}({\pi}^{\text{PCF}}_{n-1|n-1} ),f ) \right\rvert \nonumber\\
& \leq \left\lvert  (\pi_{n-1|n-1}- {\pi}^{\text{PCF}}_{n-1|n-1},P_Tf )\right\rvert 
+ \sum_{j=1}^k \parallel (P_{s_j}-Q^m_{s_j}) P_{T-t_j}f \parallel_\infty\nonumber\\
 & +\sum_{j=0}^{k-1} 
 \left\lvert  
\left({\Phi}_{\mathcal{D},(u,r)}^{m,j}({\pi}^{\text{PCF}}_{n-1|n-1})
-{\widehat{\Phi}}_{\mathcal{D},(u,r)}^{m,j}({\pi}^{\text{PCF}}_{n-1|n-1}),P_{T-t_{j}}f\right)
 \right\rvert.
\end{align}
One can use the same argument with the case of PDE problem
to obtain a higher order estimate of the PCF.
An error bound of the posterior approximation 
\begin{align}
	\label{eq:posteriorestimate}
	\left\lvert (\pi_{n|n}-{\pi}^{\text{PCF}}_{n|n},f) \right\rvert 
	& =
	\Bigg\vert 
	\frac{(\pi_{n|n-1},fg^{y_n})}{({\pi}_{n|n-1},g^{y_n})}
	-
	\frac{({\pi}^{\text{PCF}}_{n|n-1},fg^{y_n})}{({\pi}_{n|n-1},g^{y_n})} 
\nonumber \\ & \qquad  
+ \frac{({\pi}^{\text{PCF}}_{n|n-1},fg^{y_n})}{({\pi}_{n|n-1},g^{y_n})}
	-
	\frac{({\pi}^{\text{PCF}}_{n|n-1},fg^{y_n})}{({\pi}^{\text{PCF}}_{n|n-1},g^{y_n})}
	\Bigg\vert\nonumber \\
	&  \leq
	\frac{1}{(\pi_{n|n-1},g^{y_n})}
	\left\lvert (\pi_{n|n-1}-{\pi}^{\text{PCF}}_{n|n-1},fg^{y_n})\right\rvert  \nonumber \\
	& \qquad + \frac{\parallel f \parallel_\infty}{(\pi_{n|n-1},g^{y_n})}
	\left\lvert (\pi_{n|n-1}-{\pi}^{\text{PCF}}_{n|n-1},g^{y_n})\right\rvert  
\end{align}
is given in terms of an error estimate of the prior approximation.
We stress
that PCF 
does not include
any Monte-Carlo 
subroutine
and 
therefore
its error estimate 
for the weak approximation
can be 
of high order
with respect to the number of iterations $k$
in view of
Eqs.~(\ref{eq:priorerrorineq})
and (\ref{eq:posteriorestimate}).

Recall the path level operator ${\text{KLV}}^{(m)}$ 
can be replaced
by the flow level operator 
$\widetilde{\text{KLV}}^{(m)}$ 
without harming the order of accuracy.
By doing this in the PCF at the path level,
we define the 
\emph{PCF at the flow level}
by the successive algorithm that produces
$\widetilde{\pi}^{\text{PCF}}_{n|n-1}$
and
$\widetilde{\pi}^{\text{PCF}}_{n|n}$.

\subsection{Adaptive patched cubature filter (APCF)}
\label{sec:aPCF}

It would be 
worthwhile to
mention that
PCF (\ref{eq:patchedparticlefitler}),
like SIR (\ref{eq:SIR}),
performs the prior approximation
without using the next time observational data.
This naturally leads us to
develop a variant of PCF
that 
will share the common essential feature with
SISR (\ref{eq:SIS})
in the aspect that the observation process is involved in moving particles forward in time.

In order to do that, we 
first 
consider
a modification of 
the standard KLV scheme
in which
some particles 
are 
adaptively
accelerated
when it causes no significant difference 
in the integration of the test function.
If
the smoothness of the test function 
is not 
known
in advance,
the accuracy requirement of
the KLV numerical approach 
leaves no choice other than 
to let the family of particles 
forward
following the pre-specified partition
until the next observation time.
This is because,
for truly irregular test functions, 
accurate integration would require exploration of the irregularities.
However 
if the test function is smooth enough
and the less regular set is of significantly lower
dimension than the main part of the smoothness,
then 
we are allowed to let the particles
to go straight to the next observation time from some considerable distance back instead
of the step predicted in the worst case which we would otherwise have used
to terminate the algorithm.

We build this insight into the practical algorithm.
At each application of the KLV operator,
the algorithm evaluates the test function using a 
one step 
prediction 
straight
to the next observation time
and compares this with the evaluation using 
a two step (one next step and the rest step to the next observation time) prediction.
If two evaluations
agree within the error tolerance, 
then the particles immediately leap to the 
next observation time.
Otherwise the prediction will follow the original partition.

In terms of accuracy,
the approach is pragmatically rather
successful because the opportunities for two 
(or three to break certain pathological symmetries)
step prediction to produce consistent
answers by chance is essentially negligible. 
Furthermore,
the adaptive switch 
for which the KLV is employed
to move the prediction measure forward but
move a part of it straight to the observation time whenever the relevant part of the
test function 
is smooth enough
has a very significant effect of 
pruning the computation and speeding up the algorithm
due to the reduction of particles to be recombined at each iteration.

This adaptive KLV method 
of course
cannot be applied without a test function. 
Differently from the PDE problem,
the test function is not specified in the filtering problem.
Therefore
in practice we take the smooth likelihood as test function to lead the adaptation.

Recall
$\mathcal{D} = \{ 0=t_0 < t_1 < \cdots < t_k = T \}$
is a partition of $[0,T]$ with $s_j=t_j-t_{j-1}$.
We use the likelihood 
$g^{y_n}$ 
to define the splitting operator
acting on
a discrete measure ${\mu}^{j-1} = \sum_{i=1}^n \kappa_i \delta_{x^i}$
at time $t_{j-1}$.
Let
${\mu}^{j-1}_{i,21}=\text{KLV}\left(\delta_{x^i},t_j-t_{j-1}\right)$,
${\mu}^{j-1}_{i,22}=\text{KLV} \left( {\mu}_{i,21}, t_k-t_j\right)$
and
${\mu}^{j-1}_{i,1}= \text{KLV}\left(\delta_{x^i},t_k-t_{j-1}\right)$.
Let $I_\tau$ be the collection of index $i$ satisfying
$\vert  ({\mu}^{j-1}_{i,1}- {\mu}^{j-1}_{i,22},  g^{y_n}) \vert<\tau$.
Then the discrete measure ${\mu}^{j-1}$ is the union of two discrete measure
$ \mu^{j-1} = \mu^{j-1, < \tau} \sqcup \mu^{j-1, \geq \tau}$
where
$\mu^{j-1, < \tau} =  \sum_{i \in I_\tau} \kappa_i \delta_{x^i}$.
For simplicity,
$\mu^{k-1, \geq \tau}$ is defined to be the null set.
The process defines the splitting operator
\begin{equation}
  \label{eq:sorting}
\text{SPL}^{(\tau)}
\left( \mu^{j-1}, g^{y_n} \right)
\equiv
\mu^{j-1, < \tau}
\end{equation}
for $1 \leq  j \leq k$.

Define a sequence of discrete measures as follows
\begin{equation}
  \begin{split}
  \label{eq:adaptivesequenceofdiscretemeasure}
\Phi^{m,0}_{\mathcal{D},(u,r),\tau}({\mu}^0) & ={\mu}^0, \\
\widehat{\Phi}_{\mathcal{D},(u,r),\tau}^{m,j-1}({\mu}^0) & = 
\text{REC}^{(u_{j-1},r_{j-1})}\left(
{\Phi}_{\mathcal{D},(u,r),\tau}^{m,j-1}({\mu}^0) \right), \\
\widehat{\Phi}^{m,j-1,< \tau}_{\mathcal{D},(u,r),\tau}({\mu}^0) & = 
\text{SPL}^{(\tau)}\left(
\widehat{\Phi}^{m,j-1}_{\mathcal{D},(u,r),\tau}({\mu}^0), g^{y_n}\right), \\
\Phi_{\mathcal{D},(u,r),\tau}^{m,j}({\mu}^0) &= \text{KLV}^{(m)}
\left(\widehat{\Phi}_{\mathcal{D},(u,r),\tau}^{m,j-1 , < \tau }({\mu}^0),s_j \right).
  \end{split}
\end{equation}
for $1 \leq  j \leq k$.
Let 
$\widehat{\Phi}^{m,j-1}_{\mathcal{D},(u,r),\tau}({\mu}^0)
=
\widehat{\Phi}^{m,j-1,< \tau}_{\mathcal{D},(u,r),\tau}({\mu}^0)
\sqcup
\widehat{\Phi}^{m,j-1, \geq \tau}_{\mathcal{D},(u,r),\tau}({\mu}^0)$
and
\begin{equation}
\Psi_{\mathcal{D},(u,r),\tau}^{m,j-1,k}({\mu}^0) 
 = \text{KLV}^{(m)} \Big( 
\text{KLV}^{(m)} 
\left( \widehat{\Phi}_{\mathcal{D},(u,r),\tau}^{m,j-1, \geq \tau}({\mu}^0),
t_j-t_{j-1} \right),T-t_j \Big).
\end{equation}
for $1 \leq  j \leq k-1$.

We define
the \emph{adaptive patched cubature filter (APCF) at the path level} by
\begin{equation}
  \begin{split}
	\label{eq:adaptivepatchedparticlefitler}
{\pi}^{\text{APCF}}_{n|n-1} 
& = \left(\bigsqcup_{j=1}^{k-1} \Psi_{\mathcal{D},(u,r),\tau}^{m,j-1,k}
({{\pi}^{\text{APCF}}_{n-1|n-1}})\right)
\sqcup \Phi_{\mathcal{D},(u,r),\tau}^{m,k}({{\pi}^{\text{APCF}}_{n-1|n-1}}), \\
{\pi}^{\text{APCF}}_{n|n} 
& = \text{REW}\left( {\pi}^{\text{APCF}}_{n|n-1}, g^{y_n} \right),
  \end{split}
\end{equation}
for $n \geq 1$.
The algorithm 
can be stated as the following.
\begin{enumerate}
  \item One breaks the measure into patches and performs individual recombination for each one.
  \item One splits given discrete measure to lead some of the particles to the next observation time
	and the rest particles to the next iteration time using the KLV method.
  \item One performs data assimilation via bootstrap reweighting at 
	every inter-observation time 
	which might differ from the time step 
	for the numerical integration.
  \item One again applies the patched recombination.
\end{enumerate}

Via
replacing $\text{KLV}^{(m)}$ by $\widetilde{\text{KLV}}^{(m)}$,
we define the 
\emph{APCF at the flow level}
that produces
$\widetilde{\pi}^{\text{APCF}}_{n|n-1}$ and
$\widetilde{\pi}^{\text{APCF}}_{n|n}$
instead of
${\pi}^{\text{APCF}}_{n|n-1}$ and
${\pi}^{\text{APCF}}_{n|n}$.

In view of Eq.~(\ref{eq:posteriorestimate}),
the likelihood is indeed a natural choice
for the filtering problem 
in which
the posterior measure is of primary interest.
One can apply $g^{y_n}$ and $fg^{y_n}$
simultaneously
as the test function
for the SPL operator
in Eq.~(\ref{eq:adaptivesequenceofdiscretemeasure})
if one would like to obtain a posterior approximation that
accurately integrates $f$.

Note both
SISR (\ref{eq:SIS})
and 
APCF (\ref{eq:adaptivepatchedparticlefitler})
are built upon
the same philosophy - making use of the observational information
to lead the particles 
for
a more accurate approximation of the posterior
possibly at the expense of the accuracy of the corresponding prior approximation.
However
the way of modifying 
the basis algorithm is different from one another.
In particular,
while SISR leads the particles only using
the instance of the observation $y_n$,
APCF fully uses the likelihood $g^{y_n}$ to achieve the adaptation.
Furthermore,
APCF 
cares the domain of importance 
without introducing a new dynamics.

\subsection{Practical implementation}
\label{sec:implementation}
One has to
specify
the time partition and 
the way of patched recombination
for the implementation of PCF and APCF.
We here present 
adaptive partition and adaptive recombination
as alternatives to 
the Kusuoka partition 
and
the covering with fixed-size balls,
respectively.
Differently from prior suggestions,
our ones
are subject to
the test function
and thus called adaptive.

Before doing that,
we at this point mention that
the work in
\cite{crisan2013kusuoka}
also employs cubature on Wiener space
to solve the nonlinear filtering problem.
Comparing these two kinds of cubature filters,
one major difference 
is how to simplify
the support of discrete measure
between successive KLV applications
to control computational cost.
The one developed in
\cite{crisan2013kusuoka}
makes use of
the Monte-Carlo scheme
based on
branching and pruning mechanism.
The algorithm looks for a reduced measure whose distance from the original measure
is minimised in some sense.
Therefore
the simplification procedure should be applied
to the whole discrete measure all at once.
On the contrary,
PCF and APCF
take the
deterministic moment-matching
recombination strategy,
that can be applied locally in the support of measure
for an enhanced efficiency.

In addition to the algorithm characteristics,
the problem setting in
\cite{crisan2013kusuoka}
is rather different from the current paper
as the observation process is 
assumed to be not discrete but continuous 
(for more details we refer the reader to 
\cite{LL2}).
In this case,
the time integration 
of the KLV method is performed along with even partition of small intervals.
However, in case of sparse observations,
the numerical integration until the next observation time
requires multiple steps preferably 
with uneven partition of decreasing intervals
rather than even partition.
For PCF and APCF, the likelihood can serve as the test function
and we can further utilise the presence of this test function to determine time partition.
This is clearly one additional degree of freedom allowed in the cubature filter under the
scenario of
intermittent observations.

\subsubsection{Adaptive partition}

For a given test function $f$,
one can make use of the heat kernel
$P_t$ as well as $f$
to evolve the set of particles
so that one step error is
within a given degree of accuracy, i.e.,
\begin{equation}
  \begin{split}
	\label{eq:klverror}
  \parallel (P_{s_j}-Q^m_{s_j})P_{T-t_j}f \parallel_\infty  < \epsilon
  \end{split}
\end{equation}
for some $\epsilon>0$.
We define 
an adaptive partition
$\mathcal{D}({\epsilon},f) = \{t_j\}_{j=0}^k$
to be a time discretisation for which
each
$s_j=t_j-t_{j-1}$ is the supremum among
the ones
satisfying
Eq.~(\ref{eq:klverror}).
Because
$P_tf$ becomes smoother as $t$ increases,
the sequence $\{s_j\}_{j=1}^k$
tends
to decrease monotonically,
i.e.,
$s_1 \geq s_2 \geq \cdots \geq s_k$.
The upper bound of the total error 
along with the adaptive partition
is given by
\begin{equation}
\label{eq:apart}
\sup_{x}  
\left\lvert P_Tf(x) -(\Phi^{m,k}_{\mathcal{D}}(\delta_x),f) 
\right\rvert
< k \epsilon
\end{equation}
from Eq.~(\ref{eq:inequality1}).

\subsubsection{Adaptive recombination}
\label{sec:adarecomb}

Consider the condition
\begin{equation}
	\label{eq:adarecomb}
  \left\lvert  
  \left({\Phi^{m,j}_{\mathcal{D},(u,r)}} (\mu^0)
  -\widehat{\Phi}^{m,j}_{\mathcal{D},(u,r)} (\mu^0)
,P_{T-t_j}f \right)
\right\rvert < \theta
\end{equation}
given some $\theta>0$.
We define the adaptive recombination 
by the algorithm that uses 
as large
value of $u$ as possible,
for a fixed recombination degree $r$,
among the ones
satisfying Eq.~(\ref{eq:adarecomb}).
The algorithm again makes use of 
the heat kernel
$P_t$ as well as 
the test function $f$.
When the adaptive partition and the adaptive recombination are simultaneously used, 
the combination of
Eqs.~(\ref{eq:klverror})
and
(\ref{eq:adarecomb})
yields 
\begin{equation}
	\label{eq:apar}
\sup_{x}  
 \left\lvert P_Tf(x) -(\Phi^{m,k}_{\mathcal{D},(u,r)}(\delta_x),f) 
 \right\rvert 
 < k(\epsilon+\theta)
\end{equation}
from Eq.~(\ref{eq:errorineq}).
Notice,
unlike the case of Eq.~(\ref{eq:klvconv})
where the constant $C$ is not 
specified,
the upper bound of
Eq.~(\ref{eq:apar})
is
explicitly
under the control.

It deserves to mention that
the application of 
the adaptive recombination 
does not require to determine
the size (and even topology) of patches
in advance.
Given the recombination degree,
it suffices to keep 
shrinking
the size of patches until
Eq.~(\ref{eq:adarecomb}) is met.
Due to this feature,
the adaptive recombination 
can 
practically 
be 
useful in
achieving the error bound of
Eq.~(\ref{eq:apar})
when it 
is accompanied with
an efficient algorithm
that
divides 
the support of discrete measure
into local disjoint subsets.
Because the detailed algorithm of
the recombination can be found in
\cite{litterer2012high},
we conclude this section with 
one way to achieve
the adaptive recombination 
utilising 
the Morton ordering
\cite{morton1966computer}.
The methodology 
adopts
boxes,
instead of balls,
as patches
to locally cover the particles.
The algorithm is advantageous
particularly in case of high dimension.

Given a number of particles in $N$ dimension,
we perform an affine transformation to map the particles into the ones in the box $[0.5, 1)^N$.
In the following, 
we evenly divide each edge of the box by $2^n$ 
to get $2^{nN}$ sub-boxes
and 
assign the particles to these sub-boxes.
We use the
double-precision floating-point format in scientific computing:
any number $z^i \in [0.5, 1)$ is saved in terms of $\{b^i_j\}_{j=1}^{52}$
where $b^i_j$ is either $0$ or $1$
in a way that $z^i = (1/2)\times(1+\sum_{j=1}^{52}b^i_j2^{-j})$
(almost all numbers in $[0.5, 1)$ have a binary expansion of more than $52$ digits
but this reduced information is quite enough for our purpose).
In this way the point $(z^1,\cdots,z^N)$ in $N$-dimension can be expressed by $52 \times N$ binary numbers.
Interleaving the binary coordinate values yields binary values. 
Connecting the binary values in their numerical order produces 
the Morton ordering.
Then an appropriate coarse-graining leads to the subdivision of a box.
For examples, when $N=2$,
the binary value corresponding $(z^1,z^2)$ is $b^1_1b^2_1b^1_2b^2_2 \cdots b^1_{52}b^2_{52}$.
The point is 
in first quadrant if $(b^1_1,b^2_1)=(1,1)$,
in second quadrant if $(b^1_1,b^2_1)=(0,1)$,
in third quadrant if $(b^1_1,b^2_1)=(0,0)$
and
in fourth quadrant if $(b^1_1,b^2_1)=(1,0)$.
Applying this classification to a number of particles 
produces $2^2$ disjoint subsets of classified particles.
Similarly, using $b^1_1b^2_1b^1_2b^2_2$ and ignoring the rest subgrid scales
gives $4^2$ subsets when $N=2$.
Taking the inverse affine transformation,
a classification of the particles has been achieved.

The crucial point being that by sorting the one dimensional transformed points, one keeps points in a box together without ever needing to introduce the boxes and particularly empty boxes. The complexity of the clustering is no worse than  
$MN \log M$
in the number of points.
Here
$M$
is number of particles,
$N$
is dimension and 
$MN \log M$
is the cost of patching. 
Note $N \log M$
is the cost of getting those points in a patch.

\section{Numerical simulations}
\label{sec:numerical}
We perform numerical simulations 
to examine the efficiency and accuracy of the proposed filtering approaches.
We introduce the test model in subsection~\ref{sec:tmodel}
and obtain the reference solutions 
in subsection~\ref{sec:ref}.
We implement the
PCF and APCF with cubature on Wiener space of degree $m=5$ in subsection~\ref{sec:PCFaPCFd5}.
Finally, 
in subsection~\ref{sec:PCFaPCFd7},
we investigate the prospective performance of 
PCF and APCF with cubature on Wiener space of degree $m=7$.

\subsection{Test model}
\label{sec:tmodel}
It is very important to 
select
a good example
to examine the performance of the algorithms we have developed.
Here we choose a forward model and observation process
for which the analytic solution of the filtering problem 
is 
known
and can be used to measure the accuracy of the various particle approximations.

Our test model is
the Ornstein-Uhlenbeck process
\cite{uhlenbeck1930theory}
in three dimension:
	\begin{equation}
	  \label{eq:L63}
dX=
-\Lambda X
dt+g 
I_3
dW
	\end{equation}
where
$X= (x^1\; x^2\; x^3)^t$,
$\Lambda=
	\left( 
	\begin{array}{ccc}
	\sigma   & -\sigma  & 0 \\
	-\rho  & 1 & 0\\
	0 & 0 & \beta
	\end{array} 
	\right)
$,
$dW= (dW_1\; dW_2\; dW_3)^t $
and $I_3$ denotes the $3 \times 3$ identity matrix.
Here the superscript $t$ denotes the transpose.
The parameter values 
$\sigma=1$, $\rho=0.28$, $\beta=8/3$
and $g=0.5$
are chosen.
The observations
\begin{equation}
\label{eq:linearobservation}
Y_n  = X_n + \eta_n, \quad \eta_n \sim \mathcal{N}(0,R_n)
\end{equation}
are available
at every inter-observation time $T=0.5$.
We study
the cases 
in which
the covariance of observation noise is $R_n=R \times I_{3}$
for the values $R = 10^{-1}, 10^{-2}$ and $10^{-3}$.

\subsection{References}
\label{sec:ref}

\subsubsection{Kalman filter}
\label{sec:kfilter}
The conditioned measure 
for Eqs.~(\ref{eq:L63}), (\ref{eq:linearobservation})
is Gaussian 
and
$\pi_{n|n'}=\mathcal{N}(M_{n|n'},C_{n|n'}) $ 
can be obtained from the Kalman filter.
In this case, the prior covariance $C_{n|n-1}$ satisfies 
the Riccati difference equation and its solution converges
as $n$ increases
\cite{bitmead1991riccati}.
We take 
the covariance of the
initial condition $X(0)$ as the one step prediction from the limit of the Riccati
equation solution
so that
$C_{n|n-1}$ and $C_{n|n}$ do not depend on $n$
(but $M_{n|n-1}$ and $M_{n|n}$ depend on $n$).
We see that
the diagonal element of $C_{n|n-1}$ are about $10^{-1}$ for all cases of $R=10^{-1},10^{-2},10^{-3}$.
The diagonal element of $C_{n|n}$ are about $10^{-1}$ when $R=10^{-1}$,
about $10^{-2}$ when $R=10^{-2}$ and about $10^{-3}$ when $R=10^{-3}$.

\begin{figure}
\centerline{
\includegraphics[width=0.34\textwidth]{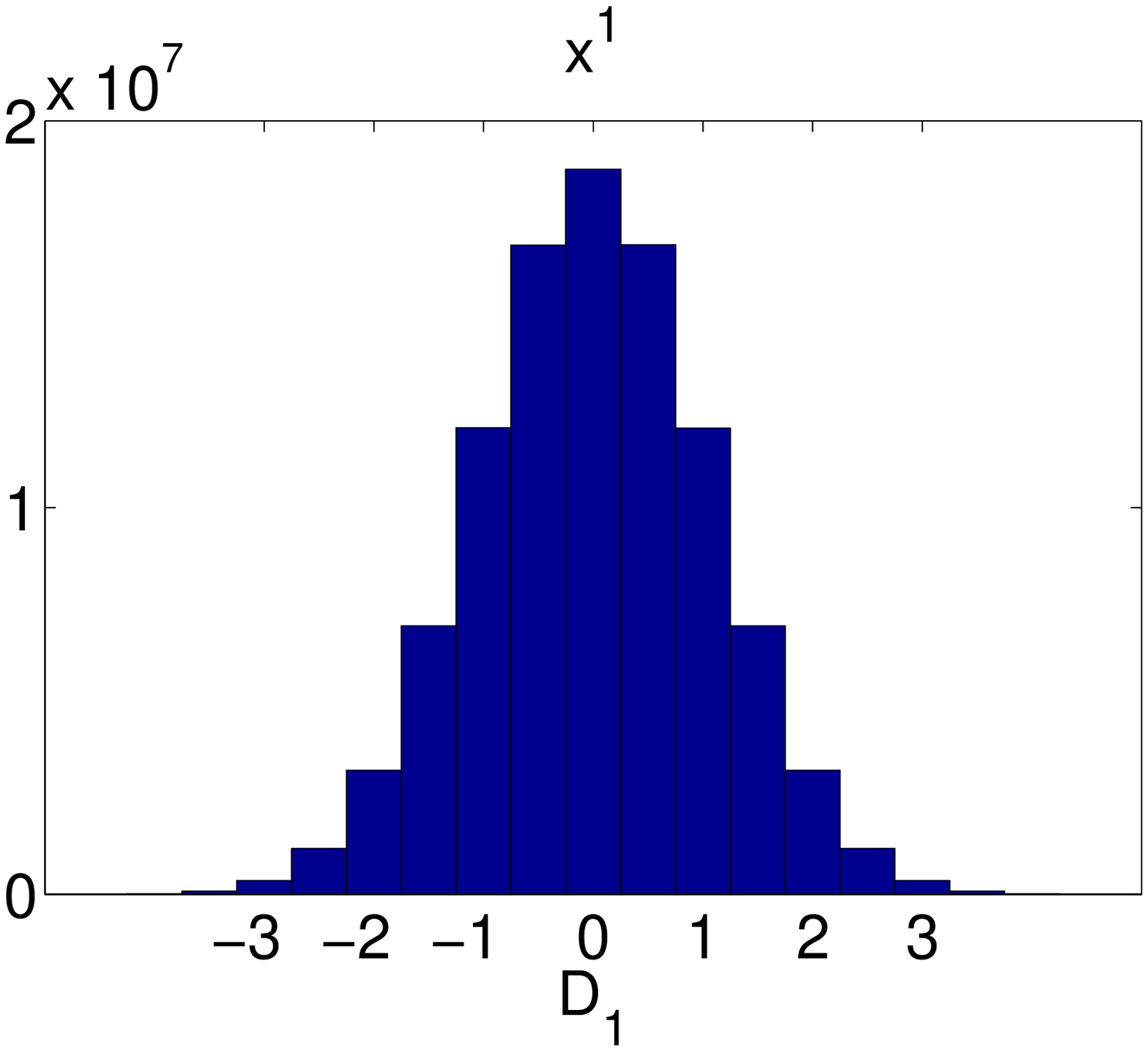}
\includegraphics[width=0.34\textwidth]{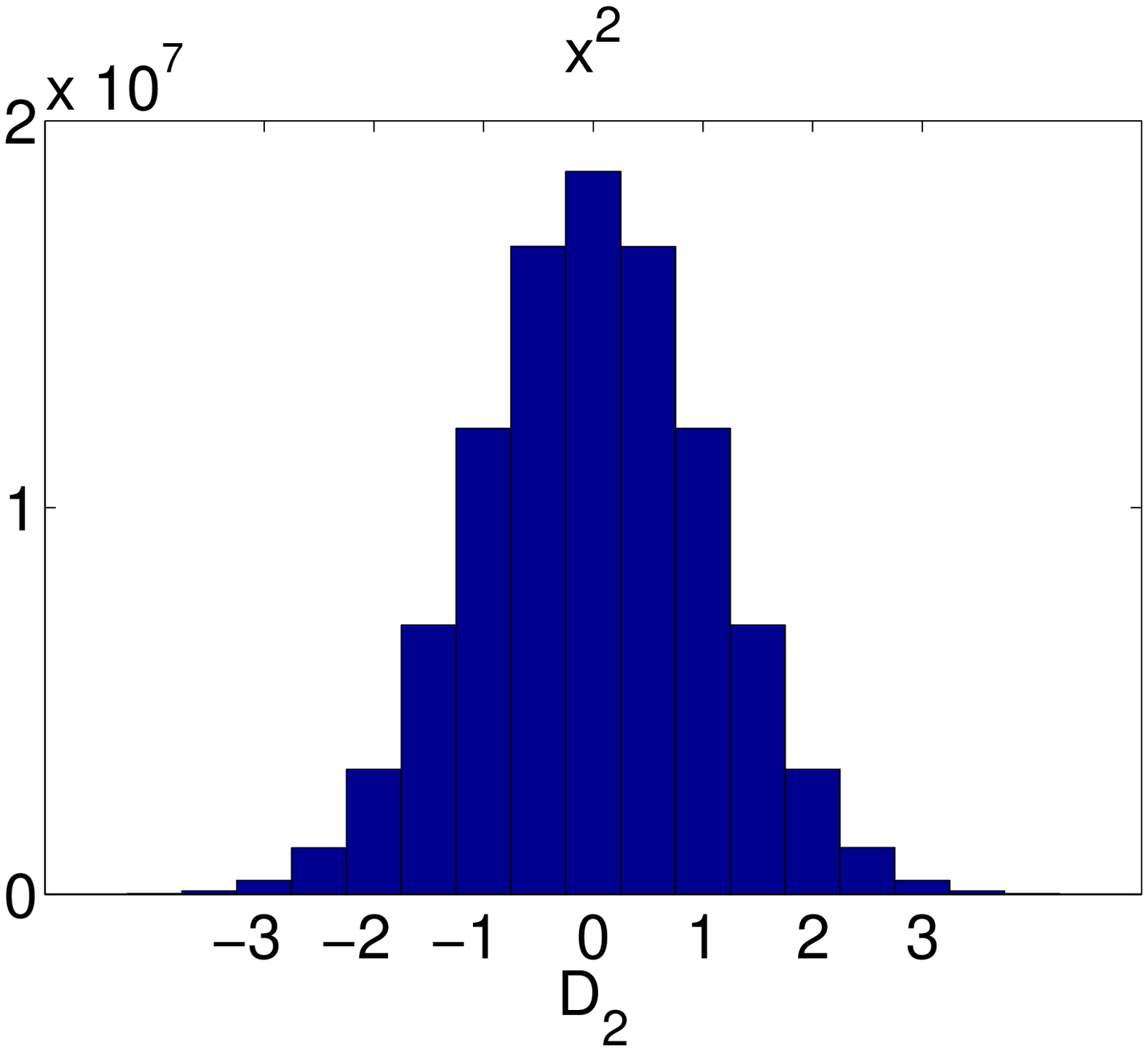}
\includegraphics[width=0.34\textwidth]{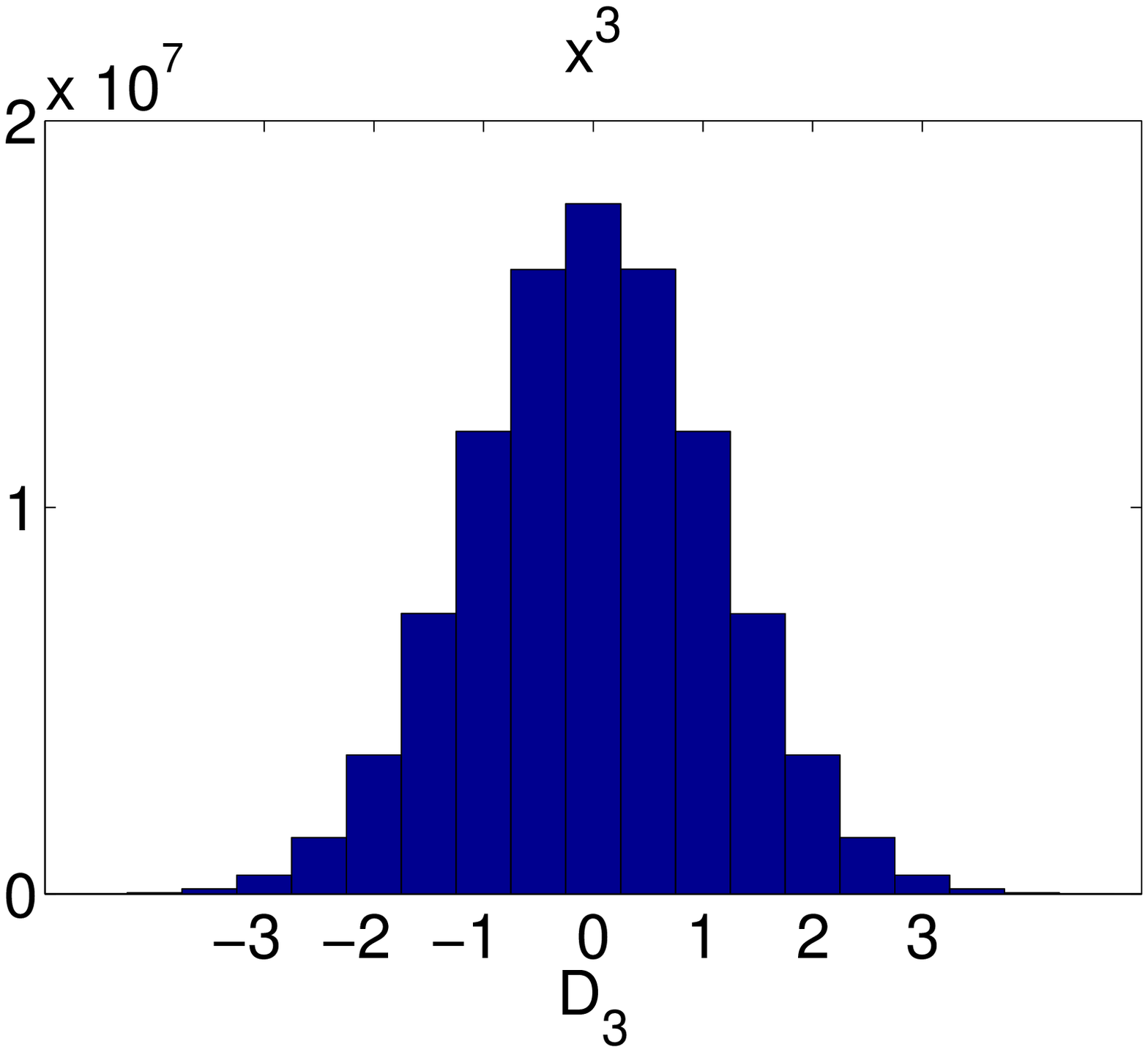}
}
\caption{
 The distribution of normalised distances between the observation and the prior mean
 when the noise covariance is $R_n = 10^{-2} \times I_3$.
}
\label{fig:observations} 
\end{figure}

In this filtering problem,
we first investigate
\emph{where are the observations}.
We apply the Kalman filter for $ 1\leq n \leq 10^8$ 
and calculate 
the values of $D_1, D_2$ and $D_3$ satisfying
$y_n={M}_{n|n-1}+(D_1\; D_2\; D_3)^t \cdot \sqrt{\text{diag}({C}_{n|n-1})}$,
where $y_n$ is determined by 
one trajectory of 
the dynamics~(\ref{eq:L63})
together with
a realisation of the
observation noise $\eta_n$.
The histograms in 
Fig.~\ref{fig:observations}
show the distribution of
these normalised distances between the observation and the prior mean
when $R=10^{-2}$
(the cases of $R=10^{-1}$ and $R= 10^{-3}$ are similar and not shown).
One can see that
most of the observations are within two times of the standard deviations from the prior
mean in each coordinate.
Among the cases of $10^8$,
there are $4,592,208$ cases for which $|D_i|>1$ for all $i=1,2,3$ at the same time.
There are $37,574$ cases for which $|D_i|>2$ for all $i$ at the same time,
and $60$ cases for which $|D_i|>3$ for all $i$ at the same time.
From the simulation,
we understand
the three cases in which
the parameter value of $D \equiv D_1=D_2=D_3$ 
is $1$, $2$ and $3$, are normal, exceptional and rare event,
respectively.

\subsubsection{$L^2$ norm of the higher order central moments}
Here
we aim to investigate 
the parameter regimes under which
our cubature filters are likely (or unlikely) to outperform.
In order to evaluate the computational error, one needs to define an error criterion relevant to
the approximations.
We realise that
unfortunately
a comparison between the evolving single trajectory 
and 
the corresponding posterior mean approximation,
which is commonly used in the filtering context,
is highly inappropriate for our purpose.
This is because the cubature approximation is basically superior 
within approximating the tail behaviour 
or higher order moments
of the probability distribution.
Therefore we instead use
the $L^2$ norm of the central moment 
to quantify the accuracy of the approximation obtained in the form of
discrete measure.

Let $\mathcal{C}^p$ be the $p$-th 
central
moment of $X=(x^1 \; x^2 \; x^3)^t$, i.e.,
\begin{equation*}
  \mathcal{C}^p_{i_1, \cdots, i_p} = 
  \mathbb{E}
  \left(
  \prod_{j=1}^p \left(x^{i_j}
  -\mathbb{E}( {x}^{i_j} )
  \right) 
  \right)
\end{equation*}
where $i_j = 1,2, 3$.
The $L^2$ norm of $\mathcal{C}^p$
is defined by
\begin{equation}
  \label{eq:momentnorm}
  \parallel \mathcal{C}^p \parallel_2
  \equiv
	\left(
	\sum_{i_1, \cdots, i_p =1 }^3 | \mathcal{C}^p_{i_1, \cdots, i_p}|^2 
	\right)^{1/2}.
\end{equation}
When $p=1$, 
Eq.~(\ref{eq:momentnorm}) is the Euclidean norm of the vector.
When $p=2$,
it is equivalent with the Frobenius norm of the matrix.
Let $\widehat{\mathcal{C}}^p$ be the $p$-th central moment of a particle approximation,
then
the relative root mean square error 
\begin{equation}
  \label{eq:rmse}
\text{rmse} \%
\equiv
\parallel \mathcal{C}^p - \widehat{\mathcal{C}}^p \parallel_2/ \parallel \mathcal{C}^p \parallel_2
\end{equation}
will be calculated to measure
the accuracy of the moment approximations.

\subsubsection{Monte-Carlo Gaussian samples}
\label{sec:montecarlosamples}
In our problem setting, the rmse errors are insensitive to
the specific time interval
between successive observations.
Taking one arbitrary time interval,
we study the cases of $D=1,2,3$
which
correspond to
normal, exceptional and rare event.
The scenario initially
may look somewhat artificial
because,
unlike the filtering in practice,
the observational data is not generated from realisations.
However we emphasise
it has been carefully designed,
while keeping the practical relevance,
in order to find the parameter regimes under which our approaches outperform
Monte-Carlo methods 
and this will eventually turn
out to be extremely helpful for a deeper understanding of the filtering problem.

We perform Gaussian sampling
to obtain three different Monte-Carlo approximations 
of the posterior measure.
For the first one,
we draw
samples from the prior measure
and subsequently 
apply
the bootstrap 
reweighting
to obtain the posterior approximation.
One can regard 
these 
bootstrap reweighted
samples from the prior 
as SIR result.
The second one is from the
SISR algorithm under the transition kernel
$\widetilde{K}(dx_n|x_{n-1},y_n)=\mathbb{P}(dx_n|x_{n-1},y_n)$,
which is
the optimal proposal in the sense of
minimising the variance of the importance weights
\cite{doucet2009tutorial}.
Finally, we draw samples
directly from
the posterior measures
as the third one.
Note, in all Monte-Carlo approximations,
neither
truncation error due to
numerical integration 
nor
resampling error
is induced
for a fair comparison.
The rmse errors
(\ref{eq:rmse})
of these Gaussian samples
are depicted in
Fig.~\ref{fig:R} 
when $R=10^{-2}$, $D=1,2,3$
and
in
Fig.~\ref{fig:D},
when $D=1$,
$R=10^{-1},10^{-2},10^{-3}$.
These results will be 
compared with the cubature filters.

\begin{table}
  \renewcommand{\arraystretch}{1.3} 
  \caption{The number of adaptive partition $k$ for KLV with $m=5$} 
  \label{tab:partm5} 
  \centering \begin{tabular}{ c | c  c cc} 
	& $\epsilon=10^{-2}$ & $\epsilon=10^{-3}$ & $\epsilon=10^{-4}$ & $\epsilon=10^{-5}$ \\ \hline
	$R=10^{-1}$ & 7 & 31 & 102 & 344  \\ 
	$R=10^{-2}$ & 10 & 29 & 101 & 330  \\ 
	$R=10^{-3}$ & 20 & 48 & 120 &  329 \\ 
  \end{tabular} 
\end{table}

\begin{figure}[t]
\centerline{
\includegraphics[width=0.49\textwidth]{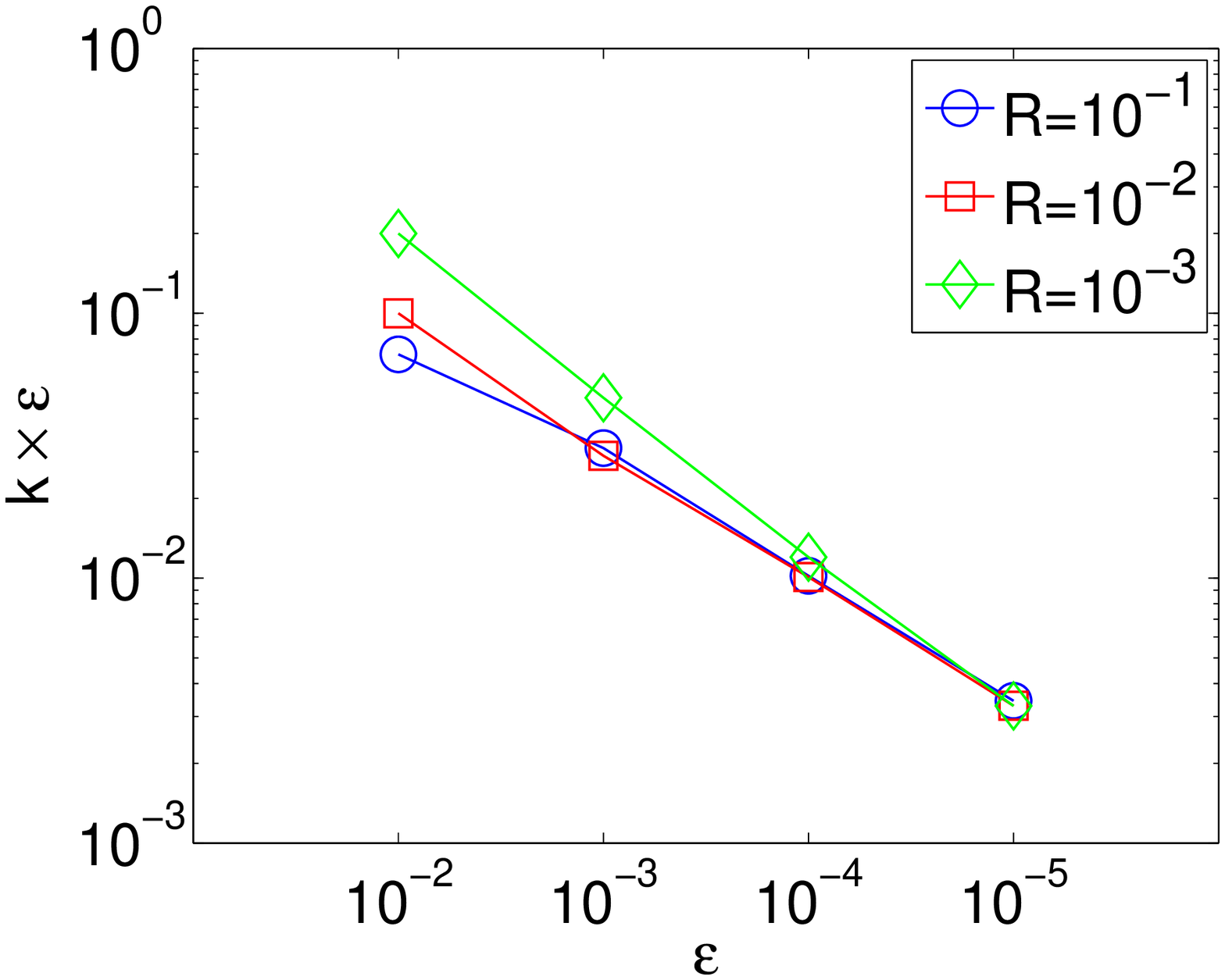}
}
\caption{
The upper bound of the total error along with the adaptive partition when $m=5$.
}
\label{fig:adaptivepartition_m5}
\end{figure}

\begin{figure}
\centerline{
  \subfigure[PCF using adaptive partition with $\epsilon=10^{-2}$]
  {\label{fig:TPCF}\includegraphics[width=0.49\textwidth]{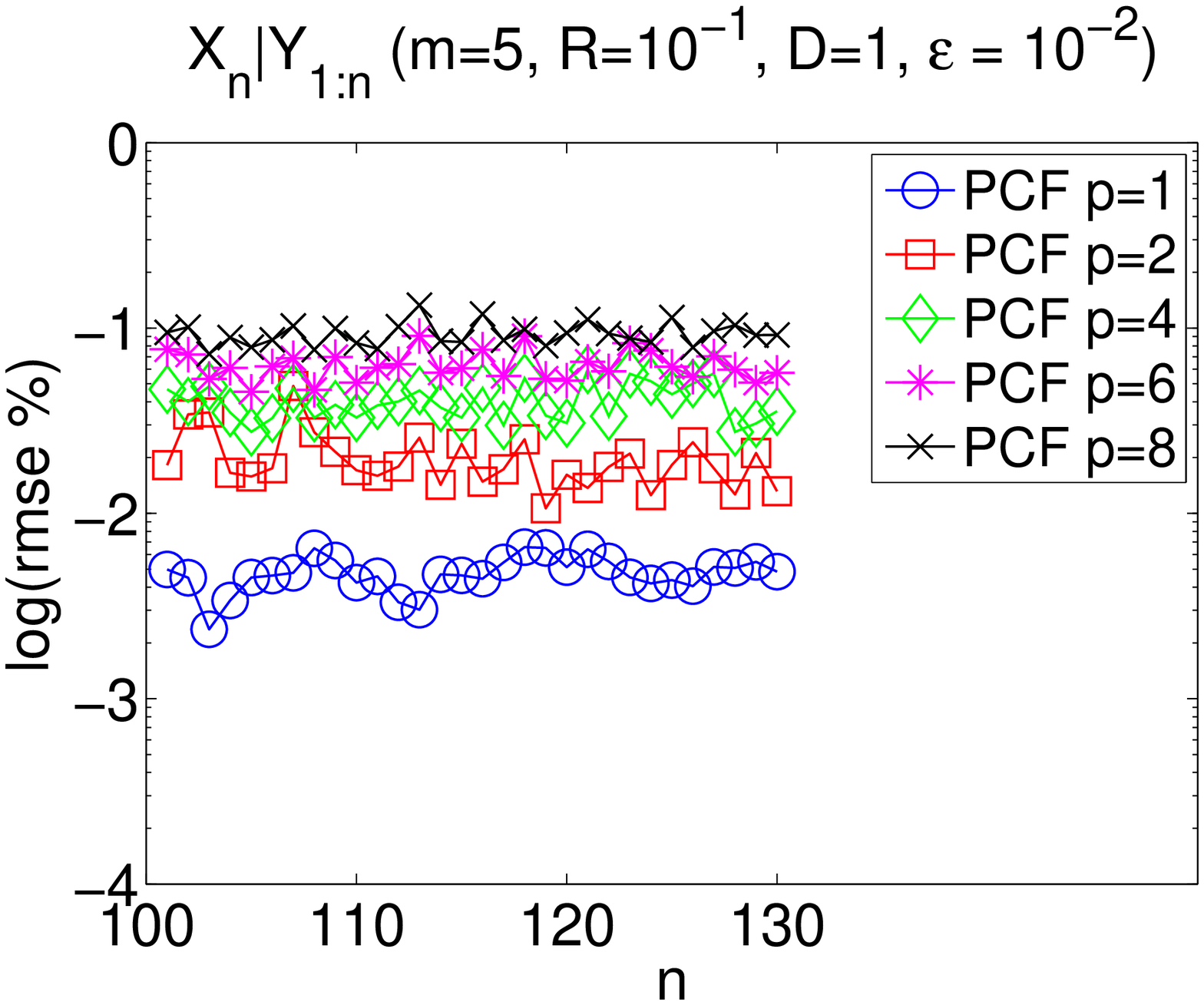}}
\subfigure[APCF using adaptive partition with $\epsilon=10^{-3}$]
{\label{fig:TaPCF}\includegraphics[width=0.49\textwidth]{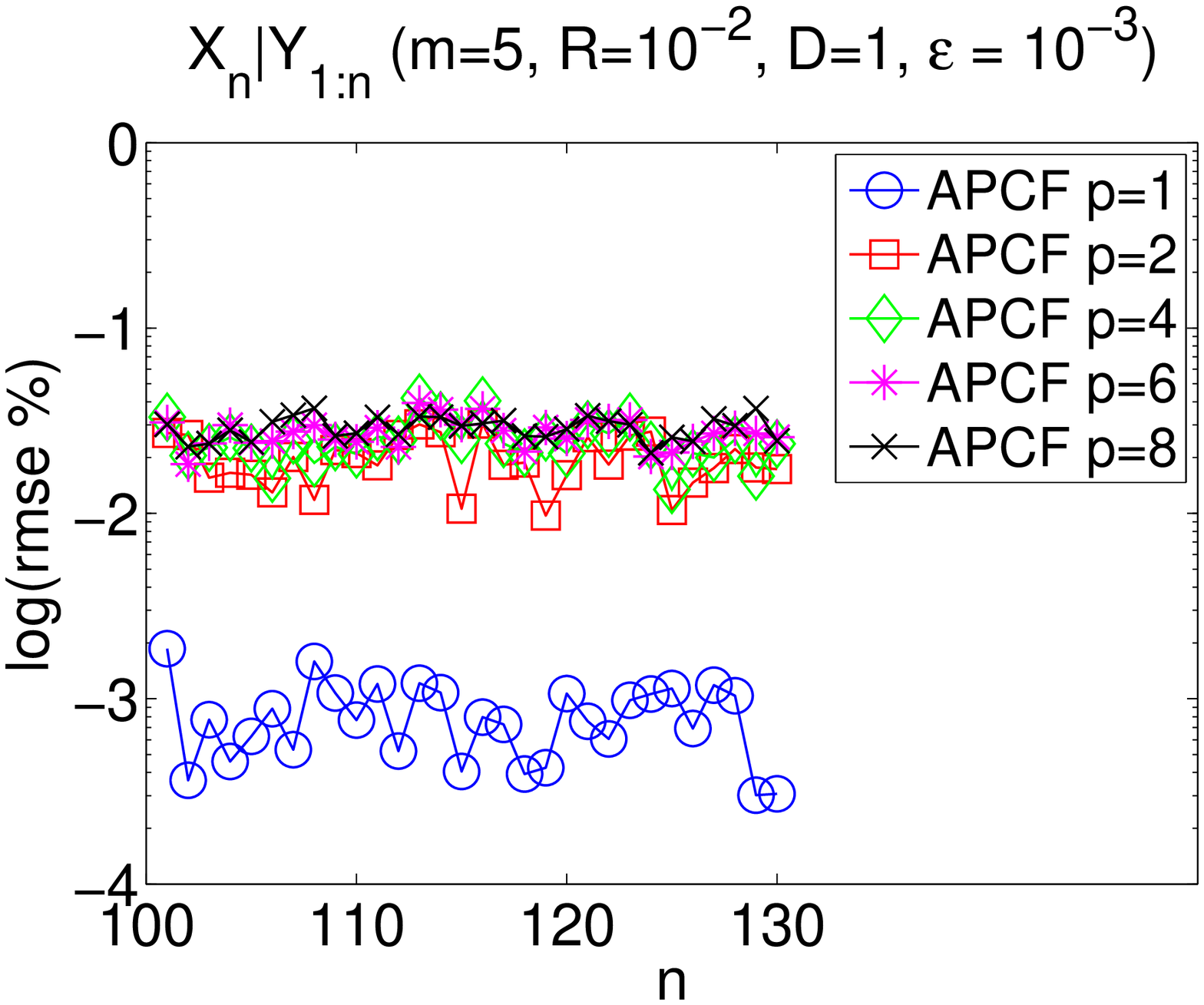}}
}
\caption{
The relative $L^2$ errors for the $p$-th moments of
the evolutionary posterior.
}
\label{fig:T}
\end{figure}

\subsection{PCF and APCF with cubature on Wiener space of degree $5$}
\label{sec:PCFaPCFd5}
\subsubsection{Choice of parameters}
We here implement the PCF and APCF at the flow level.
In case of $d=3$,
i.e.,
when the system is driven by three independent white noises,
cubature on Wiener space of degree $m=3$ and $m=5$,
with support size $n_m=6$ and $n_m=28$ respectively, are available.
We apply the KLV operator with degree $m=5$.

Using the likelihood $g^{y_n}$ as the test function $f$,
the adaptive partition
$\mathcal{D}(\epsilon,g^{y_n})$
satisfying Eq.~(\ref{eq:klverror})
with $\widetilde{Q}^m_{s_j}$ in place of ${Q}^m_{s_j}$
is analytically obtained
for the system of Eq.~(\ref{eq:L63}).
Note that the likelihood 
$g^{y_n}$ is the density function of $\mathcal{N}(y_n,R_n)$
and that the adaptive partition does not depend on $y_n$ but on $R_n$.
The number of iterations $k$ as a function of $\epsilon$ and $R$ is listed in
table~\ref{tab:partm5}.
In this case,
Fig.~\ref{fig:adaptivepartition_m5} reveals
the upper bound of
Eq.~(\ref{eq:apart})
tends to decrease as $\epsilon$ becomes smaller.
Therefore, by choosing $\theta$ to be the same order of $\epsilon$,
one can combine
the adaptive partition and the adaptive recombination 
to
achieve a desired degree of accuracy to some extent.

For the recombination of the PCF,
Eq.~(\ref{eq:adarecomb})
with $f=g^{y_n}$ for all $y_n \in \mathbb{R}^N$, i.e.,
\begin{equation}
	\label{eq:supadarecomb}
	\sup_{y_n}  \left\lvert  
  \left({\Phi^{m,j}_{\mathcal{D},(u,r)}} (\mu^0)
  -\widehat{\Phi}^{m,j}_{\mathcal{D},(u,r)} (\mu^0)
  ,P_{T-t_j}g^{y_n} \right)
\right\rvert < \theta
\end{equation}
is met
so that the recombination does not depend on $y_n$
but on $R_n$.
We choose the recombination degree $r=5$
and simulate the PCF for the cases of $\epsilon=10^{-2}, 10^{-3}$
with $\theta = 0.3 \times \epsilon$.

For the APCF, 
the tolerance $\tau$ has to be chosen
in addition to the parameters $\{ \epsilon, \theta\}$.
The value of $\tau$ varies in each case,
but we choose it so that
the SPL operator 
in Eq.~(\ref{eq:adaptivesequenceofdiscretemeasure})
allows $1/4 \sim 1/3$ part of particles leap to the next observation time
for all iterations
except the first and last few steps.
The remaining particles are reduced by the adaptive recombination,
i.e.,
the recombination satisfies
\begin{equation}
	\label{eq:aPCFadarecomb}
  \left\lvert  
  \left({\Phi^{m,j}_{\mathcal{D},(u,r),\tau}} (\mu^0)
  -\widehat{\Phi}^{m,j}_{\mathcal{D},(u,r),\tau} (\mu^0)
  ,P_{T-t_j} g^{y_n} \right)
\right\rvert < \theta
\end{equation}
where $\mu^0 = \widetilde{\pi}^{\text{APCF}}_{n-1|n-1}$.
We again choose the recombination degree $r=5$
and simulate the APCF for the cases of $\epsilon=10^{-2}, 10^{-3}$
with $\theta = 0.3 \times \epsilon$.

While the value of $D$ being fixed,
we apply the PCF and APCF to obtain 
the values of Eq.~(\ref{eq:rmse})
for the evolving posterior meausre.
Fig.~\ref{fig:T} shows that
the performances of the
two filtering algorithms are stable and 
that the numerical error estimates 
of high order moments
are insensitive to
$n$
(the rest cases produce similar plots and are not shown).

In our numerical simulations,
the number of patches 
needed to satisfy
Eq.~(\ref{eq:supadarecomb})
in the PCF
increases as the time partition approaches to the next observation time,
eventually about $8^3 \sim 16^3$.
On the contrary,
Eq.~(\ref{eq:aPCFadarecomb}) 
in the APCF
is satisfied with
$2^3$ 
($< 10$)
patches in most cases.
As a result, APCF saves computation time significantly compared with PCF.

\subsubsection{Dependence on the observation location}

\begin{figure*}[!htp] 
\centerline{
\subfigure[unweighted prior samples]
{\includegraphics[width=0.43\textwidth]{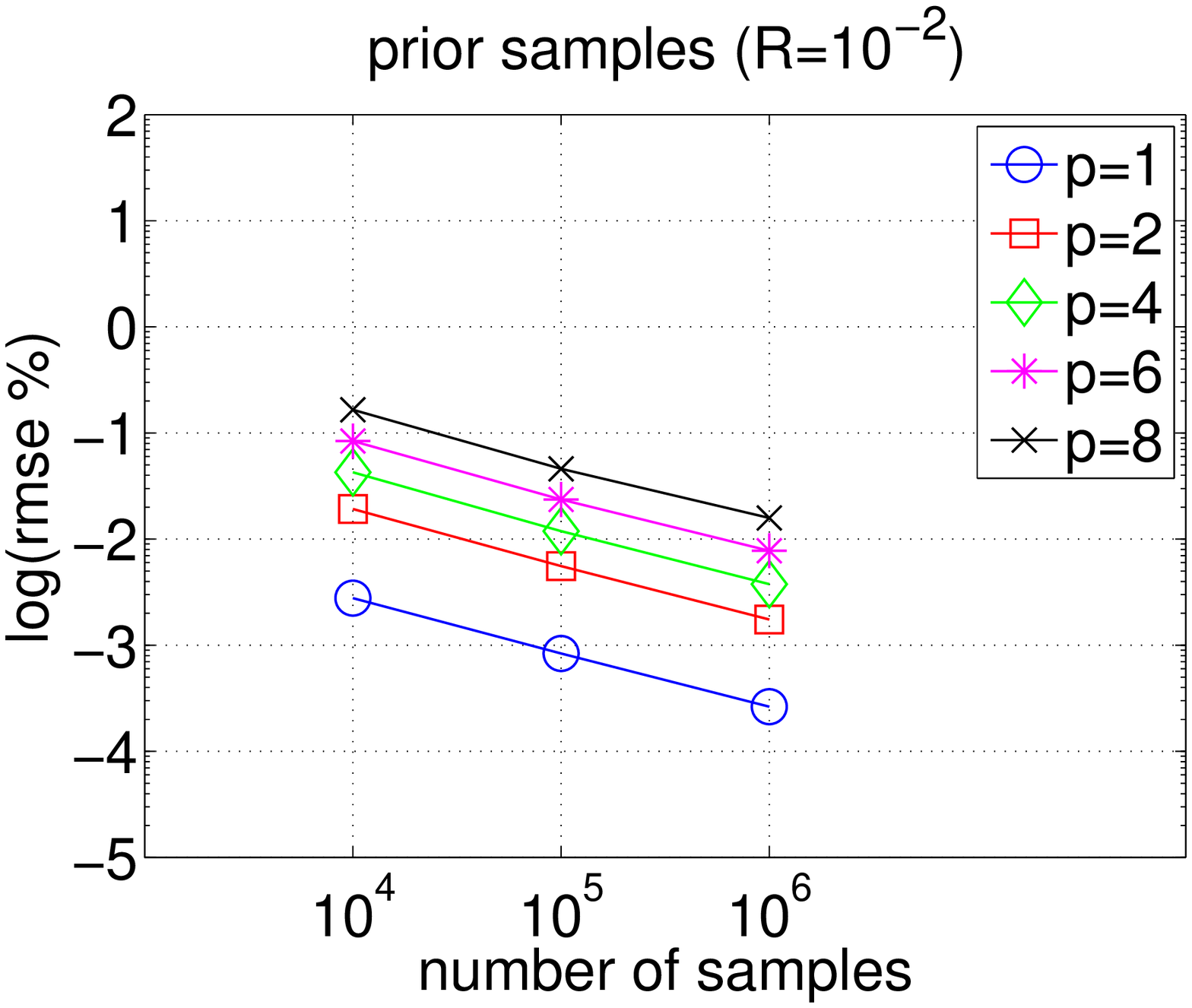} \label{fig:Ra} } 
\quad
\subfigure[cubature approximation of prior]
{\includegraphics[width=0.43\textwidth]{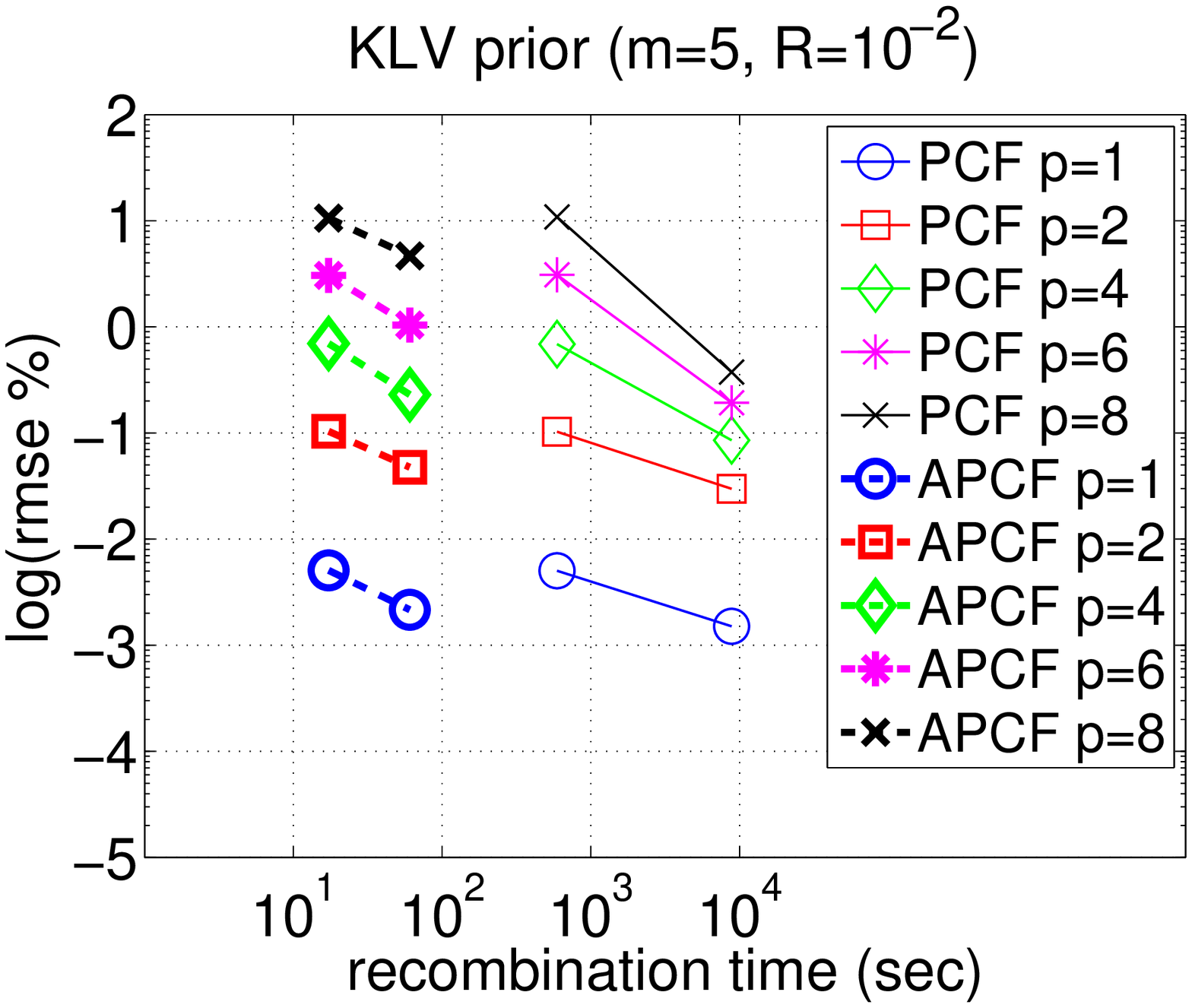} \label{fig:Rb}} 
} 

\centerline{
\subfigure[bootstrap reweighted prior samples]
{\includegraphics[width=0.43\textwidth]{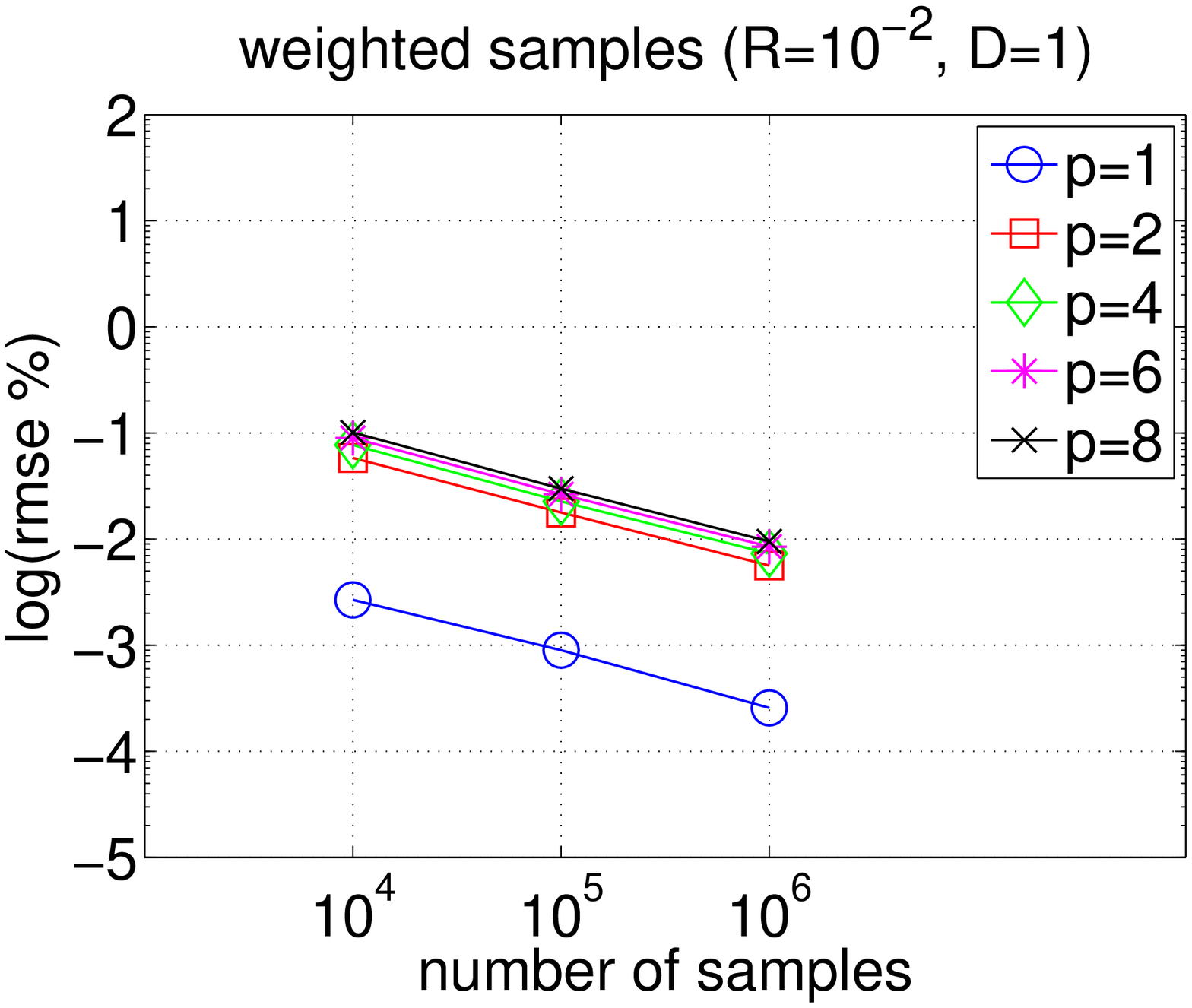} \label{fig:Rc}} 
\subfigure[bootstrap reweighted prior samples]
{\includegraphics[width=0.43\textwidth]{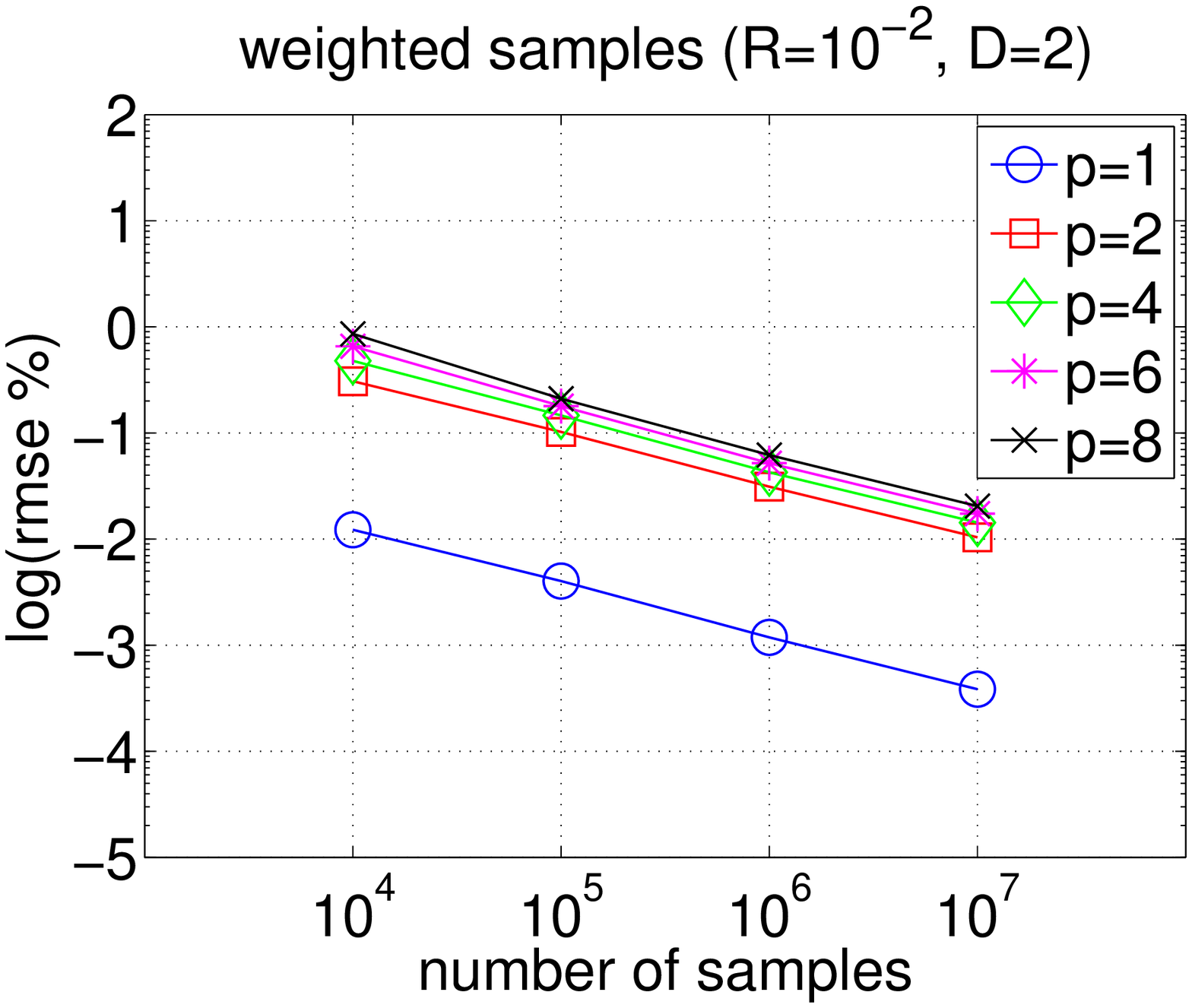} \label{fig:Re}} 
\subfigure[bootstrap reweighted prior samples]
{\includegraphics[width=0.43\textwidth]{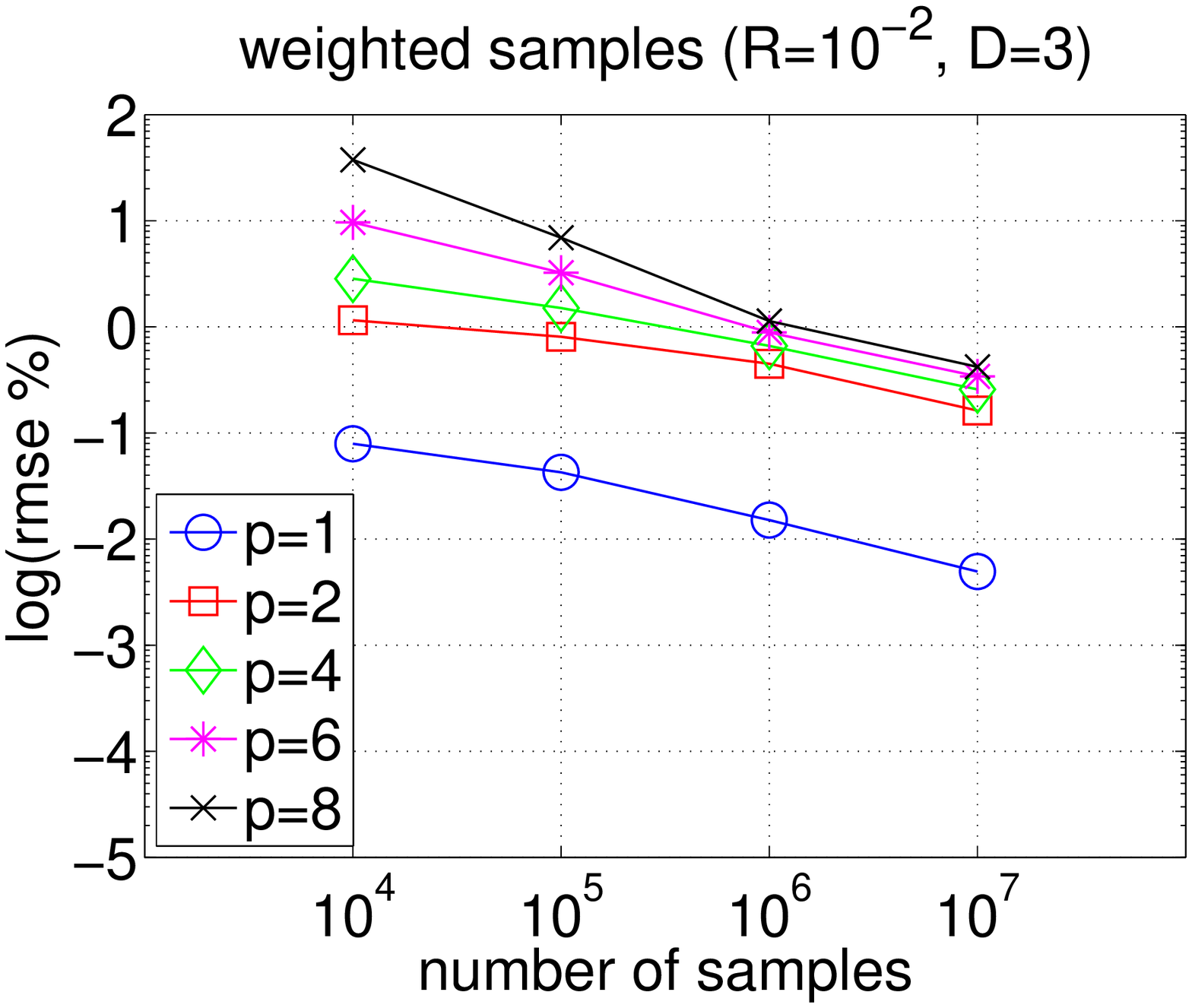} \label{fig:Rg}} 
}

\centerline{
\subfigure[weighted posterior samples]
{\includegraphics[width=0.43\textwidth]{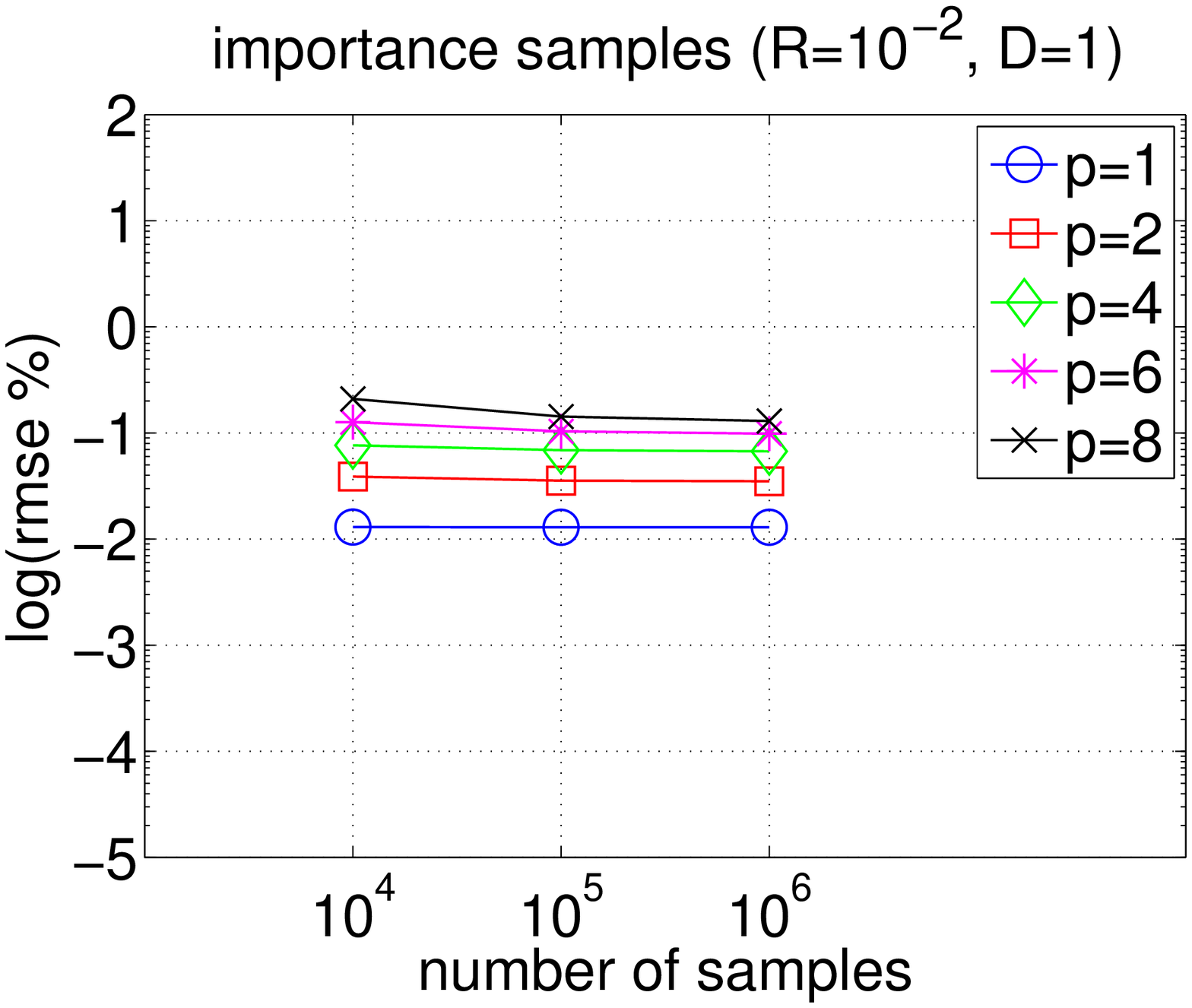} \label{fig:cd}} 
\subfigure[weighted posterior samples]
{\includegraphics[width=0.43\textwidth]{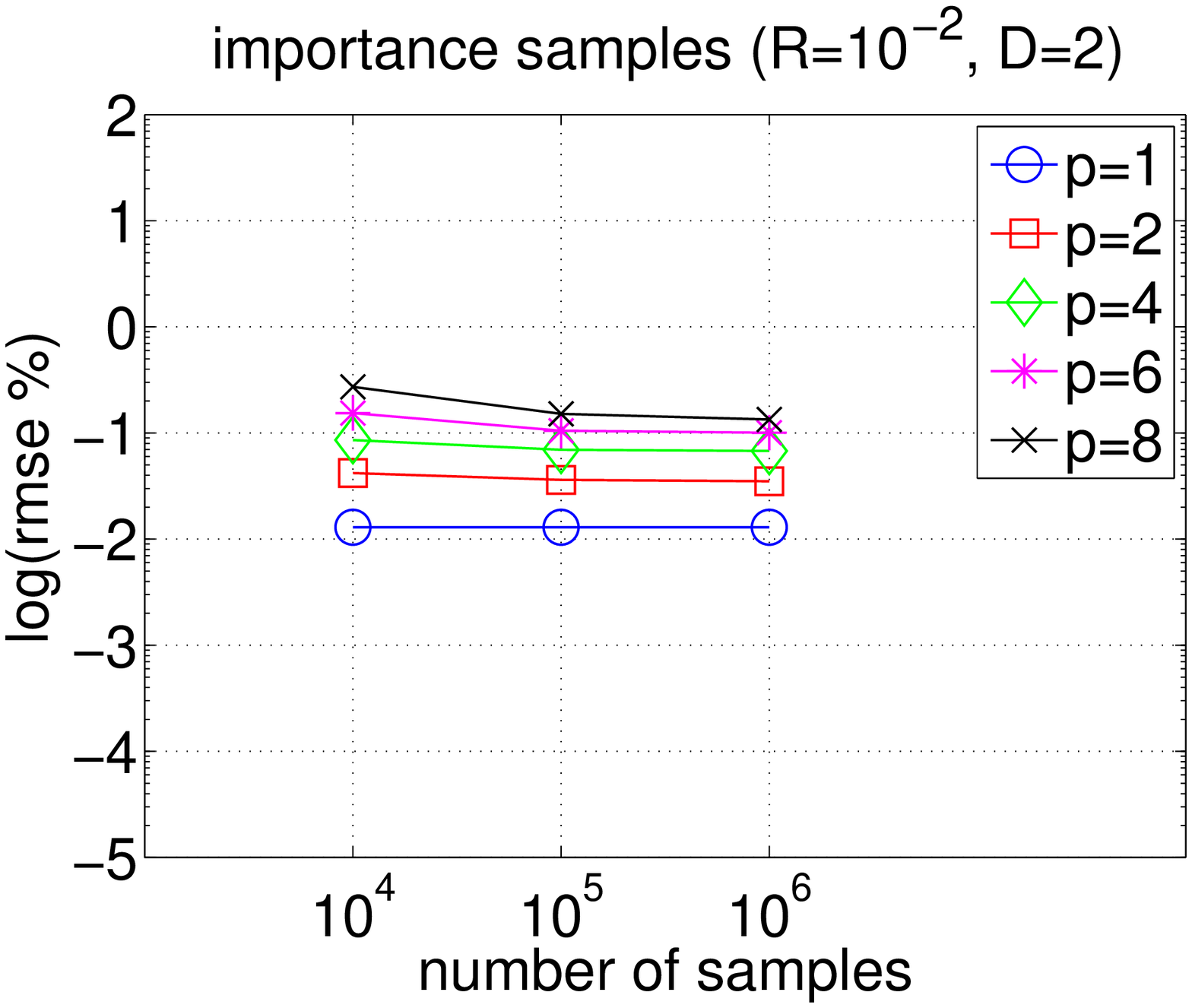} \label{fig:ef}} 
\subfigure[weighted posterior samples]
{\includegraphics[width=0.43\textwidth]{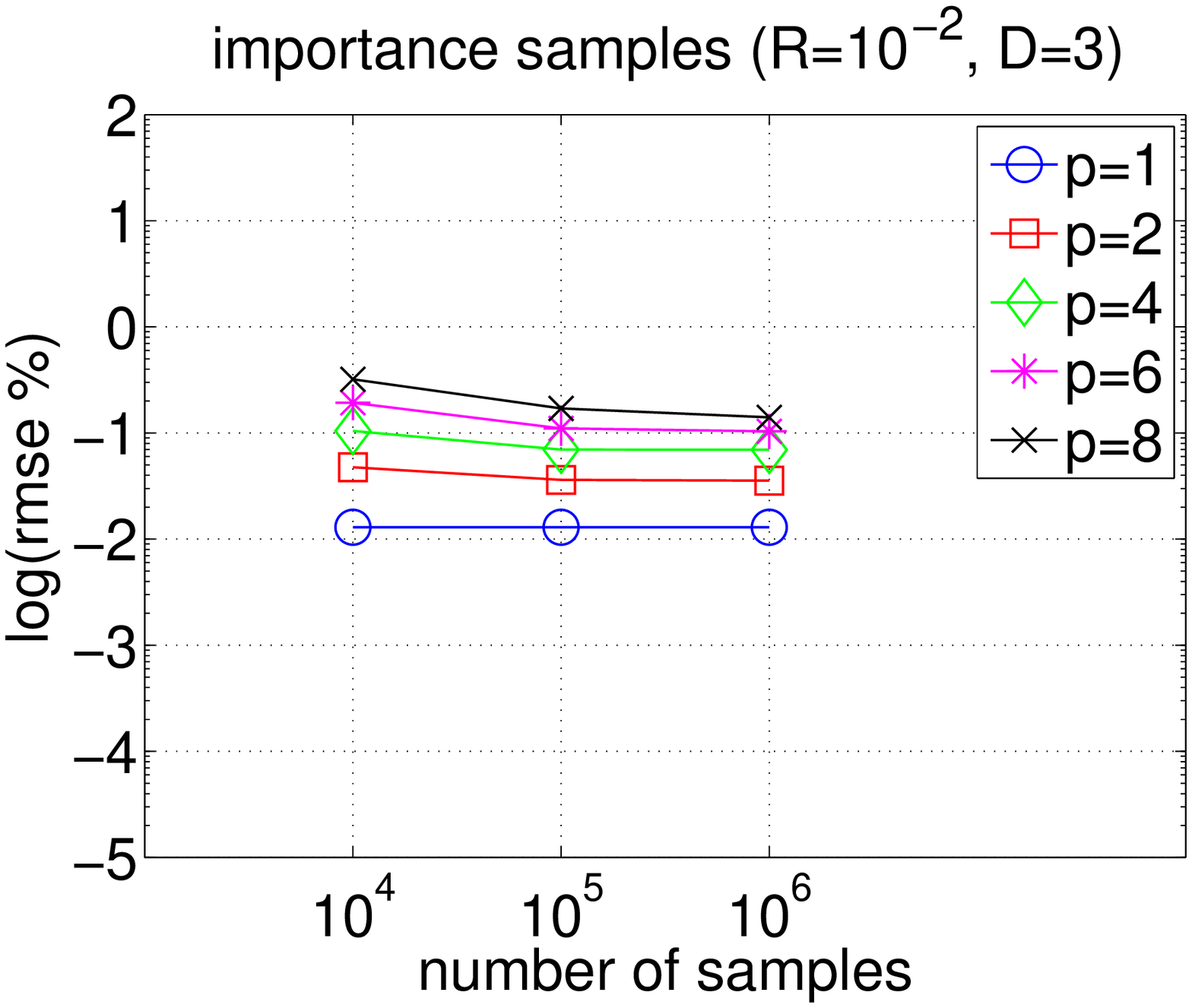} \label{fig:gh}} 
}

\centerline{
\subfigure[cubature approximation of posterior]
{\includegraphics[width=0.43\textwidth]{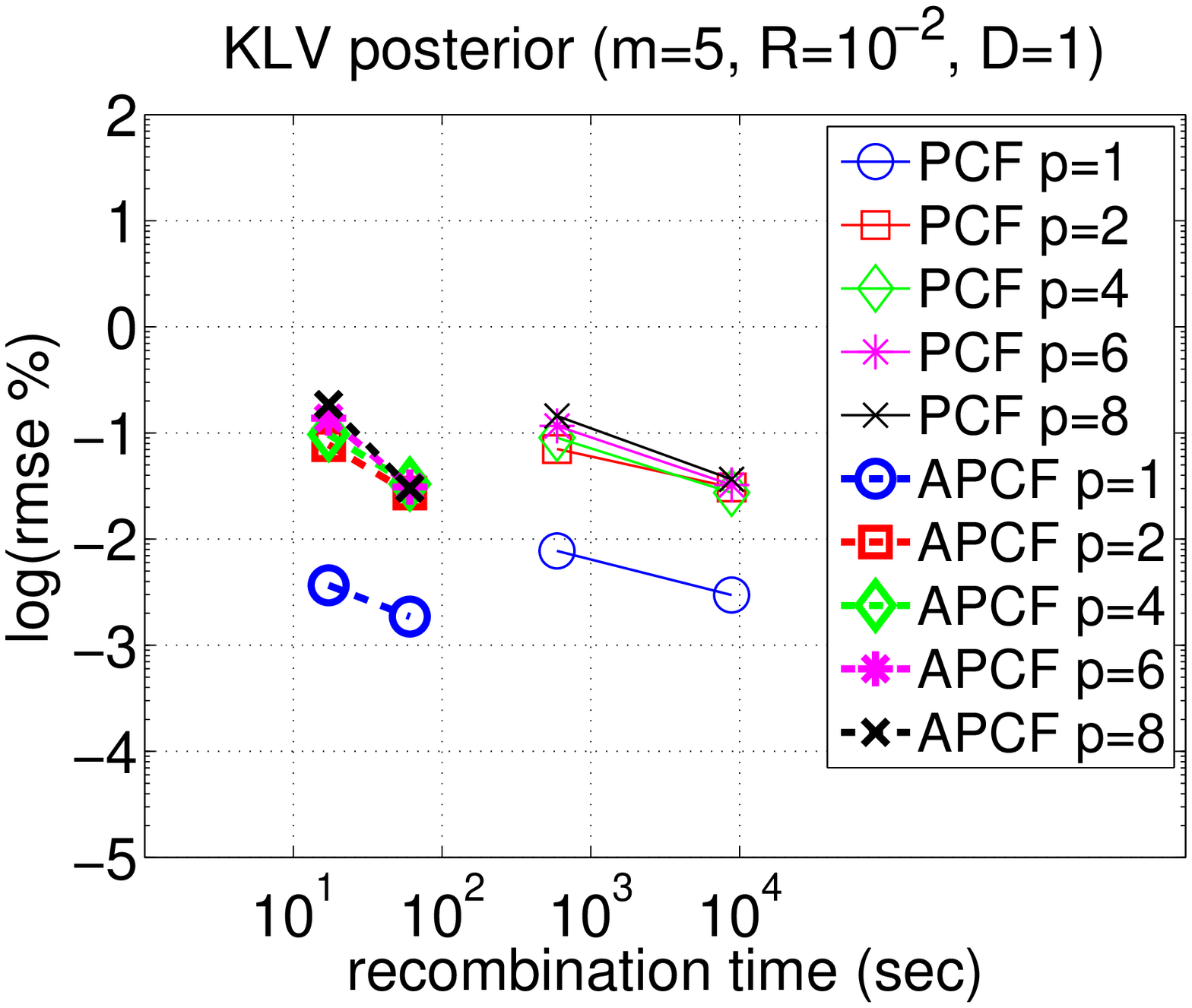} \label{fig:Rd}} 
\subfigure[cubature approximation of posterior]
{\includegraphics[width=0.43\textwidth]{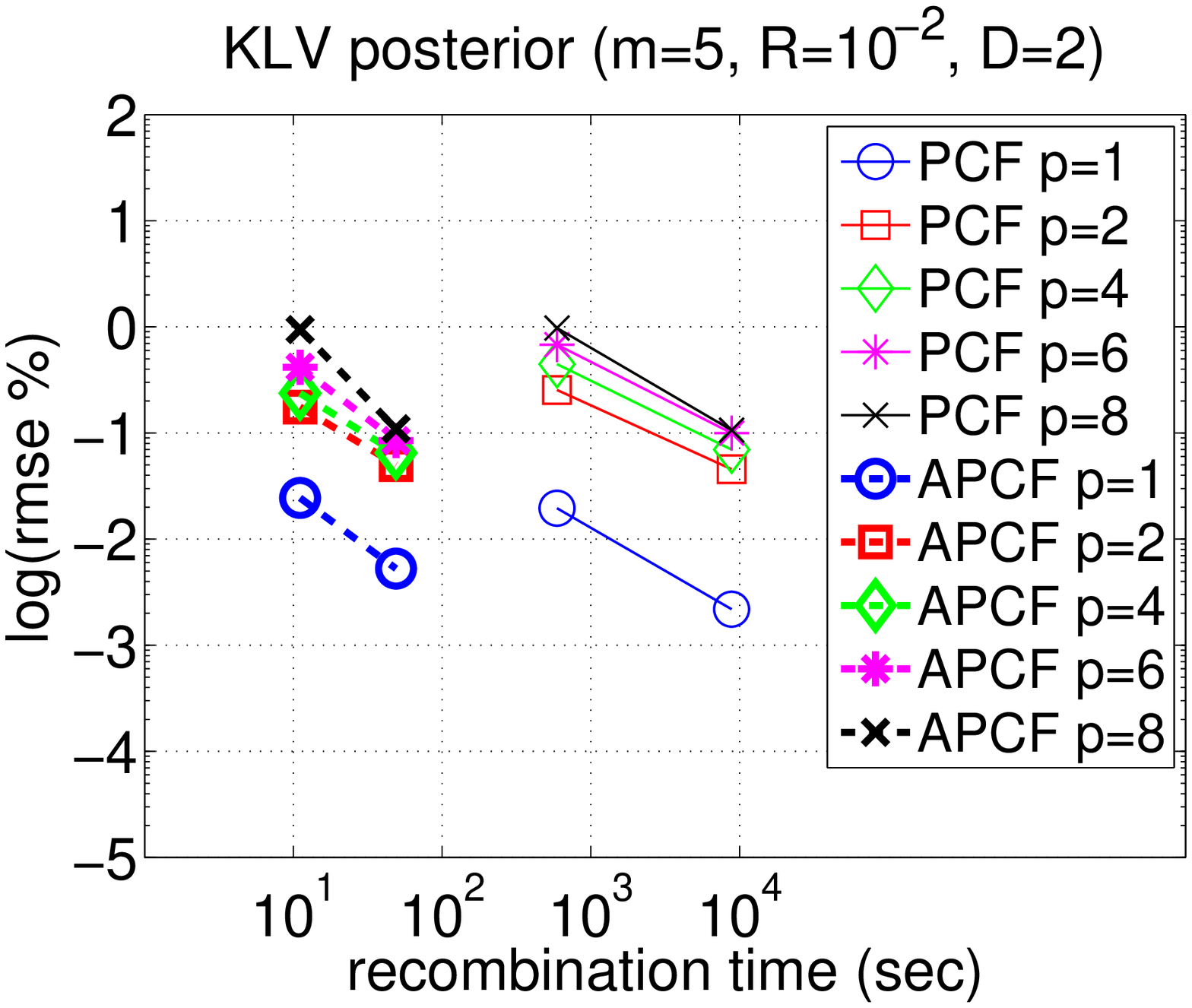} \label{fig:Rf}} 
\subfigure[cubature approximation of posterior]
{\includegraphics[width=0.43\textwidth]{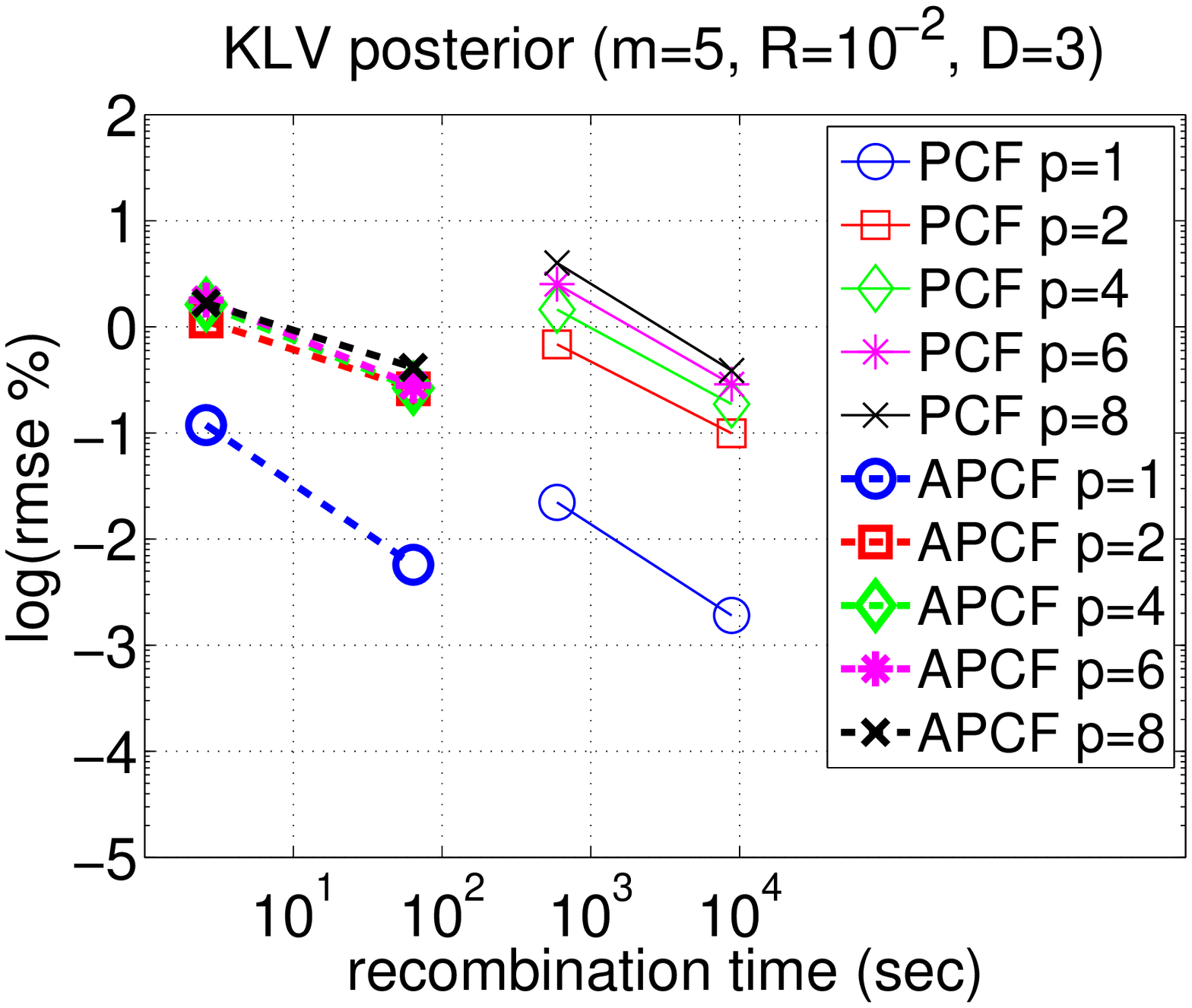} \label{fig:Rh}} 
}
\caption{
The prior and posterior approximations 
when $R=10^{-2}$ is fixed and $D=1,2,3$ varies. 
The top row is for the prior and the rest bottom three rows are for the posterior.
The second row (SIR) and 
the third row (SISR) are
from Monte-Carlo samples.
The last bottom row is from cubature approximation when $\epsilon=10^{-2}, 10^{-3}$.
} 
\label{fig:R} 
\end{figure*}

\begin{figure}
\centerline{
\subfigure[unweighted posterior samples]
{\includegraphics[width=0.43\textwidth]{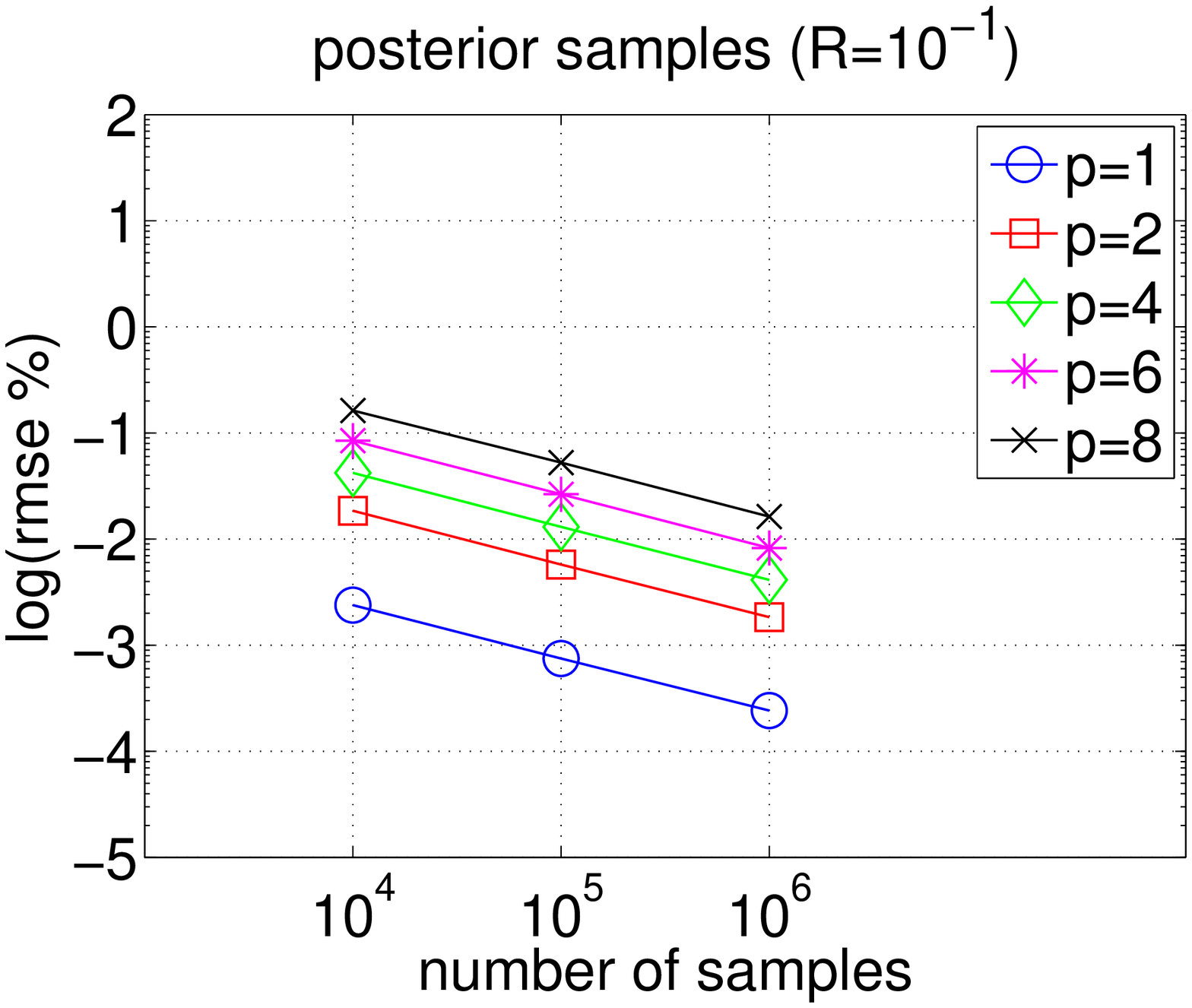} \label{fig:Da} } 
\subfigure[unweighted posterior samples]
{\includegraphics[width=0.43\textwidth]{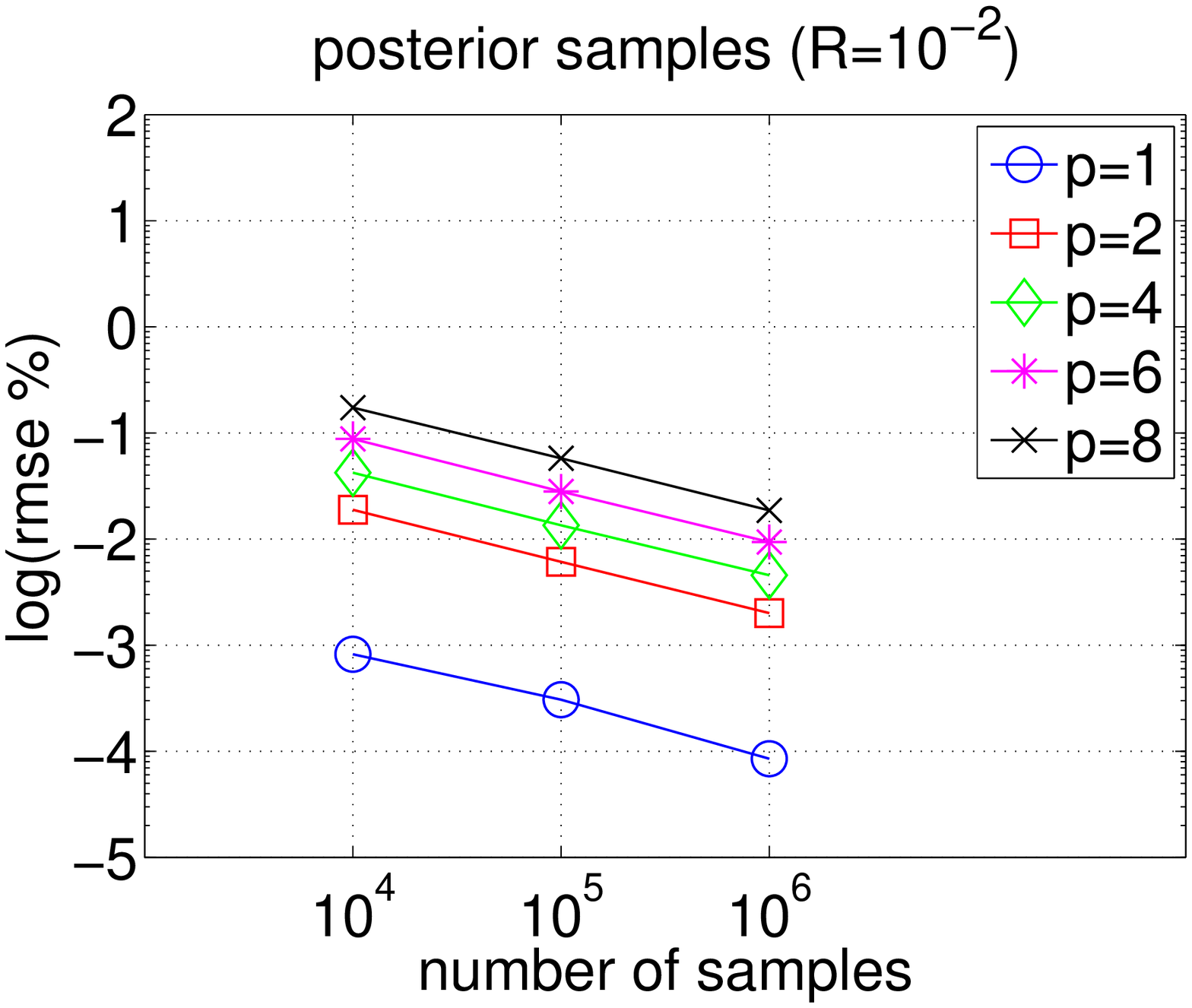} \label{fig:Dd}} 
\subfigure[unweighted posterior samples]
{\includegraphics[width=0.43\textwidth]{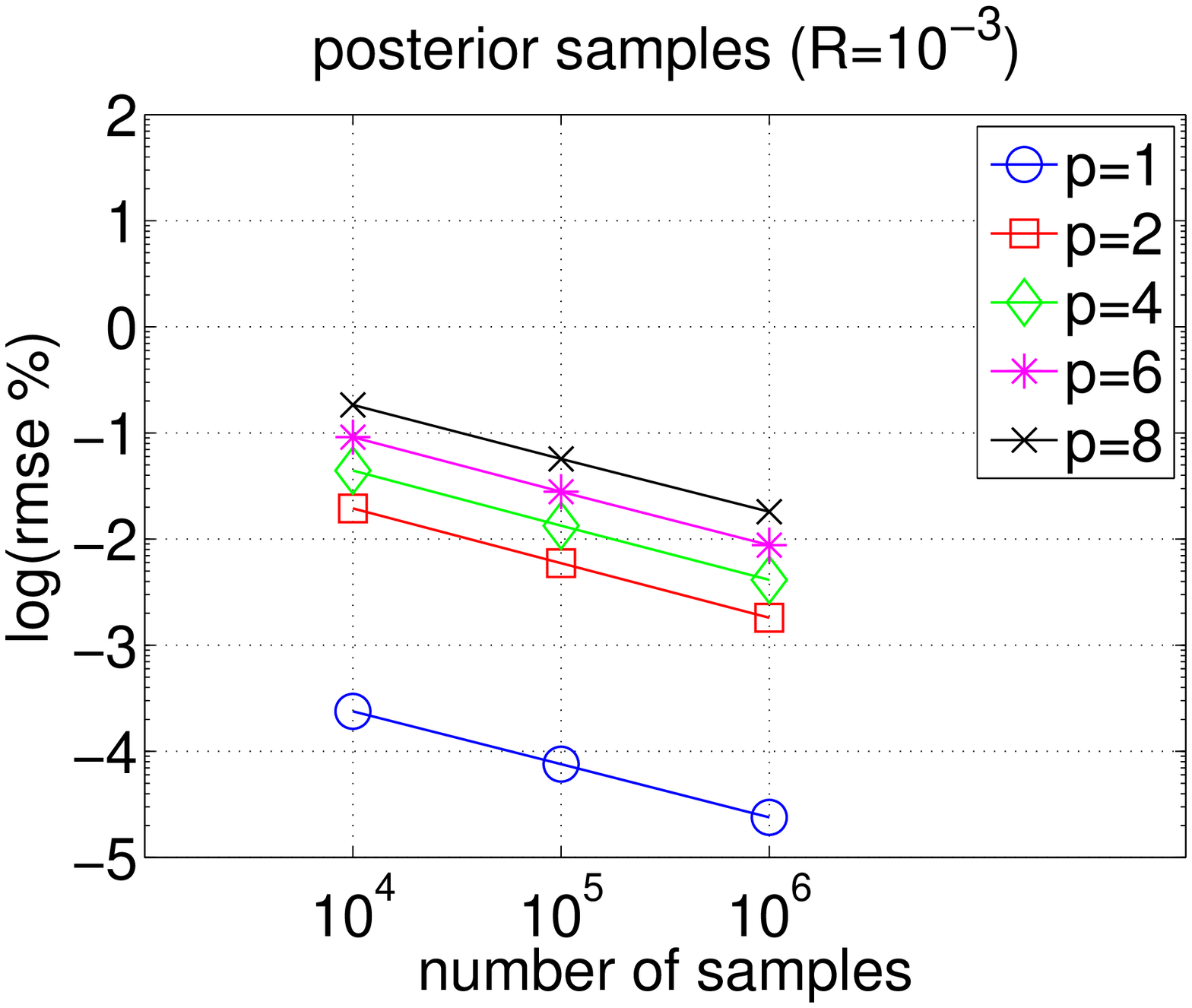} \label{fig:Dg}} 
} 

\centerline{
\subfigure[bootstrap reweighted prior samples]
{\includegraphics[width=0.43\textwidth]{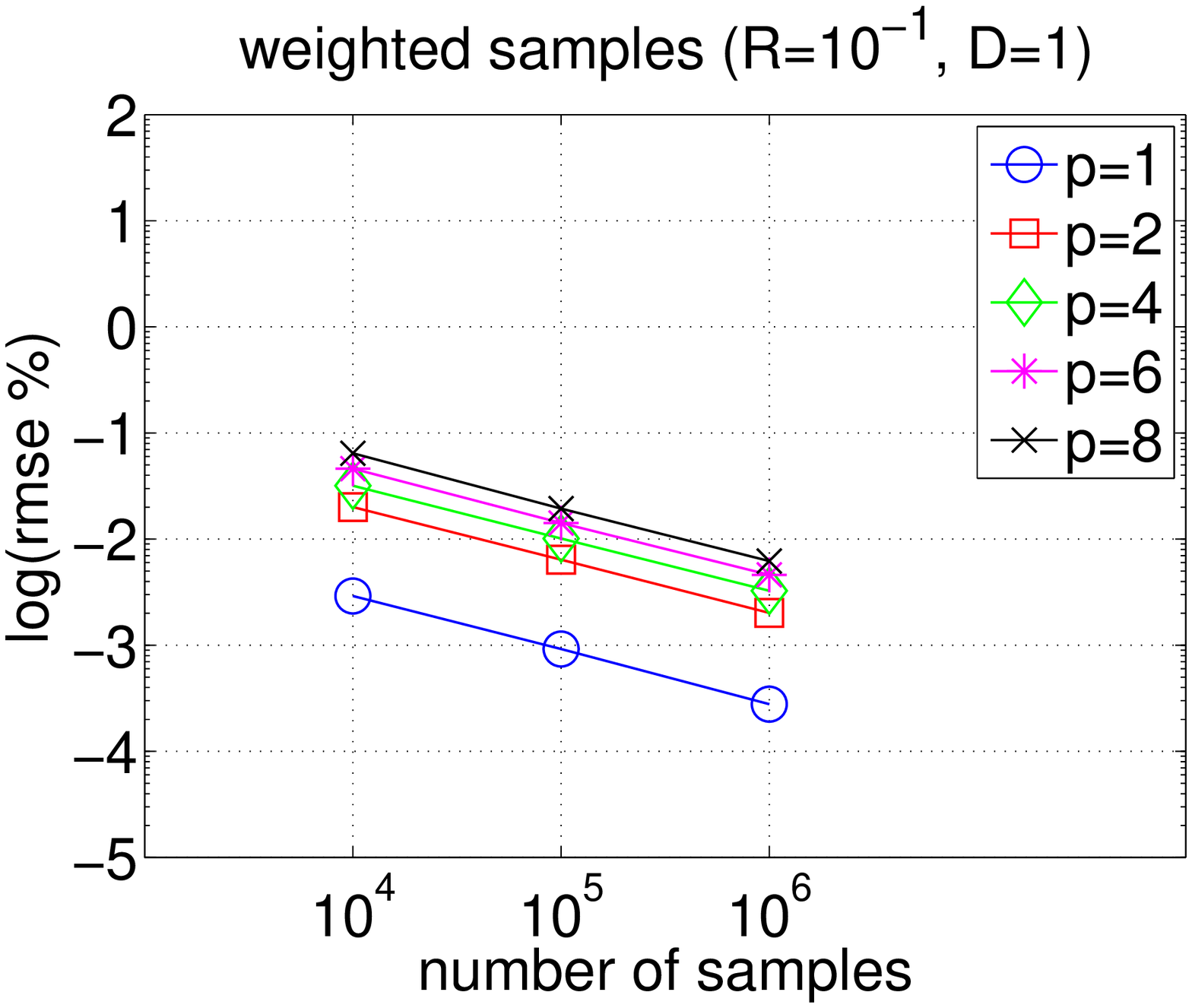} \label{fig:Db} } 
\subfigure[bootstrap reweighted prior samples]
{\includegraphics[width=0.43\textwidth]{weightedpriorsamples_Rp01D1.eps} \label{fig:De}} 
\subfigure[bootstrap reweighted prior samples]
{\includegraphics[width=0.43\textwidth]{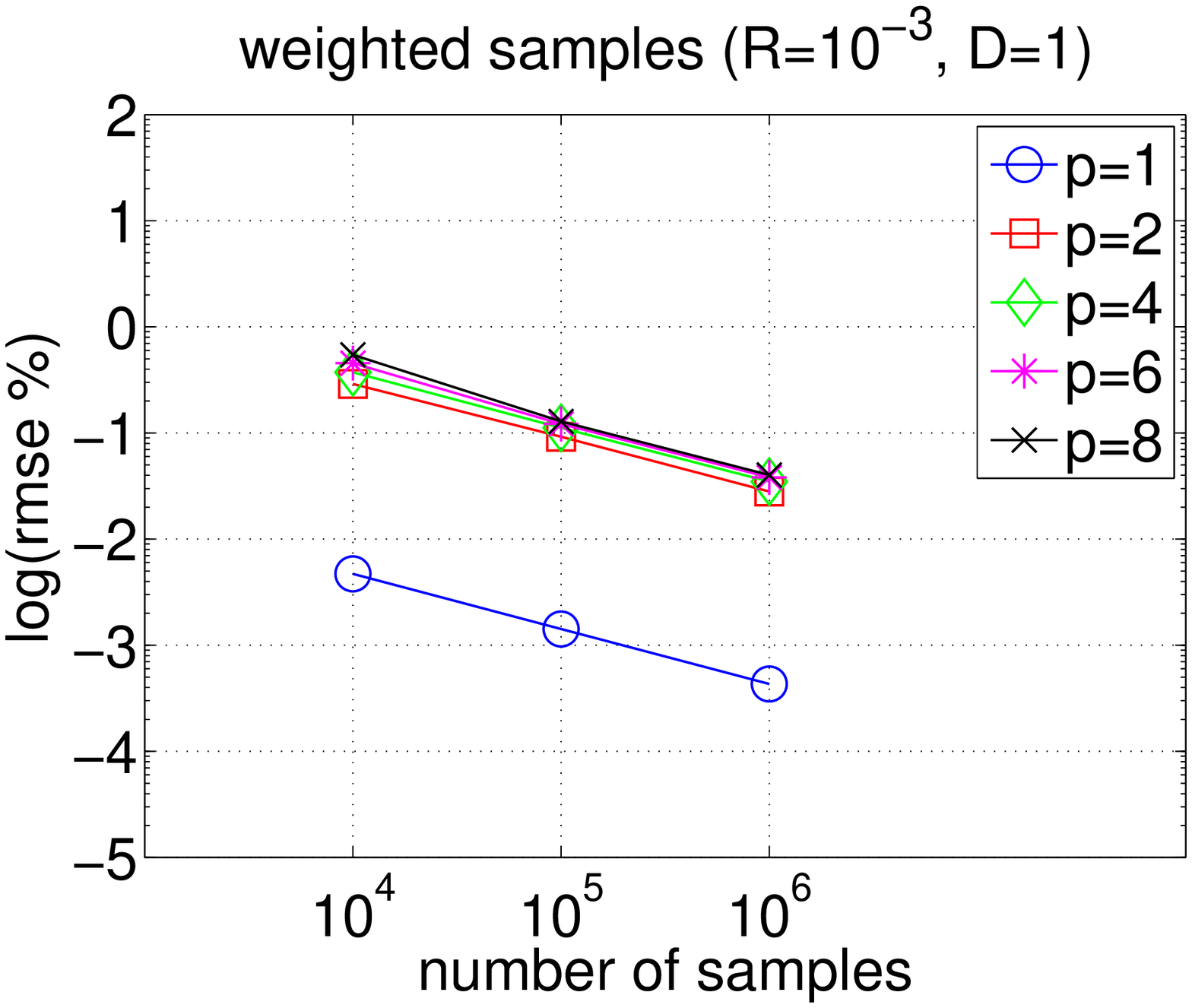} \label{fig:Dh}} 
}

\centerline{
\subfigure[weighted posterior samples]
{\includegraphics[width=0.43\textwidth]{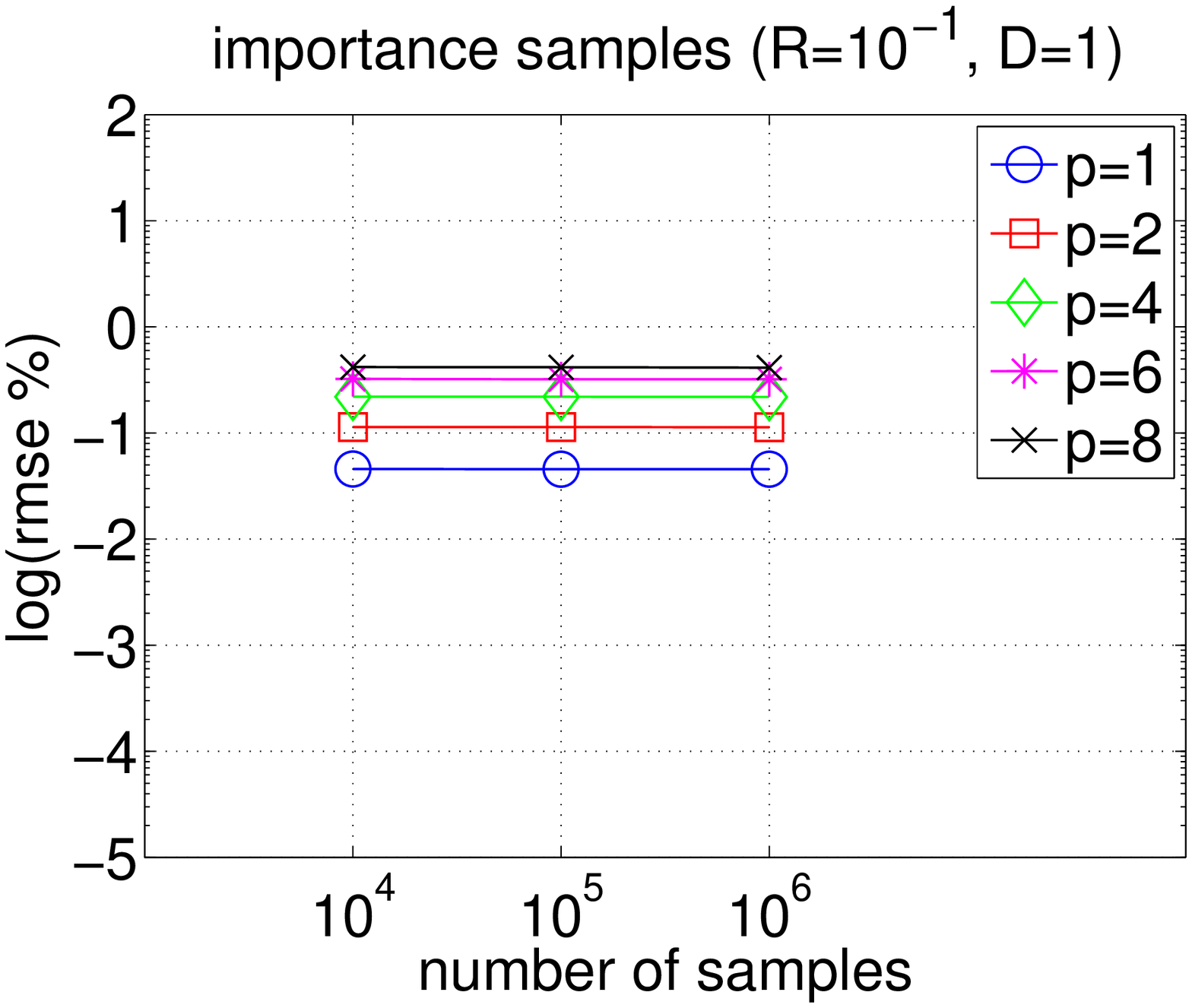} \label{fig:iDa} } 
\subfigure[weighted posterior samples]
{\includegraphics[width=0.43\textwidth]{importancesamples_Rp01D1.eps} \label{fig:iDb} } 
\subfigure[weighted posterior samples]
{\includegraphics[width=0.43\textwidth]{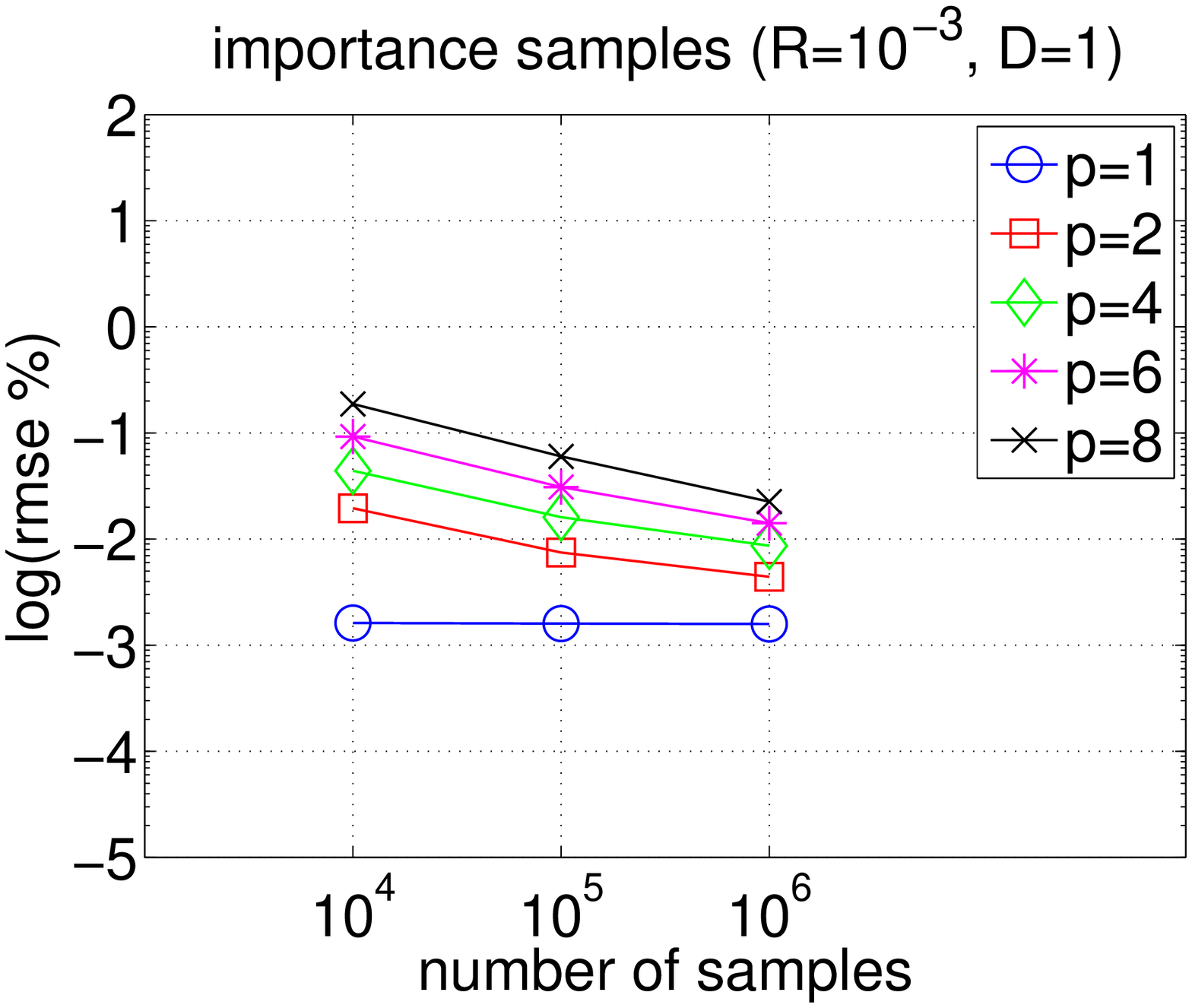} \label{fig:iDc} } 
}

\centerline{
\subfigure[cubature approximation of posterior]
{\includegraphics[width=0.43\textwidth]{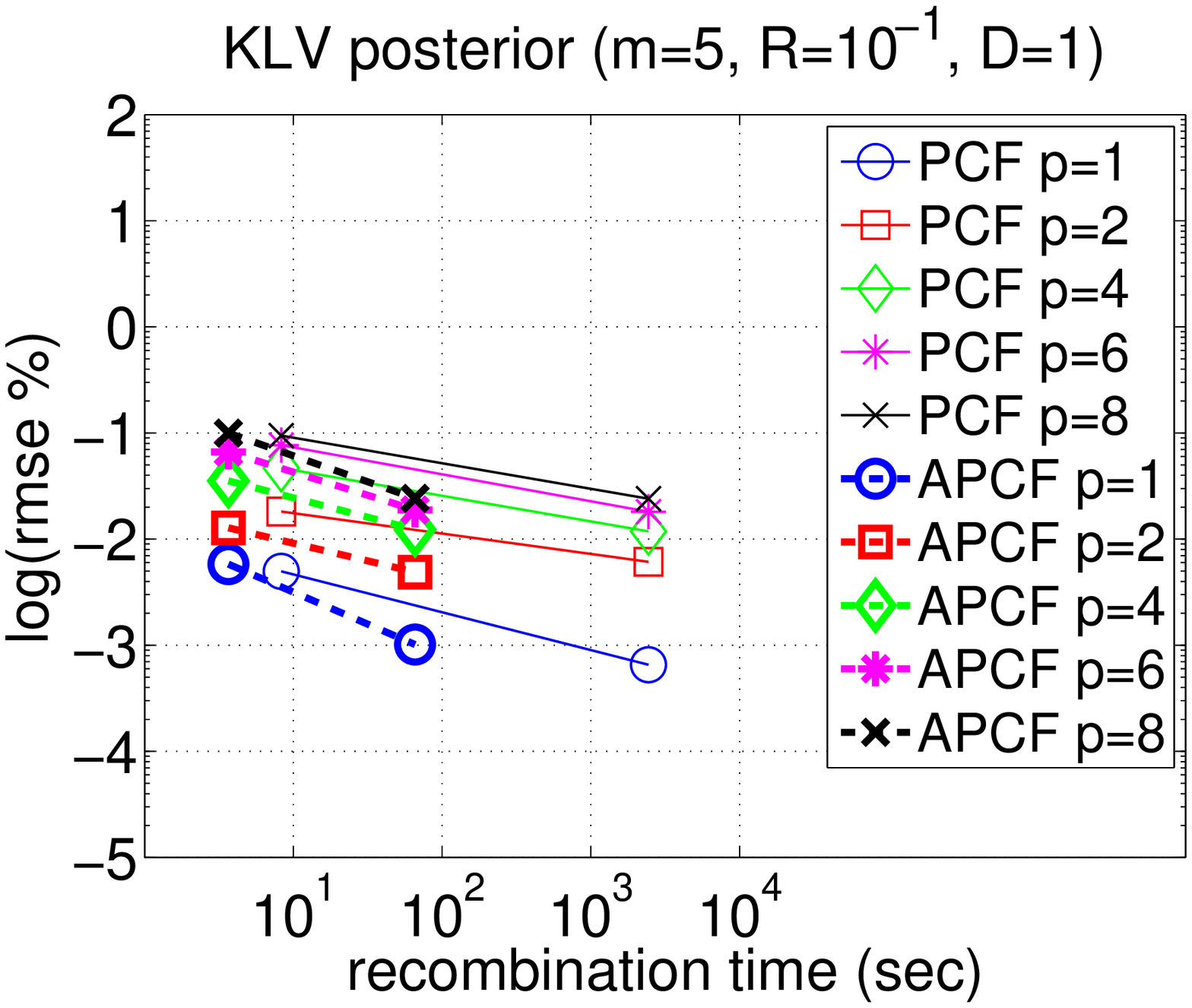} \label{fig:Dc}} 
\subfigure[cubature approximation of posterior]
{\includegraphics[width=0.43\textwidth]{cubpos_rp01d1.eps} \label{fig:Df}} 
\subfigure[cubature approximation of posterior]
{\includegraphics[width=0.43\textwidth]{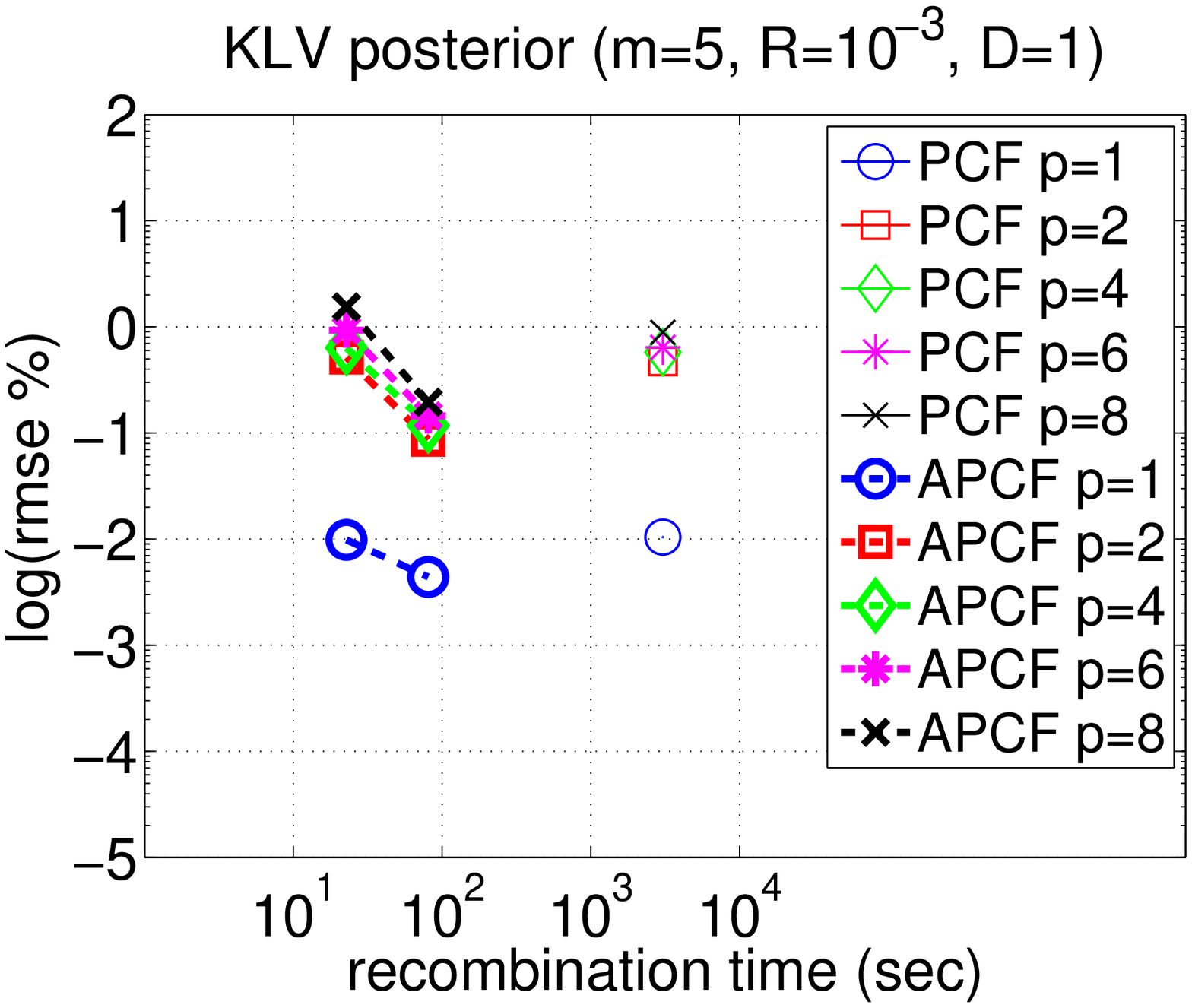} \label{fig:Di}} 
}
  \caption{
The posterior approximations when $D=1$ is fixed and $R=10^{-1},10^{-2},10^{-3}$ varies.
The top three rows are from Monte-Carlo samples.
The first row is from direct sampling of posterior,
the second is from SIR and the third is from SISR.
The last bottom row is from cubature approximation when $\epsilon=10^{-2}, 10^{-3}$.
In Fig.~\ref{fig:Di}, the PCF with $\epsilon = 10^{-3}$ is not shown.
} 
\label{fig:D} 
\end{figure}

When $R=10^{-2}$ is fixed and $D=1,2,3$ varies,
the relative $L^2$ errors of the $p$-th moments of PCF and APCF
are shown in
Figs.~\ref{fig:Rb},
\ref{fig:Rd},
\ref{fig:Rf},
\ref{fig:Rh}.
We have implemented two cases of
$\epsilon=10^{-2}$
and 
$\epsilon=10^{-3}$.
The recombination times are measured
using Visual Studio
with Intel $2.53$ GHz processor
(the autonomous ODEs are solved analytically).
Fig.~\ref{fig:R} reveals the following.
\begin{itemize}
  \item
	The prior approximation of PCF with $\epsilon=10^{-3}$ shows similar
	accuracy with $10^4$ Monte-Carlo sampling (Figs.~\ref{fig:Ra}, \ref{fig:Rb}).
  \item The accuracy of the APCF prior approximation is in general worse than PCF especially for higher order
	moments (Fig.~\ref{fig:Rb}). 
  \item 

As the observation is located further far from the prior mean,
i.e., as $D$ increases,
the posterior approximation obtained from Monte-Carlo bootstrap reweighting 
(SIR)
becomes 
less accurate 
(Figs.~\ref{fig:Rc}, \ref{fig:Re}, \ref{fig:Rg}).
As the number of samples $M$ increases, the error reduction
asymptotically
scales as $M^{-1/2}$
in all cases.

\item 
Unlike the case of SIR,
the accuracy of the importance samples (SISR) is not significantly influenced by 
the observation location
as well as the number of samples
(Figs.~\ref{fig:cd}, \ref{fig:ef}, \ref{fig:gh}).
This sample size insensitivity
is presumably because SISR duplicates the samples in this parameter regime
(compare with Fig.~\ref{fig:iDc}).

\item The accuracy of the APCF posterior approximation
is similar to PCF but APCF significantly reduces the recombination time
which is insensitive to $D$
(Figs.~\ref{fig:Rd}, \ref{fig:Rf}, \ref{fig:Rh}).

\item The accuracy of the PCF and APCF posterior approximations with $\epsilon=10^{-2}$
 is similar to $10^4$ Monte-Carlo reweighted samples (SIR) when $D=1,2$ 
(Figs.~\ref{fig:Rc}, \ref{fig:Rd}, \ref{fig:Re}, \ref{fig:Rf})
 and to $10^5$ reweighted samples (SIR) when $D=3$
(Figs.~\ref{fig:Rg}, \ref{fig:Rh}).

\item The accuracy of the PCF and APCF posterior approximations with $\epsilon=10^{-3}$
 is similar to $10^5$ Monte-Carlo reweighted samples when $D=1$
(Figs.~\ref{fig:Rc}, \ref{fig:Rd}),
to $10^6$ reweighted samples when $D=2$
(Figs.~\ref{fig:Re}, \ref{fig:Rf})
and 
to $10^7$ reweighted samples when $D=3$
(Figs.~\ref{fig:Rg}, \ref{fig:Rh}).

\item 
The accuracy of the PCF and APCF posterior approximations with $\epsilon=10^{-3}$
is superior to 
$10^6$
importance samples (SISR) when $D=1$
(Figs.~\ref{fig:cd}, \ref{fig:Rd}),
comparable to SISR when $D=2$
(Figs.~\ref{fig:ef}, \ref{fig:Rf}),
inferior to SISR when $D=3$
(Figs.~\ref{fig:gh}, \ref{fig:Rh}),
in approximating higher order moments.

\end{itemize}

There is an important insight to be gained from this experimental analysis.
Though PCF
produces a more accurate description of the
prior measure
than APCF, 
the one from this naive approximation of the prior 
is not better at approximating the posterior. 
The point is that one needs an
extremely accurate representation of the prior in certain localities. 
While
APCF
delivers
this without undue cost,
the 
PCF
method would have to deliver this accuracy uniformly and
well out into the tail of the prior.
As a result,
for the posterior approximation,
APCF can achieve a similar accuracy with PCF
but using significantly less computational cost.

In this example,
the computational cost (recombination time)
of PCF and APCF 
is uniform and irrespective of $D$ for given 
$\epsilon=10^{-2}, 10^{-3}$.
However when one uses SIR to achieve the
accuracy due to APCF with $\epsilon=10^{-3}$
in approximating higher order moments,
one needs more computational resources 
(large number of particles)
as $D$ becomes bigger.
One also cannot expect an accuracy improvement from SISR except the rare event case
($D=3$).
Therefore,
in the reliability aspect,
APCF is clearly advantageous over sequential Monte-Carlo methods.

\subsubsection{Dependence on the observation noise error}
When $D=1$ is fixed and $R = 10^{-1},10^{-2},10^{-3}$ varies,
the values of 
Eq.~(\ref{eq:rmse})
for PCF and APCF
are shown in
Figs.~\ref{fig:Dc}, \ref{fig:Df}, \ref{fig:Di}.
We have implemented two cases of
$\epsilon=10^{-2}$
and 
$\epsilon=10^{-3}$.
Fig.~\ref{fig:D} 
reveals the following.
\begin{itemize}
  \item 
	The high order moment approximations errors due to Monte-Carlo Gaussian samples 
(direct sampling of the posterior)
are insensitive to its covariance
	(recall the diagonal element of $C_{n|n}$ is of the same order with the value of $R$)
(Figs.~\ref{fig:Da}, \ref{fig:Dd}, \ref{fig:Dg}).

\item 
As the likelihood becomes narrower,
i.e., as $R$ decreases,
the posterior approximation obtained from Monte-Carlo bootstrap reweighting 
(SIR)
becomes 
less accurate 
(Figs.~\ref{fig:Db}, \ref{fig:De}, \ref{fig:Dh}).

\item The accuracy of importance samples (SISR) 
tends to increase as $R$ decreases
(Figs.~\ref{fig:iDa}, \ref{fig:iDb}, \ref{fig:iDc}).
In particular, when $R=10^{-3}$, 
the moment approximations of SISR is comparable with those from direct sampling of the posterior
except the mean
(Figs.~\ref{fig:Dg}, \ref{fig:iDc}).

\item 
	As $R$ decreases,
  the recombination time 
  needed
  to achieve a given degree of accuracy
  becomes bigger for PCF
  but this is not the case for APCF, i.e., 
  the recombination time for APCF is insensitive to $R$
(Figs.~\ref{fig:Dc}, \ref{fig:Df}, \ref{fig:Di}).
\end{itemize}

The simulation shows that 
APCF again achieves a similar accuracy with PCF in all cases but,
as the observation noise error decreases,
APCF becomes more competitive than PCF
for the solution of the intermittent data assimilation problem.
It further shows that APCF is of higher order with respect to the recombination time
and can achieve the given degree of accuracy with lower computational cost.

Although $Y_n$ is \emph{there and measurable} it is sometimes the case 
that it is actually computationally very expensive to compute and 
that actually the thing one can compute is the evaluation of likelihood for a number of locations.
For example, consider a tracking problem for an object of moderate intensity 
and diameter that does a random walk and is moving against a slightly noisy background 
and is observed relatively infrequently. Its influence is entirely local. The likelihood function 
will be something like the Gaussian centred at the position of object but completely uninformative 
elsewhere in the space. The smaller the object, the tighter or narrower the Gaussian the harder 
the problem of finding the object becomes. 
One can compute the likelihood at any point in the space, 
but only evaluations at the location of the particle are informative. In that way one sees that
\begin{enumerate}
  \item 
The $Y_n$ is \emph{observable} but only partially observed - and with low noise is very expensive to
observe accurately as one has to find the particle.
\item The likelihood can be observed at points in the space.
\end{enumerate}
In this sort of example it would be quite wrong to assume that, 
if we know the prior distribution of $X_n$ then  just because $Y_n = X_n + \eta_n$ 
we know the posterior distribution at zero cost. 
For sequential Monte-Carlo methods, 
bootstrap reweighting would seem to give a much better approach.

\subsection{Prospective performance PCF and APCF with cubature on Wiener space of degree $7$}
\label{sec:PCFaPCFd7}
A cubature formula on Wiener space of degree $m \geq 7$ is currently not 
available 
when $d=3$.
However, 
in our problem setting,
we are able to emulate
a prospective
performance
of higher order cubature formula
using 
Gauss-Hermite quadrature.

For the linear dynamics
satisfying
\begin{equation*}
X(\Delta)=F_\Delta X(0) + \nu_\Delta, \quad \nu_\Delta \sim \mathcal{N}\left( 0, Q_\Delta \right)
\end{equation*}
where $F_\Delta \in \mathbb{R}^{3\times 3}$ is a matrix,
we define the forward operator 
\begin{equation}
  \label{eq:GHC}
{\text{GHC}}^{(m)}
\left(\sum_{i=1}^n \kappa_i \delta_{x^i}, \Delta \right)
\equiv \sum_{i=1}^n \sum_{j=1}^{n_m} \kappa_i \lambda_j \delta_{ F_\Delta x^i+ z^j}
\end{equation}
where $\{\lambda_j,z^j\}_{j=1}^{n_m}$ is a Gauss-Hermite cubature of degree $m$
with respect to the law of $\nu_\Delta$.
The authors have seen that
the performance of GHC 
is similar to KLV on the flow level when $m=3, 5$
and that
Eq.~(\ref{eq:GHC})
can be used as an alternative to 
Eq.~(\ref{eq:KLVflow})
in the application of 
PCF and APCF to the test model.

\begin{table}
  \renewcommand{\arraystretch}{1.3} 
  \caption{The number of adaptive partition $k$ for GHC with $m=7$} 
  \label{tab:partm7} 
  \centering \begin{tabular}{ c |  c  c cc} 
	& $\epsilon = 10^{-2}$ & $\epsilon =10^{-3}$ & $\epsilon =10^{-4}$ & $\epsilon =10^{-5}$  \\ \hline
	$R=10^{-1}$  & 2 & 4 & 6 & 10  \\ 
	$R=10^{-2}$  & 5 & 9 & 16 & 28  \\ 
	$R=10^{-3}$  & 9 & 17 & 30 & 54  \\ 
  \end{tabular} 
\end{table}

\begin{figure}[t]
\centerline{
\includegraphics[width=0.49\textwidth]{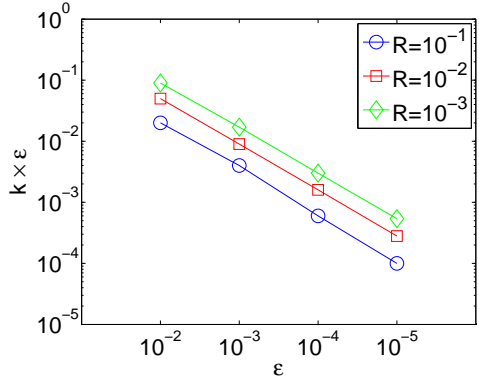}
}
\caption{
The upper bound of the total error along with the adaptive partition when $m=7$.
}
\label{fig:adaptivepartition_m7}
\end{figure}

\begin{figure}[t]
\centerline{
\subfigure[bootstrap reweighted prior samples]
{\includegraphics[width=0.43\textwidth]{weightedpriorsamples_Rp01D2.eps} \label{fig:D2a}} 
\subfigure[cubature approximation of posterior]
{\includegraphics[width=0.43\textwidth]{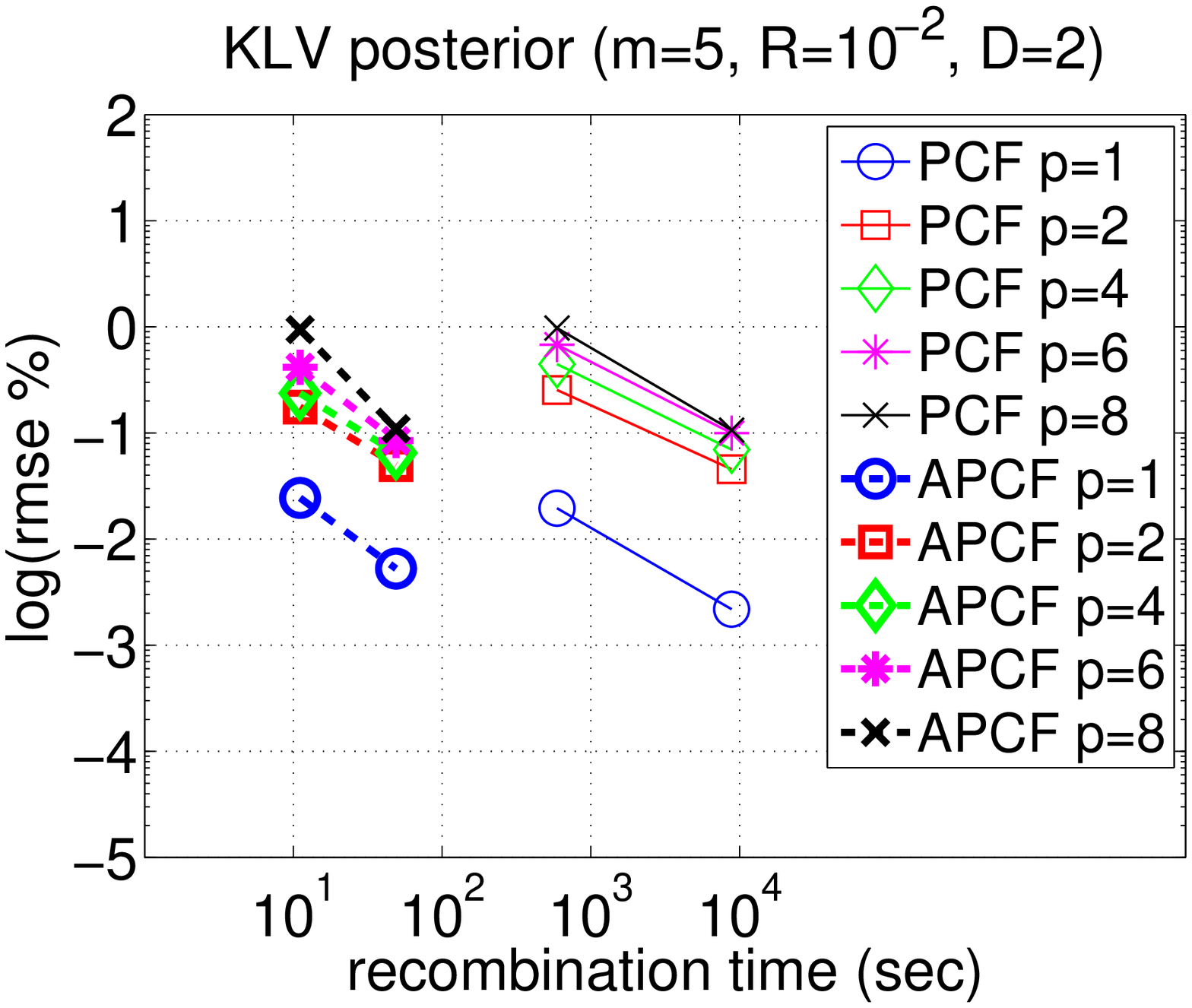} \label{fig:D2b}} 
\subfigure[cubature approximation of posterior]
{\includegraphics[width=0.43\textwidth]{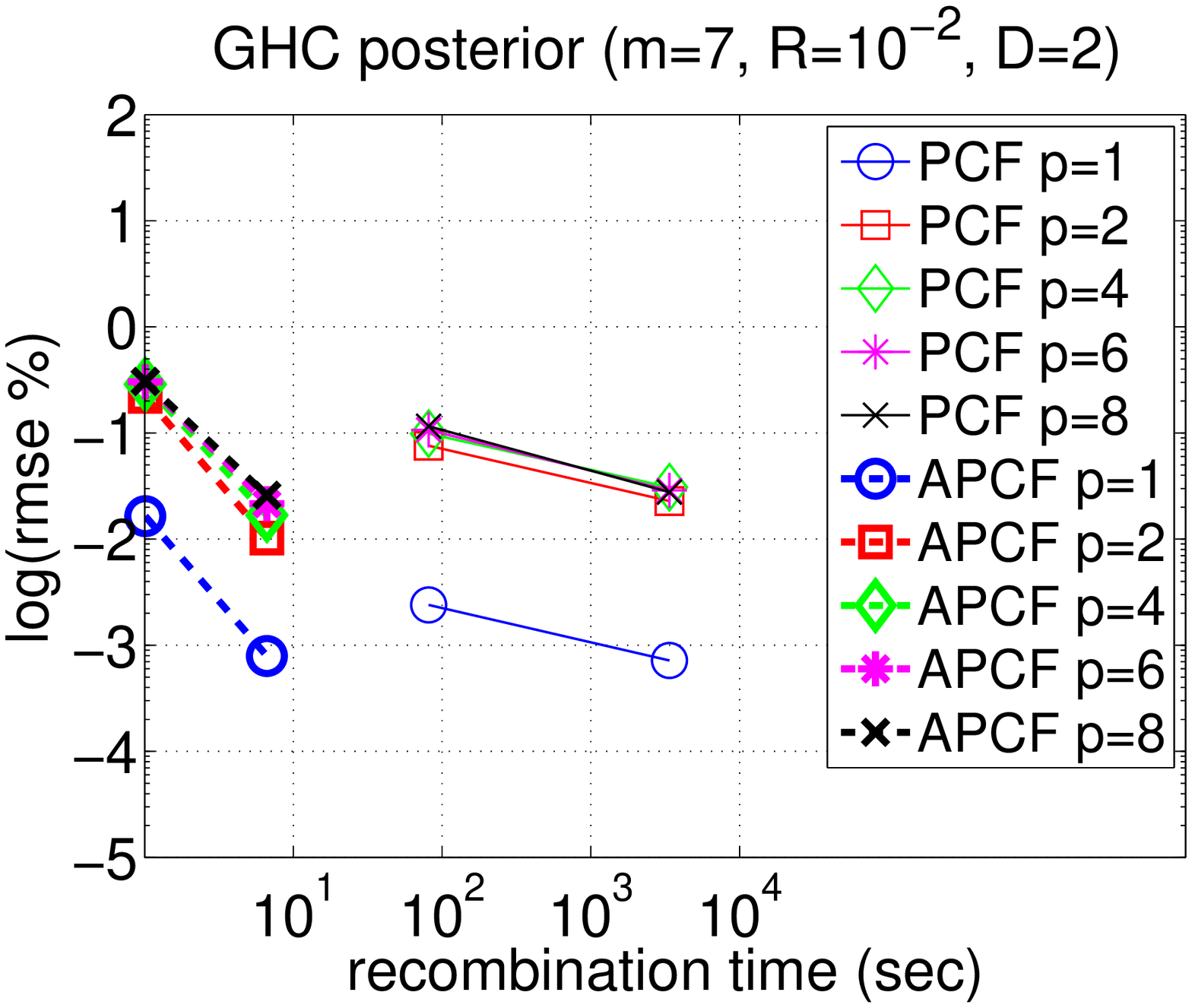} \label{fig:D2c}} 
} 
  \caption{
The posterior approximations when $R=10^{-2}$ and $D=2$.
The left is from Monte-Carlo samples 
and the middle and right is from cubature approximation when $\epsilon=10^{-2}, 10^{-3}$.
} 
\label{fig:D2} 
\end{figure}

The number of iterations $k$ in the adaptive partition,
obtained from using 
GHC
with
Gauss-Hermite cubature of degree $m=7$ whose support size is
$n_m = 64$
in place of $Q^{m}_{s_j}$,
is shown in 
table~\ref{tab:partm7}.
Here
Fig.~\ref{fig:adaptivepartition_m7}
corresponds to 
Fig.~\ref{fig:adaptivepartition_m5}
and shows an enhanced accuracy.
We apply GHC with degree $m=7$ 
to obtain a prior and posterior approximation,
where
the recombination degree $r=5$ and $\theta = 0.2\times \epsilon$ is used.
Our choice of $\tau$ is again such that $1/4 \sim 1/3$ of the particles
are allowed to leap to the next observation time.
The rmse errors 
(\ref{eq:rmse})
in the case of $R=10^{-2}$, $D=2$ and $\epsilon=10^{-2}, 10^{-3}$
are shown
in Fig.~\ref{fig:D2c}
and this
can be viewed as 
a result from
PCF and APCF with cubature on Wiener space of degree $m=7$.
Its performance is in fact one higher order improvement for both accuracy and recombination time
in view of Figs.~\ref{fig:D2a}, \ref{fig:D2b}.
From the simulation,
we expect
APCF with higher order cubature formula
can outperform Monte-Carlo approximations in any parameter regimes
including the ones
for which it used to be not so successful when it uses a low order cubature formula.
This further
highlights the strong necessity to find out cubature formula on Wiener space of degree $m \geq 7$ 
in order to solve the PDE or filtering problem
with high accuracy in a moderate dimension.

\section{Discussion}
\label{sec:discussion} 
In this paper we introduce a hybrid
methodology for the numerical resolution of the filtering problem which we
named the adaptive patched cubature filter (APCF). We explore some of its
properties and we report on a first attempt at a practical implementation.
The APCF combines many different \emph{methods}, 
each of which addresses a different part of the problem and has
independent interest. At a fundamental level all of the methods use high
order approaches to quantify uncertainty (cubature), and also to reduce the
complexity of calculations (recombination based on heavy numerical linear
algebra), while retaining explicit thresholds for accuracy in the individual
computation. The thresholds for accuracy in a stage are normally achievable
in a number of ways (e.g., small time step with low order, or large time step
with high order) and the determination of these choices depends on
computational cost. Aside from this use of the error threshold and choices
based on computational efficiency there are several other points to observe
in our development of this filter.

\begin{enumerate}
\item One feature is the surprising ease with which one can adapt the
computations to the observational data and so avoid performing unnecessary
computations. In even moderate dimensions (we work in $3+1$) this has a huge
impact for the computation time while preserving the accuracy we achieve for
the posterior distribution 
(Figs.~\ref{fig:Rd}, \ref{fig:Rf}, \ref{fig:Rh}, Figs.~\ref{fig:Dc}, \ref{fig:Df}, \ref{fig:Di}). 
It is an 
automated 
form of 
deterministic
high order importance
sampling which has wider application than the one explored in this paper,
for instance
it is used to deliver
accurate answers to PDE problems with piecewise smooth 
test function
in the example developed in
\cite{litterer2012high}.
\item Another innovation allowing a huge reduction in computation is the
ability to efficiently \emph{patch the particles} in the multiple dimensional scenario. Although
the problem might at first glance seem elementary, it is in fact the problem
of data classification. To resolve this problem we introduce an efficient
algorithm for data classification based on extending the Morton order to
floating point context. This method has now also been used effectively for
efficient function extrapolation 
\cite{wei}.
\item The KLV algorithm is at the heart of a number of successful methods
for solving PDEs in moderate dimension 
\cite{ninomiya2008weak}.
In each
case, something has to be done about the explosion of scenarios after each
time step; this in turn has to rely on and understanding the errors. In this
paper we take a somewhat different approach to the literature
\cite{lyons2004cubature}
in the way
we use higher order Lipschitz norms systematically to understand how well
functions have been smoothed, and to measure the scales on which they can be
well approximated by polynomials. This has the consequence that one can be
quite precise about the errors one incurs at each stage in the calculation.
In the end this is actually quite crucial to the logic of our approach since
an efficient method requires optimisation over several parameters -
something that is only meaningful if there are (at least in principle)
uniform estimates on errors. As a result of this perspective, we do not
follow the time steps and analytic estimates introduced 
by Kusuoka
in
\cite{kusuoka2004approximation}
although
we remain deeply influenced by balancing the smoothing properties of the
semigroup with the use of non-equidistant time steps.
\item The focus on Lipschitz norms makes it natural to apply an adaptive
approach to the recombination patches as well as to the prediction process.
In both cases we can be lead by the local smoothness of the likelihood
function as sampled on our high order high accuracy set of scenarios.
\item We have focussed our attention on the quality of the tail distribution
of the approximate posterior we construct. This is important in the
filtering problem because a failure to describe the tail behaviour of the
tracked object implies that one will lose the trajectory all together at
some point. These issues are particularly relevant in high dimensions as the
cost of increasing the frequency of observation can be prohibitive. If one
wishes to ensure reliability of the filter in the setting where there is a
significant discrepancy between the prior estimate and the realised outcome
over a time step then our APCF with cubature on Wiener space of degree $5$
already shows in the three dimensional example that it can completely
outperform sequential importance resampling Monte-Carlo approach. The
absence, at the current time, of higher order cubature formulae is in this
sense very frustrating as the evidence we give suggests that higher degree
methods will lead to substantial further benefits for both computation and
accuracy.
\end{enumerate}

In putting this paper together we have realised that there are many branches
in this algorithm that can be improved, in particular some parts of the
adaptive process and also the recombination (a theoretical improvement in
the order of recombination has recently been discovered
\cite{maria}).
There are also large parts that can clearly be parallelised.
We believe that there is ongoing scope for increasing the performance of the
APCF.

\section*{Acknowledgment}
The authors would like to thank the following institutions for their
financial support of this research. Wonjung Lee : NCEO project NERC and King
Abdullah University of Science and Technology (KAUST) Award No.
KUK-C1-013-04. Terry Lyons : NCEO project NERC, ERC grant number 291244 and
EPSRC grant number EP/H000100/1. The authors also thank the Oxford-Man
Institute of Quantitative Finance for its support.
The authors thank the anonymous referees for their helpful comments and suggestions,
which indeed contributed to improving the quality of the publication.

\bibliographystyle{siam.bst}
\bibliography{apcf}

\end{document}